%% file: arxiv-v3.tex
\documentclass{article}
\usepackage[a4paper, total={6in, 10in}]{geometry}
	
\newcommand{\R}{\mathbb{R}} 

\usepackage{placeins}
\usepackage{amsmath, amsthm, amssymb}
\usepackage{graphicx}
\usepackage{algorithm}
\usepackage{relsize}

\RequirePackage{etoolbox}
\usepackage{natbib}
\usepackage[colorlinks,citecolor=blue,urlcolor=blue,filecolor=blue,backref=page]{hyperref}
\usepackage{graphicx}
\usepackage[utf8]{inputenc}
\usepackage{enumerate}
\usepackage{verbatim}
\usepackage{booktabs}
\usepackage{mathrsfs}
\usepackage{algpseudocode}
\usepackage{color}
\usepackage{bbm,dsfont}
\usepackage{float}
\usepackage[utf8]{inputenc}
\usepackage{amsmath, amsthm, amssymb}
\usepackage{enumerate}

\usepackage{stackengine}
\stackMath
\newcommand\tsup[2][2]{%
	\def\useanchorwidth{T}%
	\ifnum#1>1%
	\stackon[-.5pt]{\tsup[\numexpr#1-1\relax]{#2}}{\scriptscriptstyle\sim}%
	\else%
	\stackon[.5pt]{#2}{\scriptscriptstyle\sim}%
	\fi%
}
\usepackage{verbatim}
\usepackage{graphicx}
\usepackage{algorithm}
\usepackage{algpseudocode}
\usepackage{listings}
\usepackage{subcaption}
\usepackage{hyperref}
\usepackage{float}
\usepackage{tikz}
\usepackage{authblk}

\usepackage[capitalize]{cleveref} 
\crefname{assumption}{assumption}{assumptions} 
\Crefname{condition}{Condition}{Conditions}
\Crefname{remark}{Remark}{Remarks}
\crefname{innercustomthm}{Proposition}{Propositions}

\usepackage{float}
\usepackage{tikz}
\usepackage{authblk}

\input{shortcuts}
\graphicspath{{../}}

\newtheorem{thm}{Theorem}
\newtheorem{definition}{Definition}

\newtheorem{lemma}{Lemma}
\newtheorem{assumption}{Assumption}

\newtheorem{innercustomthm}{Proposition}
\newtheorem{remark}{Remark}
\newtheorem{example}{Example}

\DeclareMathOperator*{\argmax}{arg\,max}

\date{}
\author[1]{Emilia Pompe\thanks{email: emilka.pompe@gmail.com}}
\author[2]{Mikołaj J. Kasprzak\thanks{kasprzak@essec.edu}}
\author[3]{Pierre E. Jacob\thanks{pierre.jacob@essec.edu}}
\affil[1]{\small  Department of Statistics, University of Oxford}
\affil[2,3]{\small Department of Information Systems, Data Analytics and Operations, ESSEC Business School}

\title{Asymptotics of cut distributions and robust modular inference using Posterior Bootstrap}

\begin{document}
\maketitle
\begin{abstract}
Bayesian inference provides a framework to combine various model components
with shared parameters, allowing joint uncertainty estimation and the use of
all available data sources.  Unfortunately, misspecification of any part of the
model might propagate to all other parts and can lead to unsatisfactory
results.  Cut distributions have been proposed as a remedy, where the
information is prevented from flowing along certain directions. We study cut
distributions from an asymptotic perspective and obtain a Bernstein-von Mises theorem, as well as a Laplace approximation with quantitative bounds.  We then propose an algorithm based on the Posterior
Bootstrap that delivers credible regions with the nominal frequentist asymptotic
coverage. The proposed methods are illustrated with numerical experiments in a variety of examples,
including causal inference with propensity scores.
\end{abstract}

\setcounter{tocdepth}{2}
\tableofcontents

\section{Introduction}\label{section:introduction_ch4}

\subsection{Cutting feedback in Bayesian models}

Statistical models are often built of different components or \emph{modules}, which
can represent different sources of information about the studied phenomenon.
Classically, Bayesian parameter inference relies on samples from the posterior distribution under the joint model combining all modules under consideration.
We refer to this as the standard Bayesian, or joint model approach.
This achieves a type of optimality when
the model is well-specified \citep{bissiri2016general}.  In
many applications, however, the analyst may have reasons to depart from the classic Bayesian approach.
This article focuses on a stepwise approach to parameter inference, termed \emph{modular inference}, 
in which some parameters are estimated within a module, and then plugged into the other modules. 
This approach aims at cutting the detrimental feedback that some modules may have onto others, while retaining 
a probabilistic approach to uncertainty propagation as in standard Bayesian inference.
Modular inference, and the associated \emph{cut posterior distributions}, have often been 
employed by practitioners and are receiving growing attention
since the observations of 
\citet{liu2009modularization,plummer2015cuts,jacob2017better}. Although still considered arguably 
unorthodox within Bayesian statistics, cut posterior distributions are a natural probabilistic counterpart to \emph{two-step} or \emph{two-stage} estimators which
are standard in econometrics \citep{vuong_1984,murphy2002estimation,newey1994large}.

Causal inference is a setting where modular Bayesian inference
is common. Consider the estimation of an average treatment effect from observational data. 
The statistician may account for confounding variables by adjusting
for the estimated probabilities of treatment assignment, referred to as propensity scores.
In a randomized trial, treatment would be assigned 
before the realization of the outcome, therefore it seems natural to estimate treatment assignment
probabilities without feedback from the outcome variable.
A first module thus estimates propensity scores given
covariates without using the outcome.
A second module estimates  the effect of treatment on the outcome, adjusting for
propensity scores estimated in the first module; see Section~\ref{section:causal_inference_example_v2}
for concrete examples. As discussed by \citet{zigler2013model,
saarela2016bayesian,StephensetalBA2023}, a joint modeling approach
may lead to biased results, whereas a modular approach provides some mitigation.

Similar situations arise in missing data problems, where
a first module imputes the missing values, and a second module
performs inference on the completed data set; see the archeological
example of \citet{carmona2020semi,yu2021variational}. Bayesian inference after multiple imputation,
which is accepted as a standard approach to missing data problems \citep{zhou2010note},
is indeed an example of a modular approach, as opposed to joint modeling.  
Modular Bayesian inference is also often applied to PK/PD
models, where the PK (pharmacokinetics) parts of the model are usually
believed to be better estimated without feedback from the PD
(pharmacodynamics) components \citep{lunn2009combining}.  Modular Bayesian inference has
also been used throughout epidemiology  \citep[e.g][]{nicholson2022interoperability}, 
in studies combining models of air pollution with models relating pollution
to adverse health outcomes \citep{blangiardo2011bayesian}, and in the analysis of computer
models \citep{liu2009modularization}.

Echoing its widespread use in practice,
modular Bayesian inference has received growing attention in the theoretical and methodological statistics literature. 
\citet{LiuGoudie2025} propose a formalism for inference with models made of arbitrary numbers of modules.
\citet{jacob2017better,FrazierNott2025} consider the question of choosing whether to employ modular or joint model inference.
Progress has been made on the computation of cut posteriors
using MCMC 
\citep{plummer2015cuts,liu2020stochastic,jacob2020unbiased} and 
variational inference \citep{carmona2020semi,yu2021variational}.
Recent articles investigate modular inference in specific important problems, such as semiparametric hidden Markov models in \citet{moss2024efficient}, or copula models in \citet{Smith2025}.

Cutting feedback is a way of addressing model misspecification, in situations where the user has some understanding of the potential flaws of their model components. 
Cutting feedback can be used in combination with other approaches
to handle model misspecification and data contamination \citep[e.g.][for a review]{NottDrovandiFrazier2024}.
For example, cutting feedback can be combined with generalized Bayesian inference
as in \citet{FrazierNott2024,tan2025optimization};
see also \cref{section:example_section_nonasymptotic}. Our theoretical framework accommodates general additive functions instead of log-likelihoods. We emphasize that cutting feedback has a distinct effect compared to modifying loss functions: its goal is to modify the information flow between modules.

\subsection{Main questions\label{sec:mainquestions}}

We consider models made of two modules; the setting is generalized to arbitrary numbers of modules in \citet{LiuGoudie2025}.
\begin{figure}[ht]
  \centering
  \includegraphics[width=0.35\textwidth]{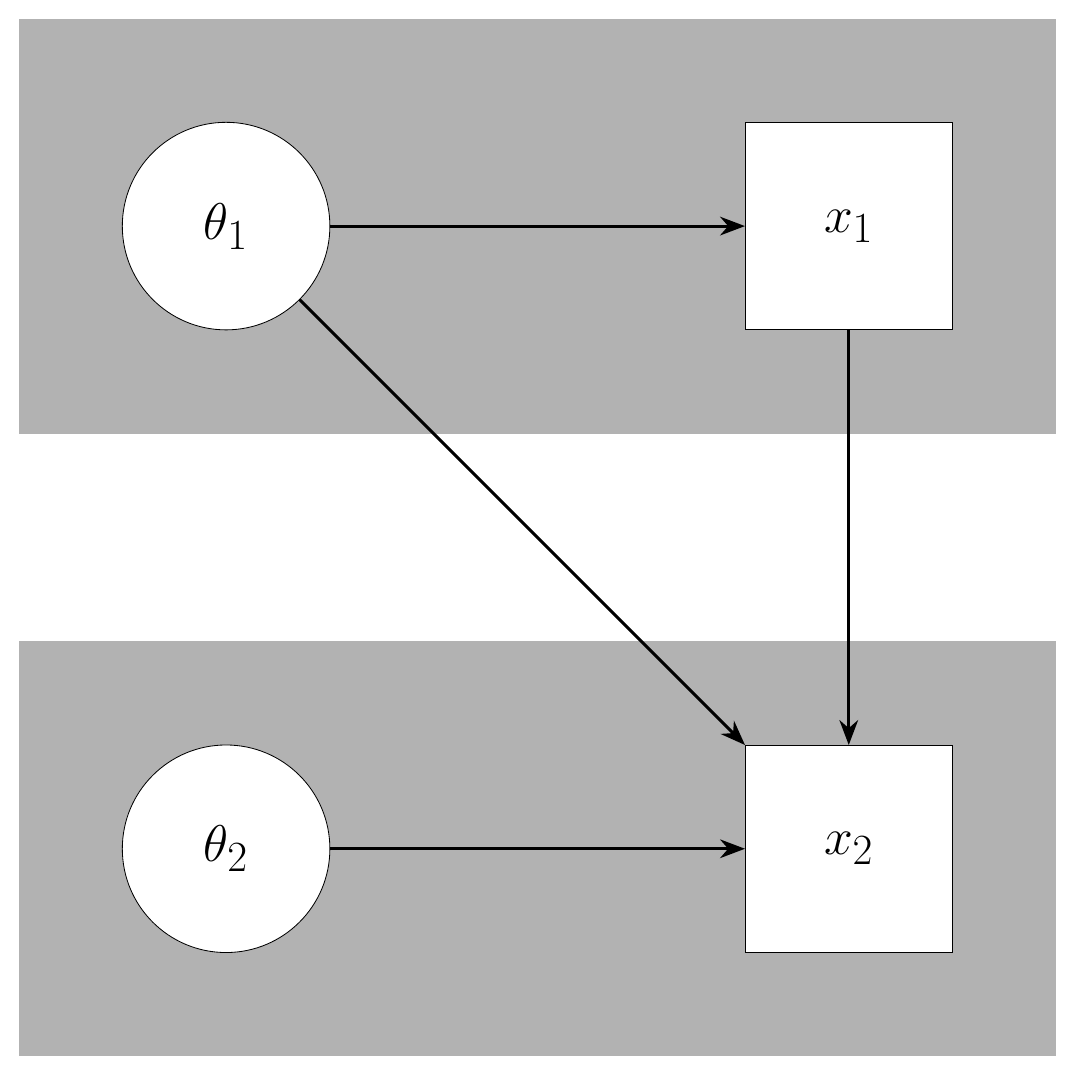}
  \caption{Graphical representation of the model \eqref{eq:general_model}, made of two modules. The parameters are $\theta_1$ and $\theta_2$, and the data sets are $x_1$ and $x_2$.}
  \label{fig:modeldiagram}
\end{figure}
Figure~\ref{fig:modeldiagram} represents a DAG with $\theta_1\in\Theta_1$ and $\theta_2\in\Theta_2$ as parameters, $x_{1}\in \mathbb{X}_1^{n_1}$ and $x_{2} \in \mathbb{X}_2^{n_2}$ as data sets composed of the measurements of $n_1$ and $n_2$ units respectively. The edges represent the possible dependencies.
Prior distributions $\pi_i$ are specified on $\theta_i$, for $i = 1,2$, and likelihood functions are denoted by $\mathcal{L}_i$. 
Thus the analyst specifies the following model, which may 
differ from the data-generating distribution denoted by $p^*$:
\begin{equation}
\label{eq:general_model}
\begin{aligned}
x_{1} & \sim  \mathcal{L}_1(x_1|\theta_1),    &\theta_1 &\sim \pi_1(\theta_1), \quad \theta_1 \in \Theta_1 \subseteq \R^{d_1}, \quad x_1 \in \mathbb{X}_1^{n_1},\\
x_{2} & \sim  \mathcal{L}_2(x_2|\theta_1, \theta_2, x_1),   &\theta_2 &\sim \pi_2(\theta_2),\quad \theta_2 \in \Theta_2 \subseteq \R^{d_2}, \quad x_2 \in \mathbb{X}_2^{n_2}.
\end{aligned}
\end{equation}

The full posterior distribution associated with the joint model \eqref{eq:general_model} is given by
\begin{align}\label{eq:full_posterior}
\pi(\theta_1, \theta_2  |x_{1}, x_{2}) & \propto   \pi_1(\theta_1)  \mathcal{L}_1(x_{1}|\theta_1) \; \pi_2(\theta_2) \mathcal{L}_2(x_{2}| \theta_1, \theta_2, x_{1})\\
&  \propto \pi(\theta_1|x_{1},x_{2}) \pi(\theta_2|\theta_1,x_{1},x_{2}).\nonumber
\end{align}
The marginal posterior distribution of $\theta_1$ under \eqref{eq:full_posterior} can be rewritten
\begin{align}
\pi(\theta_1|x_{1}, x_{2}) &\propto \pi(\theta_1|x_{1})\, \pi(x_{2}|\theta_1), \\
\text{with} &\quad  \pi(\theta_1|x_{1}) \propto \pi_1(\theta_1)\mathcal{L}_1(x_1|\theta_1), \quad \pi(x_{2}|\theta_1)= \int \mathcal{L}_2(x_{2}| \theta_1, \theta_2, x_{1})\, \pi_2(\mathrm{d}\theta_2).
\end{align}
The last term is called the feedback term of the second module on the first parameter $\theta_1$. 

The modular approach defines the \emph{cut posterior} \citep{liu2009modularization,plummer2015cuts} as 
\begin{equation}\label{eq:cut_posterior}
\pi_{\text{cut}}(\theta_1, \theta_2) \propto  \frac{\pi(\theta_1, \theta_2|x_{1}, x_{2})}{\pi(x_{2}|\theta_1)} \propto \pi(\theta_1|x_1) \pi(\theta_2|\theta_1,x_1,x_2).
\end{equation}
Under the cut posterior, the inference on $\theta_1$ is based only on the first module, so the feedback from the second module to $\theta_1$ is cut: 
$\pi_{\text{cut}}(\theta_1) = \pi(\theta_1|x_{1}) \neq \pi(\theta_1 | x_{1}, x_{2})$.
That first marginal is a standard posterior distribution. On the other hand, the joint distribution of $(\theta_1, \theta_2)$ under the cut posterior
is not standard, and its asymptotic behavior can be markedly different, starting with the fact that the cut posterior \eqref{eq:cut_posterior} and the standard posterior \eqref{eq:full_posterior} may concentrate on different regions of the parameter space.

\subsection{Contributions\label{subsec:contribution}}

Our contributions concern the asymptotic behavior of the cut posterior, 
with consequences on the frequentist coverage of associated credible sets, and two numerical methods:
the Laplace approximation and the Posterior Bootstrap.
In Section~\ref{section:asymptoticsmodularinference} we describe 
the setting, with i.i.d. data and two parametric modules. Similar questions arise with dependent data and semi- or non-parametric settings are of interest \citep[e.g.][]{moss2024efficient},
but we consider a simpler setting here, and leave extensions to future work.
We state a Bernstein-von Mises (BvM) theorem for cut posteriors along with an explicit expression of the asymptotic variance. 
This result contributes to the literature on Bayesian asymptotics under model misspecification \citep[e.g.][]{kleijn2012bernstein,bochkina2019bernstein,miller2021asymptotic}. 
In Section~\ref{section:nonasymptotic_control} we provide a Laplace approximation to the cut posterior, which we refer to as Cut-Laplace, along with a quantitative control,
echoing recent developments made in the standard setting of joint model inference \citep{JMLR:v26:24-0619,katsevich:2025}.
In Section~\ref{section:pbmi} we propose Posterior Bootstrap for Modular Inference (PBMI), an alternative numerical approach to 
modular Bayesian inference
in the vein of the Weighted Likelihood Bootstrap \citep{newton1994approximate}
and its recent extensions
\citep[e.g.][]{lyddon2019general,fong2019scalable,
nie2023bayesian,pompe2021introducing}.
PBMI relies only on the ability to perform optimization of the log-posterior densities in each module.
The method produces a distribution on parameter space that concentrates on the same region 
as the cut posterior, but with a different asymptotic covariance.
Section~\ref{section:illustrations_modified} compares cut posteriors, Laplace approximations and PBMI
in various settings. Section~\ref{section:discussion}
provides guidelines on the choice of modular inference method.

\section{Asymptotics of modular inference}\label{section:asymptoticsmodularinference}

\subsection{Setting and notation\label{section:two_scenarios}}

To describe asymptotic phenomena in their simplest form, 
we assume that $x_1$ and $x_2$ are both made of $n_1 = n_2 = n$ units,
and are realizations of a process $\{(X_{1,j},X_{2,j})\}_{j=1}^{\infty}$ generated i.i.d. from a probability distribution $p^*$, with marginals $p_1^*$ and $p_2^*$. 
We write $X_1=\{X_{1,j}\}_{j=1}^{\infty}$ and $X_2=\{X_{2,j}\}_{j=1}^{\infty}$ and denote by $P^*$ the law of $(X_1,X_2)$.
Section~\ref{section:scenario_indep} will extend the results to an alternative setting where $n_1 \neq n_2$, assuming that the two data sets are independent under both the model and the data-generating process.

The model specifies a prior $\pi_1(\theta_1)\pi_2(\theta_2)$ and assumes generalized log-likelihoods:
\begin{align}\label{eq:loglikelihoods}
   \loglikelihood_1(\theta_1|x_1)=\sum_{j=1}^{n}\log f_1(x_{1,j}|\theta_1);\qquad\loglikelihood_2(\theta_1,\theta_2|x_{1},x_2)=\sum_{j=1}^{n}\log f_2(x_{2,j}|\theta_1,\theta_2,x_{1,j}),
\end{align}
where $-\log f_1$ and $-\log f_2$ could be negative log-likelihoods, in which case $\mathcal{L}_i$ in \cref{sec:mainquestions}
corresponds to exponential of $\loglikelihood_i$; we may also consider loss functions $L_i$ that are not log-likelihoods, as in \citet{miller2021asymptotic}.
For any $\theta_1\in \mathbbm{R}^{d_1}, \theta_2\in \mathbbm{R}^{d_2}$, we let
\begin{equation}\label{eq:cutloglikelihood}
  \loglikelihood(\theta_1,\theta_2|x_1,x_2)=\loglikelihood_1(\theta_1|x_1)+\loglikelihood_2(\theta_1,\theta_2|x_1,x_2)-F(\theta_1|x_1,x_2),
\end{equation}
where $
F(\theta_1|x_1,x_2)=\log\int\exp[\loglikelihood_2(\theta_1,t|x_1,x_2)]\pi_2(t)dt$ is the log-feedback term. 

We allow $f_2$ to depend on $x_{1,j}$ in order to cover common situations where a first module generates a tabular data set, for example
by pre-processing columns, imputing missing data or generating regressors such as propensity scores, 
and a second module handles the analysis of the generated data. 
This is typical of two-stage estimation settings in econometrics \citep{newey1994large} and causal inference
\citep[e.g.][]{zigler2013model}. For example, the second module can be a regression 
of an outcome $(x_{2,j})$ onto some fitted values $(x_{1,j}^T \theta_1)$ obtained in the first module.

For any function $g:\mathbbm{R}^{d_1}\times\mathbbm{R}^{d_2}\to\mathbbm{R}$, we write for $i,j\in\{1,2\}$, 
\[\partial_i g(\theta_1,\theta_2)=\frac{\partial g}{\partial\theta_i}(\theta_1,\theta_2);\qquad\partial_{ij} g(\theta_1,\theta_2)=\frac{\partial^2 g}{\partial\theta_i\partial\theta_j^T}(\theta_1,\theta_2).
\]
For functions $f_1$ and $f_2$ as in \cref{eq:loglikelihoods}, 
the derivatives always are with respect to $\theta_i$ (and not $x_i$).
For example, $\partial_{ij} \log f_2(x_2|\theta_1,\theta_2,x_1)$
refers to the matrix of second partial derivatives of $\log f_2$ with respect to $\theta_i$ and $\theta_j$. Primes and double primes are used for gradients and Hessian with respect to all parameters, e.g. $L''(\theta)$ is the Hessian matrix of $L$ with respect to $\theta=(\theta_1,\theta_2)$.

\subsection{Asymptotics of two-step M-estimators\label{sec:asymptoticMestimators}}

The asymptotic behavior of two-step maximum likelihood estimators, or more generally two-step M-estimators, has
been well-studied. We recall some results to facilitate comparisons below.
The M-estimator in the joint model is defined as 
\begin{equation}\label{eq:joint_M}
\theta_{\mathrm{M}} = (\theta_{1, \mathrm{M}},\theta_{2, \mathrm{M}}) = \argmax_{(\theta_1, \theta_2)\in \Theta_1\times \Theta_2} \{\loglikelihoodone(\theta_1|x_{1}) + \loglikelihoodtwo (\theta_1, \theta_2| x_{1}, x_{2})\}.  
\end{equation}
Under regularity conditions that can be checked in a variety of parametric statistical models, the M-estimator converges to a vector $\theta^0$, 
and it is asymptotically normal with mean $\theta^0$ and covariance matrix given by the sandwich formula \citep{huber1967behavior}, see \citet[Chapter 5]{van2000asymptotic} for some sufficient conditions.
Under similar conditions the posterior distribution \eqref{eq:full_posterior} concentrates around the M-estimator,
and, denoting by $\theta$ a draw from the posterior distribution, the distribution of $\sqrt{n}(\theta - \theta_{\rm M})$ converges to a Normal distribution with mean 0 and covariance matrix $(J^{0})^{-1}$, where $J^0$ is
the Fisher information matrix evaluated at the limit $\theta^0$ \citep[e.g.][]{muller2013risk,miller2021asymptotic}.

Our primary interest is in understanding how these phenomena manifest in modular settings. 
We define the two-step M-estimator (2SM) $(\hat{\theta}_1,\hat{\theta}_2)$ as
\begin{align}\label{eq:2sM}
\hat{\theta}_1 = \argmax_{\theta_1 \in \Theta_1} \loglikelihoodone(\theta_1|x_{1}), \quad \hat{\theta}_2 = \argmax_{\theta_2 \in \Theta_2}\loglikelihoodtwo (\hat{\theta}_1, \theta_2| x_{1}, x_{2}).
\end{align}
We define the pseudo-true parameter $\theta^*$ as follows:
\begin{align}\label{eq:pseudotrue}
{\theta}_1^* = \argmax_{\theta_1\in\Theta_1} \mathbbm{E}_{x_1 \sim p_1^*} \loglikelihoodone(\theta_1|x_{1}), \quad {\theta}_2^* = \argmax_{\theta_2\in\Theta_2} \mathbbm{E}_{(x_1,x_2)\sim p^*}  \loglikelihoodtwo (\theta_1^*, \theta_2| x_{1}, x_{2}).
\end{align}
In general, $\hat{\theta} \neq \theta_M$ and $\theta^* \neq \theta^0$.
The behavior of the 2SM estimator has been well-studied \citep{murphy2002estimation,vuong_1984}. A simple way to obtain its asymptotic distribution, as argued in \citet[Section 6]{newey1994large}, is to assume differentiability of the log-likelihoods and to view 2SM as a regular Z-estimator,
stacking the two defining equations of the form $\Psi_1(\theta_1) = 0$
and $\Psi_2(\theta_1,\theta_2) = 0$ in a single equation,
and then employing a classical result on the asymptotic normality of 
Z-estimators under model misspecification \citep[e.g.][Section 5.3]{van2000asymptotic}.
This gives 
\begin{equation}
  \sqrt{n}(\hat\theta - \theta^*) \xrightarrow[n\to\infty]{d} \mathrm{Normal}(0, \Sigma), \quad \text{with}\quad \Sigma = \begin{pmatrix}
  \Sigma_{1,1}  & \Sigma_{1,2}\\
    \Sigma_{1,2}^T & \Sigma_{2,2}
  \end{pmatrix},\label{eq:asymptotic2sM}
\end{equation}
where $\Sigma_{1,1} = J_1^{*-1} I_1^* J_1^{*-1}$, 
$\Sigma_{1,2} =  (J_1^{*-1}R_I^* - \Sigma_{1,1} R_J^*) J_2^{*-1}$, 
$S_2 = {R_I^*}^T (J_1^*)^{-1} R_J^* + {R_J^*}^T (J_1^*)^{-1} R_I^*$,
and 
$\Sigma_{2,2} =  (J_2^*)^{-1} \{I_2^* + {R_J^*}^T \Sigma_{1,1} R_J^* - S_2\}(J_2^*)^{-1}$, with the following definitions: 
\begin{align}
    &I_1^* = \mathbbm{E}_{X_{1,1}\sim p_1^*}\left[\partial_1\log f_1(X_{1,1}|\truepar_1) \cdot \partial_1 \log f_1(X_{1,1}|\truepar_1)^T   \right]\notag,\\
    &I_2^* = \mathbbm{E}_{(X_{1,1},X_{2,1})\sim p^*}\left[\partial_2\log f_2(X_{2,1}|\,\truepar_1, \truepar_2, X_{1,1}) \cdot \partial_2\log f_2(X_{2,1}|\,\truepar_1, \truepar_2, X_{1,1})^T   \right]\notag,\\
    &R_I^* = \mathbbm{E}_{(X_{1,1},X_{2,1})\sim p^*}\left[\partial_1\log f_1(X_{1,1}|\truepar_1) \cdot \partial_2\log f_2(X_{2,1}|\,\truepar_1, \truepar_2, X_{1,1})^T   \right]\notag,\\
    &J_1^*=-\mathbbm{E}_{X_{1,1}\sim p_1^*}\left[\partial_{11}\log f_1(X_{1,1}|\truepar_1)   \right]\notag,\\
    &J_2^*=-\mathbbm{E}_{(X_{1,1},X_{2,1})\sim p^*}\left[\partial_{22}\log f_2(X_{2,1}|\truepar_1,\truepar_2,X_{1,1})   \right]\notag,\\
    &R_J^*=-\mathbbm{E}_{(X_{1,1},X_{2,1})\sim p^*}\left[\partial_{12}\log f_2(X_{2,1}|\truepar_1,\truepar_2,X_{1,1})   \right].\label{eq:j_defs}
\end{align}

\subsection{Bernstein-von Mises for the cut posterior}\label{section:both_scenarios_bayesian}

Theorem~\ref{thm:bvm_cut} presents the equivalent of the
Bernstein-von Mises (BvM) theorem for cut posteriors, under the assumption of i.i.d. observations and regular conditions on the parametric modules.
These standard assumptions are fully stated in Appendix~\ref{suppl:assumptions_bvm}.
In short, they include the following: the prior densities and loss functions are differentiable in a neighborhood of the 2SM estimator, which is consistent, the global mode of the cut posterior density is well-separated, and Hessians of the loss functions are negative positive definite at the 2SM estimator in the limit.
Appendix~\ref{suppl:verifying_assumptions} discusses the verification of these assumptions in a simple parametric model and \cref{section:example_section_nonasymptotic} discusses a more mathematically intricate model satisfying the assumptions.
An informal but intuitive proof is provided in Appendix~\ref{suppl:informal_proof_bvm} and a formal proof is in Appendix~\ref{suppl:proof_bvm}-\ref{suppl:assumptions_bvm}-\ref{suppl:bvm_interpretable_and_proofs}.

\begin{thm}\label{thm:bvm_cut}
Let $(\mle_1,\mle_2)$ be the 2SM-estimator in \eqref{eq:2sM} and
suppose that $\mle\xrightarrow{n\to\infty}\theta^*$ in $P^*$-probability, 
where $\theta^*$ is defined in \eqref{eq:pseudotrue} and lies in the interior of $\Theta_1\times\Theta_2$.
Assume \ref{assump:priors_asymptotic}-\ref{assump:integrability_asymptotic}. Define
\begin{align}
  H&=
\begin{pmatrix}
J_1^*+R_J^*(J_2^*)^{-1}(R_J^*)^T & R_J^*\\
(R_J^*)^T & J_2^*
\end{pmatrix}\nonumber\\
&=\begin{pmatrix}
      (J_1^*)^{-1}& - (J_1^*)^{-1}R_J^*(J_2^*)^{-1}\\
      - (J_2^*)^{-1}(R_J^*)^T(J_1^*)^{-1} & (J_2^*)^{-1}+ (J_2^*)^{-1}(R_J^*)^T(J_1^*)^{-1}R_J^*(J_2^*)^{-1}
  \end{pmatrix}^{-1},\label{eq:Hinverse}
\end{align}
where $J_1^*, J_2^*, R_J^*$ are given by \eqref{eq:j_defs}. Then,
denoting by $\tilde{\pi}_{\rm cut}$ the law of $\sqrt{n}(\theta-\hat{\theta})$ where $\theta \sim \pi_{\rm cut}$ is defined in \eqref{eq:cut_posterior}, and by TV the total variation distance,
we have
\[
\mathrm{TV}\left(\tilde{\pi}_{\rm cut},\mathrm{Normal}(0,H^{-1}) \right) \xrightarrow[n\to\infty]{P^*}0.
\]
\end{thm}

The top left entry of $H^{-1}$ is $(J_1^*)^{-1}$ as expected, since the first marginal of the cut posterior corresponds to the standard posterior in the first model for which BvM is well-established \citep{miller2021asymptotic}. The other entries of $H^{-1}$ are not a priori obvious.

The asymptotic variance $H^{-1}$ in Theorem~\ref{thm:bvm_cut} is in general
different from the asymptotic variance $\Sigma$ of the 2SM estimator in \eqref{eq:asymptotic2sM}.
Equality occurs when $I_1^* = J_1^*$, $I_2^* = J_2^*$ and $R_I^* = 0$, 
e.g. when $X_{1,1}$ and $X_{2,1}$ are independent under $p^*$.
In that case, $H^{-1}$ matches $\Sigma$,
and thus credible regions derived from the cut posterior at level $1-\alpha$ have 
the nominal $1-\alpha$ frequentist coverage for $\theta^*$.
In general, the coverage might
be smaller or larger than the nominal value.
Even if both $I_1^* = J_1^*$ and $I_2^* = J_2^*$, the presence of $R_I^*$ in \eqref{eq:asymptotic2sM} incurs a difference between $H^{-1}$ and $\Sigma$, both in the covariance between $\theta_1$ and $\theta_2$, and in the variance of $\theta_2$.
By design, inference on $\theta_1$ with the cut posterior is robust to misspecification of the second module; in other words, the condition $I_1^* = J_1^*$ is sufficient for the cut posterior credible regions for $\theta_1$ to have nominal frequentist coverage.

\citet{FrazierNott2024} consider the BvM theorem for the second module given a fixed value of $\theta_1$.
Their Normal approximation has both mean and variance that depend on $\theta_1$. On the other hand, the joint Normal limit in Theorem~\ref{thm:bvm_cut} 
has a conditional distribution of $\theta_2$ given $\theta_1$ which
 variance which does not depend on $\theta_1$. 
\citet{FrazierNott2024} view this as a limitation of joint 
BvM theorems, but it is arguably a feature of multivariate Normal distributions. The apparent discrepancy is resolved if the 
$\theta_1$-dependent conditional variance of \citet{FrazierNott2024} 
is asymptotically equivalent to one that does not depend on $\theta_1$. Higher-order asymptotic expansions could make the $\theta_1$-dependency explicit, and the dependency elicited in \citet{FrazierNott2024} could correspond to these higher-order terms, but this remains an open question.

Finally we comment on the bottom right entry of $H^{-1}$ in \eqref{eq:Hinverse}, which corresponds to the asymptotic variance of the second module marginally.
The uncertainty of the first module, captured by  $(J_1^*)^{-1}$, feeds into the second parameter's asymptotic variance. If $(J_1^*)^{-1}$ approaches zero, 
the entry would reduce to $(J_2^*)^{-1}$, the standard asymptotic variance in a model where $\theta_1$ is fixed at $\theta_1^*$. The additional variability induced by estimating $\theta_1$ is captured by the extra term $(J_2^*)^{-1}(R_J^*)^T (J_1^*)^{-1} R_J^* (J_2^*)^{-1}$. That term is also zero when $R_J^* = 0$, which occurs when the loss function of the second module is constant in $\theta_1$.

\subsection{Independent data sets of different sizes\label{section:scenario_indep}}

We consider a variant of the setting of \cref{section:two_scenarios},
where the two datasets $x_{1,1:n_1}$ and $x_{2,1:n_2}$ are of different sizes, and are assumed independent under $p^*$. Assume here that the loss function $f_2$ does not depend on $x_1$, and thus $L_2(\theta_1,\theta_2|x_1,x_2) = L_2(\theta_1,\theta_2|x_2)$.
We adapt the definitions in \eqref{eq:j_defs},
and note that $R_I^* = 0$ by independence and by the definition of $\theta^*$. 
Assume that $n_1$ and $n_2$ diverge at a certain relative rate: $n_1/n_2 \to \alpha$ as $n_1,n_2\to\infty$, for some $\alpha>0$. We obtain a version of Theorem~\ref{thm:bvm_cut} with asymptotic variance:
\begin{align}
  \tilde{H}^{-1}&=\begin{pmatrix}
      (\alpha J_1^*)^{-1}& - (\alpha J_1^*)^{-1}R_J^*(J_2^*)^{-1}\\
      - (J_2^*)^{-1}(R_J^*)^T(\alpha J_1^*)^{-1} & (J_2^*)^{-1}+ (J_2^*)^{-1}(R_J^*)^T(\alpha J_1^*)^{-1}R_J^*(J_2^*)^{-1}
  \end{pmatrix}.\label{eq:Hinverse_n1equalsn2}
\end{align}
This can be seen directly in the case where $\alpha$ or $1/\alpha$ is an integer. Indeed, if $n_1 = \alpha n_2$ with $\alpha \in \mathbb{N}$, rewrite the first module with data $\tilde{x}_{1,1:n_2}$ where each $\tilde{x}_{1,i}$
corresponds to $\alpha$ successive elements of $(x_{1,i})$, and the loss is $\tilde{f}_1(\tilde{x}_{1,i}|\theta_1) = \sum_{j=1}^\alpha f_1(x_{1,(i-1)\alpha + j}|\theta_1)$. Then $J_1^*$ is replaced by $\alpha J_1^*$ in \eqref{eq:Hinverse}, and the matrices $J_2^*$ and $R_J^*$ are unchanged, giving \eqref{eq:Hinverse_n1equalsn2}.

\begin{example}\label{ex:biased_data}
    This covers the biased data setting of \citet{liu2009modularization}, 
    in which we have access to a small set of good quality data,
$x_{1,1:n_1} \sim \mathrm{Normal}(\theta_1,1)$, and the interest is in estimating the location $\theta_1$. We
also have access to a large but poor quality dataset $x_{2,1:n_2}$. Since we
suspect that these measurements are biased, we specify
$x_{2,1:n_2} \sim \mathrm{Normal}(\theta_1 + \theta_2,1)$, where
$\theta_2$ represents the suspected bias. Depending on the priors 
on $\theta_1$ and $\theta_2$, the estimation of $\theta_1$ can be more accurate by ignoring the second module \citep{liu2009modularization,jacob2017better}.
\citet[Section 4]{nicholson2022interoperability} encounter such a situation in disease surveillance using both randomized and observational data.
\end{example}

\section{Laplace approximation and its error\label{section:nonasymptotic_control}}

The BvM in Theorem~\ref{thm:bvm_cut} suggests a computationally convenient
approximation of the cut posterior by a multivariate Normal distribution, which is known as the Laplace approximation.
We construct the approximation in \cref{section:nonasymptotic_control:construction}, and
provide a non-asymptotic bound on the error
in \cref{section:nonasymptotic_control:statement}. The bound is made explicit with a concrete example in \cref{section:example_section_nonasymptotic}. In this section, we simplify notation by dropping $x_1,x_2$, e.g. we rewrite $L(\theta_1,\theta_2|x_1,x_2)$ in \eqref{eq:cutloglikelihood} as $L(\theta_1,\theta_2)$.

\subsection{Construction of the Laplace approximation\label{section:nonasymptotic_control:construction}}

The Laplace approximation is defined as the multivariate Normal distribution
with mean given by the 2SM-estimator $\hat{\theta}$ in \cref{eq:2sM} 
and covariance matrix $J_n(\mle)^{-1}/n$, where $J_n(\mle)$ is obtained in \cref{eq:j_n_simplified} by plugging empirical approximations of $J_1^*, R_J^*, J_2^*$ in the expression of $H^{-1}$ in \eqref{eq:Hinverse}, and by additionally  taking into account the contribution of the prior.
Using the notation $\logposterior_1(\theta_1):=\loglikelihood_1(\theta_1)+\log\pi_1(\theta_1)$, $\logposterior_2(\theta_1,\theta_2):=\loglikelihood_2(\theta_1,\theta_2)+\log\pi_2(\theta_2)$,  we define
\begin{align}
     J_{n}(\mle)=&-\frac{1}{n}
    \begin{bmatrix}
\partial_{12}\logposterior_2(\mle)\left(\partial_{22}\logposterior_2(\mle)\right)^{-1}\partial_{12}\logposterior_2(\mle)+\partial_{11}\logposterior_1(\mle_1)&\partial_{12}\logposterior_2(\mle)\\
    	\partial_{12}\logposterior_2(\mle)^T&\partial_{22}\logposterior_2(\mle)
    \end{bmatrix}.\label{eq:j_n_simplified}
\end{align}

The following theorem provides a theoretical justification for the approximation of the cut posterior $\pi_{\text{cut}}$ in \eqref{eq:cut_posterior} by $\mathrm{Normal}(\mle,J_n(\mle)^{-1}/n)$. We remark that the intractability in the feedback term in \cref{eq:cutloglikelihood} does not prevent us from obtaining a Laplace approximation
using only derivatives of $\bar{L}_1$, $\bar{L}_2$. Specifically, we upper-bound the difference between minus the Hessian of $-\log\pi_{\rm cut}$ defined in \cref{eq:cut_posterior} and $J_{n}(\mle)$ in \cref{lem:covariance_comparison_new} in \cref{sec:non_asymptotic_gradient}.


\subsection{Error of the Laplace approximation\label{section:nonasymptotic_control:statement}}

We have the following result, proved in \cref{sec:proof_of_non_asymptotic_simplified}.

\begin{thm}\label{thm:non_asymptotic_simplified}
Suppose that $d_1$ and $d_2$ are independent of $n$ and $J_{n}(\mle)$ is given by \cref{eq:j_n_simplified}.
Suppose that, for some fixed and deterministic $\delta_1,\delta_2,\delta_3,m,\lambda,K,K',K''>0$, the probability that \cref{assump:der_integrability_new_simplified,assump:fisher_inform_simplified,assump:m_estimator_simplified,assump:priors11_simplified,assump:third_der_modules_simplified,assump:third_der_simplified} hold jointly 
is at least $\tilde{p}(n)\in[0,1].$ Moreover, suppose that  \cref{assump:tails_simplified} is satisfied. 
Then, there exist $\bar{\delta}_2,\tilde{\delta}_1>0$ and a fixed $C(d_1,d_2)>0$, independent of $n$, such that, 
\begin{align}\label{eq:tv_bound_simplified}
    P^*\left[\mathrm{TV}\left(
        \tilde{\pi}_{\rm cut}\, ,\,
            \mathrm{Normal}\left(0,J_{n}(\mle)^{-1}\right)
    \right)\leq C(d_1,d_2)n^{-1/2}\right]\geq p(n,d_1,d_2),
\end{align}
where $P^*$
is the probability with respect to the data-generating process, $
p(n,d_1,d_2)=\tilde{p}(n)+p_1(\tilde{\delta}_1,n)+p_2(\bar{\delta}_2,n)-2,
$
and $p_1(\tilde{\delta}_1,n)$, $p_2(\bar{\delta}_2,n)$ are given in \cref{assump:tails_simplified}.
\end{thm}

The result implies that, if $\tilde{p}(n)$, $p_1(\tilde{\delta}_1,n)$
and $p_2(\bar{\delta}_2,n)$ go to one as $n\to\infty$,
then the error measured in TV between the cut posterior and the Laplace approximation goes to zero in probability, at rate $n^{-1/2}$, with a constant $C(d_1,d_2)$ that depends on the dimension but not on $n$.
The main appeal of the Laplace approximation lies in its low computational cost relative to the cut posterior. Just as with the cut posterior, the Laplace approximation is expected to yield credible regions that do not have nominal frequentist coverage under model misspecification.

\begin{remark}
    \cref{thm:non_asymptotic_simplified} is a simplified version of the full non-asymptotic result in \cref{thm:non_asymptotic_final}, which involves explicit constants. When applied to a specific model and dataset,  \cref{thm:non_asymptotic_final} yields an explicit dimension dependence of the bound, as in \cref{section:example_section_nonasymptotic}.
\end{remark}

\begin{remark}
    The dimension dependence of the bound in  \cref{thm:non_asymptotic_simplified} and \cref{thm:non_asymptotic_final} depends heavily on the model. We believe that various quantities
    appearing in \cref{assump:der_integrability_new_simplified,assump:fisher_inform_simplified,assump:m_estimator_simplified,assump:priors11_simplified,assump:third_der_modules_simplified,assump:third_der_simplified,assump:tails_simplified} can only be controlled on a case-by-case basis; for example,
 the eigenvalues of matrices $-\frac{\loglikelihood''(\mle)}{n}$, $-\frac{\loglikelihood_1''(\mle_1)}{n}$ and $-\frac{\loglikelihood_2''(\mle)}{n}$ may or may not
    be lower bounded uniformly in $d_1,d_2$.
    %
    Related discussions can be found in \cite{katsevich_spokoiny2025} in the context of Bayesian inverse problems. 
\end{remark}

\begin{remark}
    In the context of Section~\ref{section:scenario_indep}, where the asymptotic variance involves a ratio $\alpha = \lim n_1/n_2$, the matrix $J_{n}(\mle)$ in \cref{eq:j_n_simplified} should be replaced by the corresponding $J_{n_2}(\mle)$ which results from replacing $n$ with $n_2$.
\end{remark}

Non-asymptotic bounds on the quality of the Laplace approximation of a standard Bayesian posterior have been obtained in \citet{JMLR:v26:24-0619,katsevich:2025}. Our bound of  \cref{thm:non_asymptotic_simplified} (and the full Theorems~\ref{thm:without_correction1} and \ref{thm:non_asymptotic_final} in the appendix)
is similar in nature, yet our proofs are more involved for the following reasons:
\begin{enumerate}[(1)]
\item With $L$ in \cref{eq:cutloglikelihood}, $L'(\mle)$ is not zero in general, and thus it must be bounded.
\item  The Hessian ${(-\log\pi_{\text{cut}})''(\mle)}/{n}$ cannot be used as the covariance in the Laplace approximation as it is intractable, due to the feedback term. We replace it with the matrix in \cref{eq:j_n_simplified} and bound the difference between the two matrices.
\item In order to make it easier to verify, our \cref{assump:tails_simplified} regulates separately the behavior of the log-likelihoods $\loglikelihood_1$ and $\loglikelihood_2 $ away from $\mle$, without regulating explicitly the behavior of  $L$ of \cref{eq:cutloglikelihood}. This means that upper bounding the tail integral of the density of $\tilde{\pi}_{\rm cut}$ is more challenging.
\end{enumerate}
Our proof techniques could also obtain bounds on distances other than the total variation distance, such as the 1-Wasserstein distance as in \citet{JMLR:v26:24-0619}.

\subsection{Explicit error bound in a biased data setting\label{section:example_section_nonasymptotic}}

We obtain an expression for the error of the Laplace approximation in a model which is misspecified in both modules and involves a generalized likelihood in the first module. We derive high-probability error bound, given in \cref{prop:biased_data}. We also show in \cref{sec:verify_assumptions_example} that the model in \cref{ex:biased} satisfies the assumptions of \cref{thm:bvm_cut}.

\begin{example}\label{ex:biased}
We revisit \cref{ex:biased_data}, replacing the first log-likelihood by the following loss:
\[
\loglikelihood_{1}(\theta_1) =
-\alpha\sqrt{1+\alpha^2}\sum_{j=1}^{n_1}\left(\sqrt{1+\left(\frac{\|x_{1,j}-\theta_1\|}{\alpha}\right)^2}-1\right),
\]
for $\alpha>0$, as in \cite{chakraborty2024}.
This loss interpolates between the Gaussian log-likelihood ($\alpha\to\infty)$ and Laplace log-likelihood ($\alpha\to 0^+$).
Let
\begin{align*}
\loglikelihood_2(\theta_1,\theta_2)=-\frac{n_2d_2}{2}\log(2\pi)-\frac{1}{2}\sum_{i=1}^{n_2}\|x_{2,i}-\theta_1-\theta_2\|^2,
\end{align*}
and consider the Gaussian priors:
\(
\pi_1(\theta_1) \propto e^{-\|\theta_1\|^2/2}
\)
and 
\(
\pi_2(\theta_2)\propto e^{-50\|\theta_2\|^2}
\).
We let $n_2=n$ and $n_1=\vartheta n$ for some fixed $0<\vartheta\leq 1$ independent of $n$. Naturally, $d_1=d_2$. Moreover, for simplicity, assume that ${d_2^{11/2}}/{n}\to 0$ and $d_2>\frac{\alpha^2}{5}$.  Furthermore, assume that the data modeled by the first module are generated i.i.d. from the standard multivariate Gaussian distribution. The data modeled by the second module are assumed to come from a sub-Gaussian distribution with mean zero and an isotropic covariance matrix. Hence, the first module is misspecified and the second one is allowed to be misspecified. 
    \end{example}

The following result specializes Theorem~\ref{thm:non_asymptotic_simplified}
to this setting, and illustrates that clean bounds on the error of the Laplace approximation can be obtained with our approach. The proof of \cref{prop:biased_data} is given in \cref{sec:proof_example_biased}.

\begin{innercustomthm}\label{prop:biased_data}
    In the setup of \cref{ex:biased}, there exist constants $c_1,c_2,c_3>0$ independent of $n$ and $d_2$, such that, for all $n\geq 1$,
    \begin{align*}
    &P^*\Bigg[\mathrm{TV}\left(
        \tilde{\pi}_{\rm cut} ,\,
            \mathrm{Normal}\left(0,J_{n}(\mle)^{-1}\right)
    \right)\leq c_1\left(n^{d_2}\exp\left[-c_2\,n\,d_2^{-9/2}\right]+d_2n^{-1/2}\right)\Bigg]
    \geq 1-\frac{c_3}{n}.
\end{align*}
\end{innercustomthm}

\begin{remark}
    We note that the bound in \cref{prop:biased_data} converges to zero if 
    \[
    n^{d_2}\exp\left[-c_2nd_2^{-9/2}\right]\xrightarrow{n_2\to\infty}0\quad\text{and}\quad d_2n^{-1/2}\xrightarrow{n_2\to\infty}0,
    \]
    which holds if and only if
    \(
    {d_2^{11/2}\log n}/{n}\xrightarrow{n_2\to\infty}0.
    \)
    If that is the case, then our bound is of order $d_2n^{-1/2}.$
    \end{remark}
\section{Posterior Bootstrap for Modular inference}\label{section:pbmi}

\subsection{Algorithm}\label{section:algorithms}

We introduce a methodology based on the Posterior
Bootstrap, which we call Posterior Bootstrap for Modular Inference (PBMI). 
The procedure, given in Algorithm~\ref{alg:pb_cut_dependent}, shares with the cut posterior
the goal of propagating uncertainty from the first to the second module
without allowing feedback. In the algorithm, $\text{Exp}(1)$ refers to the Exponential distribution with rate 1,  and $w_0,v_0$ are tuning parameters which we discuss in Section~\ref{section:incorporating_prior}.

\begin{algorithm}[!ht]
	\caption{Posterior Bootstrap for Modular Inference}\label{alg:pb_cut_dependent}
	\begin{algorithmic}[1]	
		\For {k = 1, \ldots, N} \Comment{ \textbf(in parallel)}
		\State{Draw $(w_1^{(k)}, \ldots, w_{n}^{(k)}) \sim \text{Exp}(1)$.}
		\State{ Set $\tilde{\theta}_1^{(k)} = \arg \max_{\theta_1} \left\{ \sum_{j=1}^{n} w_j^{(k)} \log f_1(x_{1,j}| \theta_1) + w_0 \log \pi_1\left(\theta_1\right) \right\}$.}
		\State{ Set $\tilde{\theta}_2^{(k)} = \arg \max_{\theta_2} \left\{\sum_{j=1}^{n} w_j^{(k)} \log f_2(x_{2,j}|  \tilde{\theta}_1^{(k)}, \theta_2,x_{1,j} )+ v_{0} \log \pi_2(\theta_2) \right\}$. 
        }
		\EndFor
		\State \Return {Sample $\{(\tilde{\theta}_1^{(k)}, \ \tilde{\theta}_2^{(k)}) \}_{k=1}^N$}.
	\end{algorithmic}
\end{algorithm}

Contrarily to the cut posterior,
PBMI provides the correct frequentist
coverage for modular inference as $n\to\infty$. It also has a number of interesting features
inherited from Posterior Bootstrap. In particular, it delivers samples following a distribution that can be different from Normal, such as skewed or multimodal distributions
(see the discussion in \citet{newton1994approximate}, and the authors'
response). An example of skewness will be encountered in the experiments of \cref{section:epidemiological_study_v2}. 

The ``for loop'' in Algorithm~\ref{alg:pb_cut_dependent} can be executed in parallel. Whereas computation of the cut posterior using MCMC is challenging,
due to intractability of the feedback term \citep{plummer2015cuts}, computation of PBMI appears simpler, with each module adding an optimization program per sample.  Finally, PBMI allows for
incorporating the  prior information through the choice of $w_0$ and $v_0$, as we
explain in Section~\ref{section:incorporating_prior}.

\subsection{Asymptotic behavior}\label{section:algorithms:asymptotics}

The following result establishes asymptotic properties of
Algorithm~\ref{alg:pb_cut_dependent}. 

\begin{thm}\label{thm:dependent}
Suppose that $(\tilde{\theta}_1,\tilde{\theta}_2)$ is drawn from Algorithm~\ref{alg:pb_cut_dependent} and let $\mle = (\mle_1,\mle_2)$ be the 2SM estimator in \eqref{eq:2sM}. Moreover, suppose that \cref{assump:log_lik,assump:identifiability,assump:log_lik_smoothness,assump:positive_definite,assump:log_prior_smoothness} hold and let $ Z \sim \mathrm{Normal}(0, \Sigma)$ for $\Sigma$ the asymptotic variance of $\mle$ given by \cref{eq:asymptotic2sM}. Then, for any Borel set $A \subseteq \R^{d_1+d_2}$ with $\mathbbm{P}(Z\in\partial A)=0$, with $\partial A$ denoting the boundary of $A$, we have
\begin{align*}
\mathbbm{P} \left\{\sqrt{n}(\tilde{\theta}_1 - \hat{\theta}_1, \tilde{\theta}_2 - \hat{\theta}_2) \in A \big| x_{1, 1:n}, x_{2, 1:n}\right\} \xrightarrow{n\to\infty} \mathbbm{P}(Z \in A)\quad \text{in } P^*\text{-probability} .
\end{align*}
\end{thm}
A proof of \cref{thm:dependent} is presented in Appendix~\ref{suppl:proof_main_thms}. 
\begin{remark}\label{rmk:modified_algorithm}
In the setting of \cref{section:scenario_indep}, 
Algorithm~\ref{alg:pb_cut_dependent} can be modified by replacing 
the inner loop with the following steps:
\begin{algorithmic}[1]
    \State Draw $(w_1^{(k)}, \ldots, w_{n_1}^{(k)}) \sim \mathrm{Exp}(1)$.
    \State Set $\tilde{\theta}_1^{(k)} = \arg\max_{\theta_1} 
        \Big\{ \sum_{j=1}^{n_1} w_j^{(k)} \log f_1(x_{1,j}\mid \theta_1) 
        + w_0 \log \pi_1(\theta_1) \Big\}$.
    \State Draw $(v_1^{(k)}, \ldots, v_{n_2}^{(k)}) \sim \mathrm{Exp}(1)$.
    \State Set $\tilde{\theta}_2^{(k)} = \arg\max_{\theta_2} 
        \Big\{ \sum_{j=1}^{n_2} v_j^{(k)} \log f_2(x_{2,j}\mid 
        \tilde{\theta}_1^{(k)}, \theta_2) + v_0 \log \pi_2(\theta_2) \Big\}$.
\end{algorithmic}
The conclusions of \cref{thm:dependent} remain valid 
for the resulting modified algorithm.
\end{remark}

The above result demonstrates an advantage of PBMI regarding the coverage of confidence sets, in line with what was found about the posterior bootstrap in \citet{lyddon2019general}, but
it is also interesting to consider the point of view of
prediction. 
\citep{fushiki2005bootstrap, fushiki2010bayesian} compared the predictive performance of various bootstrap methods with that of Bayesian inference, for a 
model corresponding to the first module of \cref{eq:general_model} considered in isolation. 
Their results imply that 
the Posterior Bootstrap yields better prediction for the first
module compared to the standard posterior for $\theta_1$, as long as $J_1^* \neq I_1^*$.
In
Appendix~\ref{suppl:prediction} we investigate prediction for the second module and show
that depending on the exact setup, PBMI 
may yield better or worse prediction 
compared to the cut
posterior.

\subsection{Incorporating prior information}\label{section:incorporating_prior}

Even though the choice of hyperparameters $w_0$ and $v_0$ does not affect the
first-order asymptotic approximation in \cref{thm:dependent},
it plays a role in calibrating the impact of the
priors.
By default, $w_0$ and $v_0$ can be set to 1 in analogy with MAP estimators.
We can also set $w_0$ and $v_0$ as 
in \citet{pompe2021introducing}, with the aim of calibrating the impact of the prior on the second-order Edgeworth
expansions for  $(\tilde{\theta}_1, \tilde{\theta}_2)$ generated by PBMI. 
This leads to setting
\begin{equation}\label{w_v_0_formula}
  w_0 = \text{diag}\left(I_1^{*1/2} J_1^{*-1} I_1^{*1/2}\right), \quad v_0 =\text{diag}\left(I_2^{*1/2} J_2^{*-1} I_2^{*1/2}\right),
\end{equation}
as optimal values of $w_0$ and $v_0$ if the priors factorize, and
\begin{equation}\label{w_v_0_formula_2}
  w_0 = \text{tr}\left(I_1^{*1/2} J_1^{*-1} I_1^{*1/2}\right), \quad v_0 =\text{tr}\left(I_2^{*1/2} J_2^{*-1} I_2^{*1/2}\right),
\end{equation}
otherwise.
For  derivations justifying the above formulas, see
\citet{pompe2021introducing}.
Since matrices $I_i^*$ and $J_i^*$ are rarely known, in practice they need to be approximated, as in the Laplace approximation. Note that the choice of $w_0$ and $v_0$ is
important for informative priors. When the prior is vague and meant to
represent a lack of prior information, one can 
set these values to 0 for simplicity.

\citet{pompe2021introducing} noticed that Posterior Bootstrap with prior
penalization cannot be used with priors that are unbounded as then the corresponding maximum in Algorithm~\ref{alg:pb_cut_dependent} would lie at the boundary. 
In such cases, an alternative way of
incorporating prior information is via so-called pseudosamples from
the prior predictive \citep{fong2019scalable}.  A meaningful calibration of the impact of the prior 
when using pseudosamples is possible for conjugate priors, yet can be problematic with non-conjugate
ones. In particular, one can expect unstable results when the pseudosamples 
are likely to be outliers with respect to the existing dataset; see \citet[Sections 4 and 6]{pompe2021introducing}. 

\input{numerics_input}

\FloatBarrier

\section{Discussion}\label{section:discussion}

Future avenues of research include extension to more than two modules,
semi-parametric and non-parametric settings,
different asymptotic regimes for the relative sample sizes associated with different modules, as well as higher-order asymptotics. 

Whether to cut the feedback or not is a vast question, discussed extensively in \citet{jacob2017better,FrazierNott2025}. If the analyst commits to a modular approach,  
there are still various possible implementations. This paper studies the asymptotic behavior of cut posteriors, provides a Laplace approximation (Cut-Laplace) and another approach based on the Posterior Boostrap (PBMI). Cut-Laplace and PBMI both require optimization of the modules' log-densities, and Cut-Laplace also requires the evaluation of Hessians in each module. When both are feasible, and depending on hardware, PBMI typically has a higher cost than Cut-Laplace, which itself has a cost comparable to variational approximations of the cut posterior \citep{carmona2020semi,yu2021variational}. 

Under our conditions, cut posterior, Cut-Laplace and PBMI concentrate on the same region of the parameter space, corresponding to the limit of the 2SM estimator. All methods also share the same rate of convergence and a Normal limit. However, the asymptotic variance differs as soon as the model is misspecified. For PBMI, the asymptotic variance corresponds to that of the 2SM estimator (see \cref{thm:dependent}), and thus credible regions constructed from PBMI have nominal frequentist coverage: they can be used to construct frequentist confidence intervals for the limit of the 2SM estimator. If that is the goal, then PBMI is the preferred method. We recommend obtaining Cut-Laplace whenever possible, and comparing with PBMI; the mismatch should asymptotically reflect the amount of model misspecification. In passing, \cref{section:epidemiological_study_v2} illustrated that PBMI can deliver skewed distributions, which is by design impossible for Cut-Laplace. 

The cut posterior can be the preferred option, for example, in small sample sizes where both Cut-Laplace and PBMI have little justification. The cut posterior has an interpretable, non-asymptotic variational interpretation \citep[Lemma 1,][]{yu2021variational}. Furthermore, \citet{carmona2020semi,nicholls2022valid} show that cut posteriors satisfy some desired properties such as \emph{order coherence} and \emph{validity} as belief updates. 
Numerically, there may be situations where optimization is, in fact, more difficult than sampling, in which case nested MCMC may be easier to implement than PBMI or even than Cut-Laplace. Finally, the uncertainty reflected in cut posteriors is entirely conditional on one realization of the data; in some settings this could be the uncertainty of interest to the analyst, as opposed to the amalgamated frequentist and Bayesian uncertainty of PBMI.

\paragraph{Acknowledgments.}
PEJ acknowledges support from 
the European Union (ERC), through the Consolidator grant `UMCMC',
project number 101171415. MJK acknowledges support from the France 2030 program. The authors thank David Nott and David Frazier for interesting discussions.








\bibliography{references}

\appendix

\numberwithin{equation}{section}

\section{Proof of modular Bernstein-von Mises theorem\label{suppl:bernstein_von_mises}}

\setcounter{assumption}{0}
\renewcommand{\theassumption}{\Alph{section}.\arabic{assumption}}

\subsection{Informal proof of Theorem~\ref{thm:bvm_cut}\label{suppl:informal_proof_bvm}}

We start with an informal proof that provides some intuition
on the specific form of the limiting covariance matrix of the cut posterior distribution.
We use the notation from Section~\ref{section:asymptoticsmodularinference}, with  $L(\theta|x_1,x_2)$ 
the utility function defined in \eqref{eq:cutloglikelihood}.
We perform a Taylor expansion of it around the 2SM estimator $(\hat{\theta}_1, \hat{\theta}_2)$ 
defined in Section~\ref{sec:asymptoticMestimators}:
\begin{align*}
  L(\hat\theta + n^{-1/2} s|x_1,x_2) &\approx 
  L(\hat\theta |x_1,x_2) + s n^{-1/2}  L'(\hat\theta |x_1,x_2)
  + \frac{1}{2n} s^T L''(\hat\theta |x_1,x_2) s.
\end{align*}
Assuming that the first derivative is negligible at the 2SM estimator, the cut pseudo-log-likelihood behaves as a quadratic form around $\hat\theta$ as $n\to\infty$ and the 
asymptotic covariance matrix is given by the inverse of the limit of $n^{-1}  L''(\hat\theta |x_1,x_2)$.
The elements of the Hessian are computed as follows:
\begin{align*}
  \partial_{11} L(\theta_1,\theta_2|x_1,x_2) & = \ldots ? \ldots\\
  \partial_{12} L(\theta_1,\theta_2|x_1,x_2) & = \partial_{12} L_2(\theta_1,\theta_2|x_1,x_2)\\
  \partial_{22} L(\theta_1,\theta_2|x_1,x_2) & = \partial_{22} L_2(\theta_1,\theta_2|x_1,x_2).
\end{align*}
The first element $\partial_{11}  L(\theta_1,\theta_2|x_1,x_2)$ is the least tractable, because of the feedback term $F(\theta_1|x_1,x_2)$, defined as the logarithm of an integral.  
On the other hand, 
we expect $n^{-1} \partial_{12} L(\theta_1,\theta_2|x_1,x_2)$ to converge to $R_J^*$, and $n^{-1} \partial_{22} L(\theta_1,\theta_2|x_1,x_2)$ to converge to $J_2^*$,
with the definitions in \eqref{eq:j_defs}. Hence, the inverse of the limiting covariance matrix should be of the form:
\[H = \begin{pmatrix}
A & R_{J}^*\\
R_J^{*T} & J_2^*
\end{pmatrix},\]
for some matrix $A$. 
To find this matrix $A$, we use the formula for inverting a block matrix,
$$
H^{-1} = \begin{pmatrix}
(A - R_J^* J_2^{*-1}R_J^T)^{-1} & -(A - R_J^* J_2^{*-1}R_J^{*T})^{-1} R_J^* J_2^{*-1}\\
-J_2^{*-1}R_J^{*T} (A - R_J^* J_2^{*-1}R_J^{*T})^{-1} & 
J_2^{*-1} + J_2^{*-1} R_J^{*T} (A - R_J^* J_2^{*-1}R_J^{*T})^{-1} R_J^* J_2^{*-1}
\end{pmatrix}.
$$
Meanwhile, we know that $\theta_1$ is distributed
as the standard posterior in the first module,
with limiting covariance matrix $J_1^{*-1}$.
To match the asymptotic variance of $\theta_1$
under the cut posterior, we must have 
$$
(A - R_J^* J_2^{*-1}R_J^{*T})^{-1} = J_1^{*-1},
$$
which gives $A = J_1^* + R_J^* J_2^{*-1}R_J^{*T}$, leading to the asymptotic variance
in \cref{thm:bvm_cut}.

\subsection{A general Bernstein-von Mises theorem for cut posteriors\label{suppl:proof_bvm}}

We proceed to proving Theorem \ref{thm:bvm_cut}.
The general strategy is to adapt \cite[Theorem 4]{miller2021asymptotic}, which 
is a Bernstein-von Mises (BvM) result for generalised posteriors, to the setting of cut posteriors. This leads to Theorem~\ref{thm:asymptotic_main} below. Then,
in Appendix~\ref{suppl:assumptions_bvm}
we provide a set of verifiable assumptions, namely \cref{assump:tails_asymptotic,assump:priors_asymptotic,assump:third_der_asymptotic,assump:finite_hessian_asymptotic,assump:positive_definite_asymptotic,assump:integrability_asymptotic,assump:der_bounds_asymptotic},  that ensure that the assumptions of \cref{thm:asymptotic_main} are satisfied. This is shown in Appendix~\ref{suppl:bvm_interpretable_and_proofs}, leading to \cref{thm:bvm_cut}.
Finally, Appendix~\ref{suppl:verifying_assumptions} discusses how  
 \cref{assump:tails_asymptotic,assump:priors_asymptotic,assump:third_der_asymptotic,assump:finite_hessian_asymptotic,assump:positive_definite_asymptotic,assump:integrability_asymptotic,assump:der_bounds_asymptotic}
 can be checked in some examples.

\subsubsection{Statement}

We start with the adaptation of
\cite[Theorem 4]{miller2021asymptotic}
to the setting of cut posteriors.
We may drop $x_1$ and $x_2$ from the notation
introduced in Section~\ref{section:asymptoticsmodularinference}.
We define the normalizing constant and the density of the cut posterior as
\begin{equation}
    z_n=\int_{\mathbbm{R}^d} e^{L(\theta_1,\theta_2)}\pi(\theta_1,\theta_2)d\theta_1 d\theta_2;\qquad\posterior(\theta_1,\theta_2)= e^{L(\theta_1,\theta_2)}\pi(\theta_1,\theta_2)/z_n.
\end{equation}
An open ball of radius $\eta>0$ around $x$ is denoted by $B_{\eta}(x)$. 
\begin{thm}\label{thm:asymptotic_main}
    Let $\theta^*\in\mathbbm{R}^d$,
    and let $\mle=(\mle_1,\mle_2)=\mle(n)\in\mathbbm{R}^d$ be a sequence such that $\mle\xrightarrow[n\to\infty]{}\theta^*$.
     Consider the following assumptions.
   \begin{enumerate}[(1)]
    \setcounter{enumi}{-1}
    \item The prior density $\pi_1$ is continuous at $\theta_1^*$ and $\pi(\theta_1^*)>0$. Similarly, assume that $\pi_2$ is continuous at $\theta_2^*$ and $\pi(\theta_2^*)>0$. 
   \item $\loglikelihood$ can be represented as 
    \begin{align}\label{eq:cutloglikelihood_representation}
\frac{1}{n}\loglikelihood(\theta) &=     \frac{1}{n}\loglikelihood(\mle)-\frac{(V_n)^T\left(\theta-\mle\right)}{n}-\frac{1}{2}\left(\theta-\mle\right)^TH_n\left(\theta-\mle\right)
-r_n\left(\theta-\mle\right),
\end{align}
for $\theta=(\theta_1,\theta_2)\in\mathbbm{R}^d$, where
\begin{itemize}
\item $V_n$ is a vector in $\mathbbm{R}^d$ that satisfies $\frac{\|V_n\|}{\sqrt{n}}\xrightarrow[n\to\infty]{}0$;
\item $H_n$ is a symmetric matrix, converging to some positive definite $H_0$;
\item $r_n:\mathbbm{R}^d\to\mathbbm{R}$ is such that there exist some constants $\epsilon_0,c_0>0$, such that for all sufficiently large $n$ and all $x\in B_{\epsilon_0}(0)$, we have $|r_n(x)|\leq c_0\|x\|^3$.
\end{itemize}

\item $\loglikelihood_2$ can be represented as
\[
\frac{1}{n}\loglikelihood_2(\theta)=\frac{1}{n}\loglikelihood_2(\mle)-s_n(\theta-\mle),
\]
where $s_n$ is such that there exist some constants $\epsilon_1,c_1>0$ such that for all sufficiently large $n$ and all $x\in B_{\epsilon_1}(0)$, we have $|s_n(x)|\leq c_1\|x\|.$

\item The following strong identifiability conditions hold:
\begin{itemize}
    \item for any $\delta_1>0$,  
    \begin{equation}
        \liminf_{n\to\infty} \inf_{\theta_1\in B_{\delta_1}(\mle_1)^c}\frac{1}{n}(\loglikelihood_1(\mle_1)-\loglikelihood_1(\theta_1)) > 0;
    \end{equation}
    \item for any $\delta_2>0$, there exists $\delta_1>0$ such that
    \begin{align}\label{eq:strong_identifiability}
    \liminf_{n\to\infty}\underset{\theta_2\in B_{\delta_2}(\mle_2)^c}{\inf_{\theta_1\in B_{\delta_1}(\mle_1)}}\frac{1}{n}(\loglikelihood_2(\mle_1,\mle_2)-\loglikelihood_2(\theta_1,\theta_2)) > 0.
    \end{align}
\end{itemize}
\end{enumerate}
Then, the cut posterior concentrates around $\theta^*$:
\begin{equation}\label{eq:thm1:consistency}
    \int_{B_{\epsilon}(\theta^*)}\posterior(\theta_1,\theta_2)d(\theta_1,\theta_2)\xrightarrow[n\to\infty]{} 1,\quad\text{for all }\epsilon>0.
\end{equation}
The normalising constant of the cut posterior satisfies
\begin{equation}\label{eq:thm1:normalising_constant}
    z_n\sim \frac{\exp(\loglikelihood(\mle))\pi(\theta^*)}{\det{H_0}^{1/2}}\left( \frac{2\pi}{n} \right)^{d/2},\qquad\text{as }n\to\infty,
\end{equation}
and letting $q_n$ be the density of $\sqrt{n}(\theta-\mle)$, where $\theta\sim\posterior$, we have
\begin{equation}\label{eq:thm1:bvm}
    \int_{\mathbbm{R}^d}\left|q_n(x)-\mathcal{N}(x\,|\,0,H_0^{-1})\right|dx\xrightarrow[n\to\infty]{}0.
\end{equation}
\end{thm}
\begin{remark}
    The convergence as $n\to\infty$ in \cref{thm:asymptotic_main} is deterministic.
\end{remark}

We next provide the proof of \cref{thm:asymptotic_main}. 
The main difference with the proof of \cite[Theorem 4]{miller2021asymptotic} is that the generalised log-likelihood is replaced by the function $L$ in \eqref{eq:cutloglikelihood}, which involves the intractable feedback term. Let
\[
g_n(x)=\exp\left[\left(\loglikelihood(\mle+x/\sqrt{n})-\loglikelihood(\mle)\right)   \right]\pi(\mle+x/\sqrt{n})\quad\text{and}\quad
g_0(x)=\exp\left(-\frac{1}{2}x^TH_0x\right)\pi(\theta^*).
\]
Throughout the proof, for any $x\in\mathbbm{R}^d$, we write
$x^{(1)}=(x_1,\dots,x_{d_1})$ and $x^{(2)}=(x_{d_1+1},\dots,x_{d_2})$.

\subsubsection{Proof of Theorem~\ref{thm:asymptotic_main}}
We show that
\begin{align}\label{eq:asymptotic_aim}
\int |g_n-g_0|\xrightarrow[n\to\infty]{}0.
\end{align}
It then follows that, for any $a_n\to a\in\mathbbm{R}$, $\int|a_ng_n-ag_0| \to 0$ as $n\to\infty$, using the triangle inequality.
We then let $1/a_n=e^{-\loglikelihood(\mle)}n^{d/2}z_n$ and $1/a=\pi(\theta^*)(2\pi)^{d/2}\det{H_0}^{-1/2}$ and note that
\begin{align}\label{eq:g_n_g_0}
\frac{1}{a_n} = \int g_n \xrightarrow[n\to\infty]{}\int g_0 = \frac{1}{a},
\end{align}
therefore $a_n\to a$. Hence, we obtain
\[
\int \left| e^{\loglikelihood(\mle+x/\sqrt{n})}\pi(\mle+x/\sqrt{n})n^{-d/2}/z_n - \frac{\det{H_0}^{1/2}}{(2\pi)^{d/2}}e^{-x^TH_0x/2}  \right|dx\xrightarrow[n\to\infty]{}0,
\]
which is \cref{eq:thm1:bvm}. 
\cref{eq:thm1:consistency} follows directly from \cite[Lemma 28]{miller2021asymptotic}, and \cref{eq:thm1:normalising_constant} follows from \cref{eq:g_n_g_0}.
The rest of the proof is devoted to showing \cref{eq:asymptotic_aim}.

\paragraph{Pointwise convergence of $g_n$ to $g$.}

For any fixed $x$, $\pi(\mle+x/\sqrt{n})\to\pi(\theta^*)$ because of Assumption (0) in \cref{thm:asymptotic_main} and the fact that $\mle\to\theta^*$. Using \eqref{eq:cutloglikelihood_representation}, we have 
\[
\loglikelihood(\mle+x/\sqrt{n})-\loglikelihood(\mle)=-\frac{(V_n)^Tx}{\sqrt{n}}-\frac{1}{2}x^TH_nx-nr_n(x/\sqrt{n})\xrightarrow[n\to\infty]{}-\frac{1}{2}x^TH_0x.
\]
Hence, $g_n\to g$ pointwise.

\paragraph{Definition of functions dominating $g_n$ for sufficiently large $n$.}

Let $\rho\in(0,\lambda)$, where $\lambda$ is the smallest eigenvalue of $H_0$. Then, let $A_n=H_n-\rho I$ and $A_0=H_0-\rho I$. Moreover, let $\delta_1,\delta_2>0$ satisfy \cref{eq:strong_identifiability} and be small enough that $\sqrt{\delta_1^2+\delta_2^2}<\rho/(2c_0)$, $\sqrt{\delta_1^2+\delta_2^2}\leq \epsilon_0$, $\delta_1<\epsilon_1$ and $\pi(\theta)\leq 2\pi(\theta^*)$ for all $\theta\in B_{2\sqrt{\delta_1^2+\delta_2^2}}(\theta^*)$. Furthermore, make $\delta_1$ so small that, for
\begin{align*}
&\tilde{\varepsilon}_2:=\frac{1}{2}\liminf_{n\to\infty}\underset{\theta_2\in B_{\delta_2}(\mle_2)^c}{\inf_{\theta_1\in B_{\delta_1}(\mle_1)}}\frac{1}{n}(\loglikelihood_2(\mle_1,\mle_2)-\loglikelihood_2(\theta_1,\theta_2))>0,
\end{align*}
$\delta_1c_1<\tilde{\varepsilon}_2$. This is possible because decreasing the value of $\delta_1$ does not decrease $\tilde{\varepsilon}_2$.
Set:
\begin{align*}
    &\tilde{\varepsilon}_1:=\frac{1}{2}\liminf_{n\to\infty} \inf_{\theta_1\in B_{\delta_1}(\mle_1)^c}\frac{1}{n}(\loglikelihood_1(\mle_1)-\loglikelihood_1(\theta_1))>0.
\end{align*}
and define
\begin{align}
&h_n(x)\notag\\
=&\begin{cases}
    \exp\left(-\frac{1}{2}x^TA_nx-(V_n)^Tx/\sqrt{n}\right)2\pi(\theta^*),&\text{if }\|x^{(1)}\|\leq\delta_1\sqrt{n} \text{ and } \|x^{(2)}\|\leq\delta_2\sqrt{n},\\
    \begin{aligned}
e^{-n\tilde{\varepsilon}_1}\exp\Big[\loglikelihood_2(\mle+x/\sqrt{n})-F(\mle_1+x^{(1)}/\sqrt{n})\\
-\loglikelihood_2(\mle)+F(\mle_1)\Big]\pi(\mle+x/\sqrt{n})
\end{aligned}, &\text{if }\|x^{(1)}\|>\delta_1\sqrt{n},\\
    \begin{aligned}
    e^{-n\tilde{\varepsilon}_2}\exp\Big[\loglikelihood_1(\mle_1+x^{(1)}/\sqrt{n})-F(\mle_1+x^{(1)}/\sqrt{n})\\
    -\loglikelihood_1(\mle)+F(\mle_1)\Big]\pi(\mle+x/\sqrt{n})
    \end{aligned}, &\text{if }\|x^{(1)}\|\leq \delta_1\sqrt{n}\text{ and }\|x^{(2)}\|> \delta_2\sqrt{n}. 
\end{cases}\label{eq:def_h_n}
\end{align}
Fix $x\in\mathbbm{R}^d$. We next show that $|g_n(x)|\leq h_n(x)$ for large enough $n$, considering the above cases.

First, consider the case where $\|x^{(1)}\|\leq\delta_1\sqrt{n}$ and $\|x^{(2)}\|\leq\delta_2\sqrt{n}$.
Since $\|\mle-\theta^*\|\xrightarrow[n\to\infty]{}0$, for large enough $n$,
$\mle+x/\sqrt{n} \in B_{2\sqrt{\delta_1^2+\delta_2^2}}(\theta^*)$, and thus $\pi(\mle+x/\sqrt{n})\leq 2\pi(\theta^*)$. 
Moreover, for sufficiently large $n$, $\|\mle-\theta^*\|<\epsilon_0$ and the bound on $r_n$ applies. Hence,
\begin{align*}
\loglikelihood(\mle+x/\sqrt{n})-\loglikelihood(\mle)
&= -\frac{(V_n)^T x}{\sqrt{n}} - \frac{1}{2} x^T H_n x - n r_n(x/\sqrt{n}) \\
&\leq -\frac{1}{2} x^T H_n x + \frac{1}{2} \rho \|x\|^2 - \frac{(V_n)^T x}{\sqrt{n}} 
= -\frac{1}{2} x^T A_n x - \frac{(V_n)^T x}{\sqrt{n}},
\end{align*}
since
\[
|nr_n(x/\sqrt{n})|\leq c_0\|x\|^3/\sqrt{n}\leq c_0\sqrt{\delta_1^2+\delta_2^2}\|x\|^2\leq \frac{1}{2}\rho\|x\|^2.
\]
We conclude that $|g_n(x)|\leq h_n(x)$ for $\|x^{(1)}\|\leq\delta_1\sqrt{n} \text{ and } \|x^{(2)}\|\leq\delta_2\sqrt{n}$.

The two other cases can be treated by writing the generalised log-likelihood ratio in $g_n$ as
\[L_1(\mle_1 + x^{(1)}/\sqrt{n}) + L_2(\mle + x /\sqrt{n}) - F(\mle_1 + x^{(1)}/\sqrt{n}) - \{L_1(\mle_1) + L_2(\mle) - F(\mle_1)\},\]
and either bounding $L_1(\mle_1 + x^{(1)}/\sqrt{n}) - L_1(\mle_1)$ by $-n\tilde{\varepsilon}_1$ if $\|x^{(1)}\|>\delta_1\sqrt{n}$,
or $L_2(\mle + x /\sqrt{n}) - L_2(\mle)$ by $-n\tilde{\varepsilon}_2$ otherwise, and leaving the other terms unchanged.

Consequently, the functions $h_n(x)$ dominate $g_n(x)$ for all sufficiently large $n$.
Finally, define
\begin{equation}\label{eq:def_h_0}
    h_0(x)=\exp\left(-\frac{1}{2}x^TA_0x\right)2\pi(\theta^*),
\end{equation}
and note that $h_n\to h_0$ pointwise. Indeed, for all $x$, when $n$ large enough,
$\|x^{(1)}\|\leq \delta_1 \sqrt{n}$ and $\|x^{(2)}\|\leq \delta_2 \sqrt{n}$, so $h_n$ takes the form
given in the first item of its definition in \cref{eq:def_h_n}.

\paragraph{Convergence of $\int h_n$ to $\int h_0$.}
Let $B_n=\{x\,:\,\|x^{(1)}\|\leq\delta_1\sqrt{n} \text{ and } \|x^{(2)}\|\leq\delta_2\sqrt{n}\}$. 
\paragraph*{Integrating over $B_n$.}
Recall that $A_n\to A_0$ and $A_0$ is positive definite. Indeed $\rho < \lambda$ so all eigenvalues of $A_0$ are positive. This means that, for all $n$ sufficiently large, $A_n$ is also positive definite. It follows that, for large enough $n$,
\begin{align*}
    \int_{B_n} h_n =& \;2\pi(\theta^*)    \int_{B_n} \exp\left(-\frac{1}{2}x^TA_nx-(V_n)^Tx/\sqrt{n}\right)dx\\
    &\hspace{-10mm}=2\pi(\theta^*) \exp\left[\frac{1}{2n}(V_n)^TA_n^{-1}V_n\right]   \int_{B_n} \exp\left[-\frac{1}{2}\left(x+A_n^{-1}V_n/\sqrt{n}\right)^TA_n\left(x+A_n^{-1}V_n/\sqrt{n}\right)\right]dx\\
    \leq &2\pi(\theta^*)\exp\left[\frac{1}{2n}(V_n)^TA_n^{-1}V_n\right] \frac{(2\pi)^{d/2}}{\det{A_n}^{1/2}} \quad \text{(completing the integral)}\\
    &\xrightarrow[n\to\infty]{} 2\pi(\theta^*)\frac{(2\pi)^{d/2}}{\det{A_0}^{1/2}}=\int h_0,
\end{align*}
because $A_n$ is positive definite and so
\[
1\leq \exp\left[\frac{1}{2n}(V_n)^TA_n^{-1}V_n\right] \leq \exp\left[\frac{1}{2n}\|V_n\|^2\|A_n^{-1}\|_{op}\right]\xrightarrow[n\to\infty]{}1.
\]

\paragraph*{Integrating over $\|x^{(1)}\|>\delta_1\sqrt{n}$.}
Here we write $x_1$ instead of $x^{(1)}$ and $x_2$ instead of $x^{(2)}$. Then,
\begin{align*}
    &\int_{\|x_1\|>\delta_1\sqrt{n}}\int_{\mathbbm{R}^{d_2}}h_n(x_1,x_2)dx_2dx_1\\
    =&e^{-n\tilde{\varepsilon}_1}\exp\left[F(\mle_1)-\loglikelihood_2(\mle_1,\mle_2)\right]\\
    &\cdot\int_{\|x_1\|>\delta_1\sqrt{n}}\frac{\int\exp\left[ \loglikelihood_2(\mle_1+x_1/\sqrt{n},\mle_2+x_2/\sqrt{n})\right]\pi_2(\mle_2+x_2/\sqrt{n})dx_2}{\exp[F(\mle_1+x_1/\sqrt{n})]}\pi_1(\mle_1+x_1/\sqrt{n})dx_1\\
    =&n^{d_2/2}e^{-n\tilde{\varepsilon}_1}\left[\int \exp\left[\loglikelihood_2(\mle_1,s)-\loglikelihood_2(\mle_1,\mle_2)\right]\pi_2(s)ds\right]\cdot\int_{\|\theta_1-\mle_1\|>\delta_1}\frac{\int\exp\left[\loglikelihood_2(\theta_1,x_2)\right]\pi_2(x_2)dx_2}{\exp[F(\theta_1)]}\pi_1(\theta_1)d\theta_1\\
    \leq& n^{d_2/2}e^{-n\tilde{\varepsilon}_1}\xrightarrow[n\to\infty]{}0.
\end{align*}
The third line is obtained by replacing $F(\mle_1)$ by its definition and a change of variables from $x_2$ to $\theta_2$ in the right-most integral.
The inequality stems from the square bracket being less than one because $\mle_2$ maximizes $\loglikelihood_2(\mle_1,\cdot)$, and the fraction in the last integral being equal to one,
the integral is that of the density function $\pi_1$ over part of the space.

\paragraph*{Integrating over $\|x^{(1)}\|\leq \delta_1\sqrt{n}\text{ and } \|x^{(2)}\|>\delta_2\sqrt{n}$.}

Recall that we have assumed that $\delta_1 < \epsilon_1$ and $\delta_1 c_1 < \tilde{\varepsilon}_2$.
Thus, we can introduce $\delta_3>0$ such that $\sqrt{\delta_1^2+\delta_3^2}<\epsilon_1$ and $(\delta_1+\delta_3)c_1<\tilde{\varepsilon}_2$. 
Then, by definition of $h_n$,
\begin{align*}
    &\int_{\|x_1\|\leq\delta_1\sqrt{n}}\int_{\|x_2\|>\delta_2\sqrt{n}}h_n(x_1,x_2)dx_2dx_1\\
    =& e^{-n\tilde{\varepsilon}_2}\left[\int \exp\left[\loglikelihood_2(\mle_1,s)\right]\pi_2(s)ds  \right]\int_{\|x_1\|\leq\delta_1\sqrt{n}}\int_{\|x_2\|>\delta_2\sqrt{n}}\frac{\exp[\loglikelihood_1(\mle_1+x_1/\sqrt{n})-\loglikelihood_1(\mle_1)]}{\int\exp[\loglikelihood_2(\mle_1+x_1/\sqrt{n},t)]\pi_2(t)dt}\\
    &\hspace{8.5cm}\cdot\pi_1(\mle_1+x_1/\sqrt{n})\pi_2(\mle_2+x_2/\sqrt{n})dx_1dx_2.
\end{align*}
After a change of variable $x \leftrightarrow \theta = \mle + x/\sqrt{n}$, we bound the integral of $\pi_2$ by 1.
Next we bound $\loglikelihood_1(\theta_1)-\loglikelihood_1(\mle_1)$ by zero, and we introduce $\exp(-L_2(\mle_1,\mle_2))$ in both numerator and denominator, and 
bound $\loglikelihood_2(\mle_1,s)-\loglikelihood_2(\mle_1,\mle_2)$ in the numerator by zero. We obtain the upper bound:
\begin{align*}
    n^{d/2} e^{-n\tilde{\varepsilon}_2} \int_{\|\theta_1 - \mle_1\|\leq \delta_1}\frac{\pi_1(\theta_1)}{\int \exp[\loglikelihood_2(\theta_1,t)-\loglikelihood_2(\mle_1,\mle_2)]\pi_2(t)dt}d\theta_1.
\end{align*}
We then restrict the integral in the denominator from $\mathbbm{R}^{d_2}$ to $\{t: \|t-\mle_2\|\leq \delta_3\}$, 
and use Assumption (2) in \cref{thm:asymptotic_main}.
Indeed, for $\|\theta_1-\mle_1\|\leq\delta_1$ and $\|t-\mle_2\|\leq \delta_3$, we have that $(\theta_1,t) - \mle$ is in $B_{\sqrt{\delta_1^2+\delta_3^2}} \subset B_{\epsilon_1}(0)$, and thus $L_2(\theta_1,t)-L_2(\mle_1,\mle_2)\geq -(\delta_1+\delta_3)c_1n$. Hence, the above displayed equation is upper bounded by
\begin{align*}
    n^{d/2} e^{-n(\tilde{\varepsilon}_2 - (\delta_1+\delta_3)c_1)} \left(\int_{\|\theta_2 - \mle_2\|\leq \delta_3}\pi_2(d\theta_2)\right)^{-1}, 
\end{align*}
and this goes to zero as $n\to\infty$ because $\tilde{\varepsilon}_2 - (\delta_1+\delta_3)c_1>0$ and using Assumption (0) on $\pi_2$.

Thus $\int h_n\xrightarrow[n\to\infty]{}\int h_0$.

\paragraph{Conclusion.}

We have shown that
1) $g_n\to g_0$ and $h_n\to h_0$ pointwise;
2) $\int h_n\to\int h_0$;
3) $g_n=|g_n|\leq h_n$ for all $n$ sufficiently large.
Furthermore,
$g_n,g_0,h_n,h_0\in L^1$ for all $n$ sufficiently large.
Indeed, 
$\int h_0$ and $\int g_0$ are finite because $H_0,A_0$ are positive definite. 
We have shown that $\int h_n\to\int h_0<\infty$, therefore $\int g_n\leq \int h_n<\infty$ for all sufficiently large $n$. 
Measurability of $g_n$ and $h_n$ follows from measurability of $\loglikelihood_1,\loglikelihood_2$ and $\pi$.
\Cref{eq:asymptotic_aim} now follows from the generalized dominated convergence theorem \cite[Exercise 2.20, 2.21]{folland2013}.
This concludes the proof of \cref{thm:asymptotic_main}.

\subsection{Assumptions on the model components\label{suppl:assumptions_bvm}}

Next, we aim to provide a set of verifiable and interpretable assumptions that ensure that \cref{thm:asymptotic_main} holds.
Recall the setting of Section~\ref{section:asymptoticsmodularinference}, 
with an i.i.d. data-generating process $p^*$, and log-losses $-\log f_1$ and $-\log f_2$.
The 2SM estimator is defined as in \cref{eq:2sM}.
Assumption~\ref{assump:tails_asymptotic} below ensures that $(\mle_1,\mle_2)$ is unique with $P^*$-probability converging to one. In \cref{thm:bvm_cut}, we assume that $(\mle_1,\mle_2)\to(\truepar_1,\truepar_2)$ in probability, for some $(\truepar_1,\truepar_2)$. We recall the definitions in \eqref{eq:j_defs}.

\subsubsection{Assumption about the priors}

The first assumption is slightly stronger than Assumption (0) in the statement of \cref{thm:asymptotic_main}. 

\begin{assumption}\label{assump:priors_asymptotic}    
    The prior density $\pi_1$ is continuous at $\theta_1^*$ and $\pi(\theta_1^*)>0$. The prior density $\pi_2$ is continuously differentiable at $\theta_2^*$ and $\pi(\theta_2^*)>0$.
\end{assumption}



\subsubsection{Assumption about the tails of the cut posterior}

The second assumption is directly (3) in the statement of \cref{thm:asymptotic_main}, but in probability.

\begin{assumption}\label{assump:tails_asymptotic}
    For all $\delta_1>0$, there exists $\gamma_1>0$, such that
    \begin{align*}
&\lim_{n\to\infty}\mathbbm{P}\left[\sup_{\theta_1 \in B_{\delta_1}(\mle_1)^c}\frac{1}{n}\left(\loglikelihood_1(\theta_1|X_1)-\loglikelihood_1(\mle_1|X_1)\right)\leq - \gamma_1\right]=1.
    \end{align*}
Moreover, for all $\delta_2>0$, there exist $\delta_1>0,\gamma_2>0$, such that
\begin{align*}
&\lim_{n\to\infty}\mathbbm{P}\left[\underset{\theta_2 \in B_{\delta_2}(\mle_2)^c}{\sup_{\theta_1 \in B_{\delta_1}(\mle_1)}}\frac{1}{n}\left(\loglikelihood_2(\theta_1,\theta_2|X_1,X_2)-\loglikelihood_2(\mle_1,\mle_2|X_1,X_2)\right)\leq - \gamma_2\right]=1.
\end{align*}
\end{assumption}

\subsubsection{Assumptions about the bulk of the cut posterior}

The next assumption directly implies (2) in \cref{thm:asymptotic_main}, 
and is also used in verifying (1).

\begin{assumption}\label{assump:third_der_asymptotic}
$L_2,L$ are three times differentiable in a neighbourhood of $\mle$, almost surely. Moreover, there exists $\delta>0$, such that, almost surely, for all sufficiently large $n$ 
\begin{align*}
&\max\left\{\sup_{s\neq 0,\,\|s\|\leq \delta}\frac{\|\loglikelihood''(\mle+s|X_1,X_2)-\loglikelihood''(\mle|X_1,X_2)\|_{op}}{n\|s\|},\right.\\
&\sup_{s\neq 0,\,\|s\|\leq \delta}\frac{|\loglikelihood_2(\mle+s|X_1,X_2)-\loglikelihood_2(\mle|X_1,X_2)|}{n\|s\|},\\
&\left.\sup_{s\neq 0,\,\|s\|\leq \delta}\frac{\|\partial_{22}\loglikelihood_2(\mle_1,\mle_2+s|X_1,X_2)-\partial_{22}\loglikelihood_2(\mle_1,\mle_2|X_1,X_2)\|_{op}}{n\|s\|}\right\}\leq \asymptthirdder,
\end{align*}
where $\asymptthirdder$ is such that there exists a fixed, deterministic $K>0$ satisfying
\[
\lim_{n\to\infty}\mathbbm{P}\left[\asymptthirdder<K\right]=1.
\]

\end{assumption}


For the convergence in probability of $H_n=-\frac{1}{n}\loglikelihood''(\mle|X_1,X_2)$, we make the following assumption.

\begin{assumption}\label{assump:finite_hessian_asymptotic}
    For all $\theta = (\theta_1,\theta_2)$ in some open ball around $\truepar = (\truepar_1,\truepar_2)$ and for all $x_1,x_2$,
    \begin{align*}
&\left|\partial_{[jk]} \log f_1(x_1|\theta_1)\right|\leq M_2(x_1),\qquad \text{for all }1\leq j\leq k\leq d_1;\\
        &\left| \partial_{[jk]}\log f_2(x_2|\theta_1,\theta_2,x_1)   \right|\leq M_3(x_1,x_2), \qquad \text{for all }1\leq j\leq k\leq (d_1+d_2).
    \end{align*}
   where in the above equations, $\partial_{[jk]}$ refers to the partial derivatives with respect to the $j$-th and $k$-th components of $\theta = (\theta_1, \theta_2)$, $M_2$ and $M_3$ do not depend on $\theta_1,\theta_2$ and $\mathbbm{E}_{X_{1,1}\sim p_1^*}M_2(X_{1,1})<\infty$,\\ $\mathbbm{E}_{(X_{1,1},X_{2,1})\sim p^*}M_3(X_{1,1},X_{2,1})<\infty$.
\end{assumption}

The following assumption is on the Hessian of the losses around the pseudo-true parameter.

\begin{assumption}\label{assump:positive_definite_asymptotic}
    The following matrices are positive definite for all $(\theta_1,\theta_2)$ inside some open ball around $(\truepar_1,\truepar_2)$,
    \begin{align*}
        J_1(\theta_1)=-\mathbbm{E}_{X_{1,1}\sim p_1^*}\left[\partial_{11}\log f_1(X_{1,1}|\theta_1)\right],\qquad J_2(\theta_1,\theta_2)=-\mathbbm{E}_{(X_{1,1},X_{1,2})\sim p^*}\left[\partial_{22}\log f_2(X_1,X_2|\theta_1,\theta_2)\right].
    \end{align*}
\end{assumption}



The following two assumptions make sure that the second derivative of the scaled log-feedback term
$n^{-1}F(\theta_1)$
converges and that the limit $H_0$ of $H_n$ is positive definite.

\begin{assumption}\label{assump:der_bounds_asymptotic}
There exists $\delta>0$, independent of $n$, such that, almost surely, for sufficiently large $n$,
    \begin{align*}
    &\max\left\{\sup_{\|s\|\leq\delta}\frac{\left\|\partial_{11}\loglikelihood_2(\mle_1,\mle_2+s|X_1,X_2)\right\|_{op}}{n}; \right.\sup_{\|s\|\leq\delta}\frac{\left\|\partial_{12}\loglikelihood_2(\mle_1,\mle_2+s|X_1,X_2)\right\|_{op}}{n};\\
    &\hspace{1cm}\sup_{\|s\|\leq\delta}\frac{\left\|\partial_1\loglikelihood_2(\mle+s|X_1,X_2)\right\|}{n}; \quad\sup_{s\neq 0,\|s\|\leq \delta}\frac{\left\|\partial_{11}\loglikelihood_2(\mle_1,\mle_2+s|X_1,X_2)-\partial_{11}\loglikelihood_2(\mle_1,\mle_2|X_1,X_2)\right\|_{op}}{n\|s\|};\\
    &\hspace{1cm}\left.\sup_{s\neq0,\|s\|\leq \delta}\frac{\left\|\partial_{12}\loglikelihood_2(\mle_1,\mle_2+s|X_1,X_2)-\partial_{12}\loglikelihood_2(\mle_1,\mle_2|X_1,X_2)\right\|_{op}}{n\|s\|}\right\}\leq \asymptdermixed,
    \end{align*}
    where $\asymptdermixed$ is bounded in probability with respect to $n$.
\end{assumption}

\begin{assumption}\label{assump:integrability_asymptotic}
There exists $m>0$, independent of $n$, such that
\[
\frac{1}{n^m}\int\left\|\partial_1L_2(\mle_1,\theta_2|X_1,X_2)\right\|^2\pi_2(\theta_2)d\theta_2\qquad\text{and}\qquad \frac{1}{n^m}\int\left\|\partial_{11}L_2(\mle_1,\theta_2|X_1,X_2)\right\|_{op}\pi_2(\theta_2)d\theta_2
\]
almost surely exist and are both upper-bounded by some $\asymptintegrals$, which is bounded in probability with respect to $n$.
 Moreover, for some $\delta>0$,
\[
\int\sup_{\|\theta_1-\mle_1\|<\delta}\left\|\partial_1L_2(\theta_1,\theta_2|X_1,X_2)\right\|^2\pi_2(\theta_2)d\theta_2
\quad\text{and}\quad 
\int\sup_{\|\theta_1-\mle_1\|<\delta}\left\|\partial_{11}L_2(\theta_1,\theta_2|X_1,X_2)\right\|\pi_2(\theta_2)d\theta_2
\]
exist with probability converging to $1$, as $n\to\infty$.
\end{assumption}


\subsection{Bernstein-von Mises for cut posteriors under interpretable assumptions\label{suppl:bvm_interpretable_and_proofs}}

We can now rephrase \cref{thm:asymptotic_main} with reference to the aforementioned assumptions
on the model components, and provide a more explicit expression for the asymptotic covariance matrix $H_0$;
this is \cref{thm:bvm_cut}.

\begin{remark}
    In the proof of \cref{thm:bvm_cut} below, we use the assumption that $\loglikelihood_1$ and $\loglikelihood_2$ take the additive forms given by \cref{eq:loglikelihoods} and the assumption that the data pairs $\{(X_{1,j},X_{2,j})\}_{j=1}^{\infty}$ are generated i.i.d. However, we use those assumptions only in order to be able to apply the uniform law of large numbers to $\partial_{11}\loglikelihood_1(\cdot|X_1)$, $\partial_{22}\loglikelihood_2(\cdot|X_1,X_2)$ and $\partial_{12}\loglikelihood_2(\cdot|X_1,X_2)$ in a neighborhood of $(\theta_1^*,\theta_2^*)$. Hence, one may relax either (or both) of those assumptions as long as the uniform law of large numbers still holds.
\end{remark}

We verify that the assumptions of \cref{thm:asymptotic_main} are satisfied (in probability); then the 
total variation convergence in \cref{thm:bvm_cut} is established (in probability) as a restatement of \eqref{eq:thm1:bvm}. 
Assumption (0) is implied by \cref{assump:priors_asymptotic}. Assumption (2) is implied by the bound on $L_2(\mle+s|X_1,X_2) - L_2(\mle|X_1,X_2)$ in \cref{assump:third_der_asymptotic}. 
Assumption (3) of \cref{thm:asymptotic_main} is satisfied in probability by \cref{assump:tails_asymptotic}. So the main task is to establish \eqref{eq:cutloglikelihood_representation}, and to identify the inverse of the limit of $H_n$ as the matrix in \eqref{eq:Hinverse}.

The expansion in \eqref{eq:cutloglikelihood_representation} can be shown via Taylor's theorem with Lagrange remainder. Indeed, let $\delta$ be as in \cref{assump:third_der_asymptotic}. We have that, for sufficiently large $n$ and for all $\theta$ such that $\|\param-\mle\|\leq\delta$, there exists $t\in\mathbb{R}^d$ such that $\|t-\mle\|\leq \|\param-\mle\|$ and
\begin{align*}
\frac{1}{n}\loglikelihood(\param|X_1,X_2)
=&\frac{1}{n}\loglikelihood(\mle|X_1,X_2) + \frac{1}{n}\loglikelihood'(\mle|X_1,X_2)(\param-\mle)-\frac{1}{2}\left(\param-\mle\right)^T\left(-\frac{\loglikelihood''(t|X_1,X_2)}{n}\right)\left(\param-\mle\right)\\
=&\frac{1}{n}\loglikelihood(\mle|X_1,X_2) + \frac{1}{n}\loglikelihood'(\mle|X_1,X_2)(\param-\mle)-\frac{1}{2}\left(\param-\mle\right)^T\left(-\frac{\loglikelihood''(\mle|X_1,X_2)}{n}\right)\left(\param-\mle\right)\\
&+\frac{1}{2}\left(\param-\mle\right)^T\left(\frac{\loglikelihood''(t|X_1,X_2)-\loglikelihood''(\mle|X_1,X_2)}{n}\right)\left(\param-\mle\right).
\end{align*}
We note that, by \cref{assump:third_der_asymptotic},
\begin{align*}
\left|\left(\param-\mle\right)^T\left(\frac{\loglikelihood''(t|X_1,X_2)-\loglikelihood''(\mle|X_1,X_2)}{n}\right)\left(\param-\mle\right)\right|
\leq &\; \asymptthirdder\|\param-\mle\|^3.
\end{align*}
Then, $\epsilon_0$ of \cref{thm:asymptotic_main} is given by $\delta$ from \cref{assump:third_der_asymptotic}, $c_0$ is given by $\frac{1}{2}\asymptthirdder$. Moreover, $H_n$ is given by
\begin{align*}
H_n=&-\frac{1}{n}\loglikelihood''(\mle_1,\mle_2|X_1,X_2)\\
=&-\frac{1}{n}\begin{bmatrix}
    \partial_{11}\loglikelihood_1(\mle_1|X_1)+\partial_{11}\loglikelihood_2(\mle_1,\mle_2|X_1,X_2)-\partial_{11}F(\mle_1|X_1,X_2)&\partial_{12}\loglikelihood_2(\mle_1,\mle_2|X_1,X_2)\\
    \partial_{12}\loglikelihood_2(\mle_1,\mle_2|X_1,X_2)^T&\partial_{22}\loglikelihood_2(\mle_1,\mle_2|X_1,X_2)
    \end{bmatrix}.
\end{align*}

It remains to control the norm of $V_n=-\loglikelihood'(\mle)$, which is done in \cref{lem:l_prime_n_asymptotic_proof} below, 
and to show that $H_n$ converges to a positive definite matrix $H_0$ with inverse as in \eqref{eq:Hinverse}.
By \cref{lem:h_n_asymptotic_proof} below, $H_n$ converges in probability to the same limit as 
\[
-\frac{1}{n}
    \begin{bmatrix}
    	\partial_{11} \loglikelihood_1(\mle_1|X_1) + \partial_{12}\loglikelihood_2(\mle|X_1,X_2)\left(\partial_2\partial_2\loglikelihood_2(\mle|X_1,X_2)\right)^{-1}\partial_{12}\loglikelihood_2(\mle|X_1,X_2)^T&\partial_{12}L_2(\mle|X_1,X_2)\\
    	\partial_{12}L_2(\mle|X_1,X_2)^T&\partial_{22} L_2(\mle|X_1,X_2)
    \end{bmatrix}.
\]\normalsize
 Hence, by the convergence of $\mle$ to $\theta^*$, by \cref{assump:finite_hessian_asymptotic} and by the uniform law of large numbers, $H_n\to H_0$ in probability; and finally $H_0$ is positive definite by \cref{assump:positive_definite_asymptotic}.

\qed

\subsubsection{Control of the norm of $L'(\mle)$}

\begin{lemma}\label{lem:l_prime_n_asymptotic_proof}
Under the assumptions of \cref{thm:bvm_cut}, we have, as $n\to\infty$,
\begin{align*}
    &\frac{1}{\sqrt{n}}\|L'(\mle|X_1,X_2)\|\\
    =&\frac{1}{\sqrt{n}}\left\|\frac{\int\exp\left[\loglikelihood_2(\mle_1,\theta_2|X_1,X_2)\right]\left[\partial_{1}\loglikelihood_2(\mle_1,\theta_2|X_1,X_2)-\partial_{1}\loglikelihood_2(\mle_1,\mle_2|X_1,X_2)\right]\pi_2(\theta_2)d\theta_2}{\int\exp\left[\loglikelihood_2(\mle_1,\theta_2|X_1,X_2)\right]\pi_2(\theta_2)d\theta_2}\right\|\xrightarrow{\mathbbm{P}}0.
\end{align*}
\end{lemma}
\begin{proof}
For simplicity, we shall write $\loglikelihood_1(\theta_1):=\loglikelihood_1(\theta_1|X_1)$,
and $\loglikelihood_2(\theta_1,\theta_2):=\loglikelihood_2(\theta_1,\theta_2|X_1,X_2)$.
Let us also take $\delta>0$, small enough that
\begin{itemize}
\item simultaneously \cref{assump:third_der_asymptotic,assump:der_bounds_asymptotic} are both satisfied; 
\item there exists a fixed $M>0$, independent of $n$, such that
\[
\lim_{n\to\infty}\mathbbm{P}\left[\forall \|s\|\leq\delta,\, \frac{1}{M}\leq \pi_2(\mle_2+s)\leq M\text{ and }\|\pi_2'(\mle_2+s)\|\leq M\right]=1,
\]
which is possible by \cref{assump:priors_asymptotic} and the assumption that $\mle_2\to\truepar_2$ in probability;
\item there exists a fixed $\kappa>0$, independent of $n$, such that
\[
\lim_{n\to\infty}\mathbbm{P}\left[-\frac{1}{n}\partial_{22}\loglikelihood_2(\mle)-\frac{\delta}{3}\asymptthirdder I_{d_2}\succcurlyeq\kappa I_{d_2}\right]=1,
\]
which is possible by \cref{assump:third_der_asymptotic,assump:positive_definite_asymptotic,assump:finite_hessian_asymptotic}, the assumption that $\mle\xrightarrow{\mathbbm{P}}\truepar$, and the uniform law of large numbers.
\end{itemize}
Moreover, let us take $\epsilon>0$, such that
\begin{align*}
&\lim_{n\to\infty}\mathbbm{P}\left[\underset{\|\theta_2-\mle_2\|>\delta}{\sup}\frac{1}{n}\left(\loglikelihood_2(\mle_1,\theta_2)-\loglikelihood_2(\mle_1,\mle_2)\right)\leq - \epsilon\right]=1,
\end{align*}
which is possible by \cref{assump:tails_asymptotic}.

Recalling \eqref{eq:cutloglikelihood}, we see that the derivative of $L$ with respect to $\theta_2$ at $\mle_2$ is $\partial_2 L_2(\mle_1,\mle_2) = 0$, since $\mle_2$ maximizes $L_2(\mle_1,\cdot)$. Thus the gradient of $L$ with respect to $\theta$ is only non-zero in its first $d_1$ components, and there it can be written 
\begin{align}
&\partial_1 \loglikelihood_1(\mle_1)+\partial_1\loglikelihood_2(\mle_1,\mle_2)-\partial_{1}\left[\log\int\exp(\loglikelihood_2(\cdot,\theta_2))\pi_2(\theta_2)d\theta_2   \right]\Bigg|_{\cdot=\mle_1}\notag\\
    =&-\partial_{1}\left[\log\int\exp\left[\loglikelihood_2(\cdot,\theta_2)-\loglikelihood_2(\cdot,\mle_2)\right]\pi_2(\theta_2)d\theta_2   \right]\Bigg|_{\cdot=\mle_1}\notag
\end{align}
where we use $\partial_1 \loglikelihood_1(\mle_1)=0$, added and removed $\partial_1 \loglikelihood_2(\mle_1,\mle_2)$ from the last term, and simplified.   
We then differentiate the logarithm, then exchange integration and differentiation, which is allowed with probability converging to one by \cref{assump:integrability_asymptotic},
to obtain
\begin{align}
    =&-\frac{\int\exp\left[\loglikelihood_2(\mle_1,\theta_2)-\loglikelihood_2(\mle_1,\mle_2)\right]\partial_{1}\left[\loglikelihood_2(\cdot,\theta_2)-\loglikelihood_2(\cdot,\mle_2)\right]_{\cdot=\mle_1}\pi_2(\theta_2)d\theta_2 }{\int\exp\left[\loglikelihood_2(\mle_1,\theta_2)-\loglikelihood_2(\mle_1,\mle_2)\right]\pi_2(\theta_2)d\theta_2} 
    =:A_1+A_2,\label{eq:l_prime_asymp_dec}
\end{align}
where $A_1$ and $A_2$ are obtained by changing the variable $\theta_2$ for $\mle_2+t/\sqrt{n}$ and splitting the integral in the numerator into two parts, over $\|t\|<\delta\sqrt{n}$ and $\|t\|>\delta\sqrt{n}$ respectively.

\subsubsection*{Controlling $A_1$}
We have, by Taylor's expansion,
for $\|t\|<\delta\sqrt{n}$,
\begin{align}\label{eq:def_r_n}
    &\loglikelihood_2(\mle_1,n^{-1/2}t+\mle_2)-\loglikelihood_2(\mle_1,\mle_2)=\frac{1}{2n}t^T[\partial_{22}\loglikelihood_2(\mle_1,\mle_2)]t+t^TR_n(t)t,
    \end{align}
    where $\|R_n(t)\|_{op}\leq \frac{\asymptthirdder}{6\sqrt{n}}\|t\|\leq \frac{\delta \asymptthirdder}{6}$. Indeed, note that
\begin{align*}
    R_n(t)=&\frac{1}{n}\left[\int_0^1(1-r)\partial_{22}\loglikelihood_2(\mle_1,\mle_2+n^{-1/2}rt)dr - \frac{1}{2}\partial_{22}\loglikelihood_2(\mle_1,\mle_2)\right] \\
    =&\frac{1}{n}\int_0^1(1-r)\left[\partial_{22}\loglikelihood_2(\mle_1,\mle_2+n^{-1/2}rt) - \partial_{22}\loglikelihood_2(\mle_1,\mle_2)\right]dr,
\end{align*}
whence, via \cref{assump:third_der_asymptotic},
\begin{align*}
    \|R_n(t)\|_{op}\leq &\frac{1}{n}\int_0^1(1-r)\left\|\partial_{22}\loglikelihood_2(\mle_1,\mle_2+n^{-1/2}rt) - \partial_{22}\loglikelihood_2(\mle_1,\mle_2)\right\|_{op}dr\\
    \leq &\asymptthirdder\int_{0}^1(1-r)n^{-1/2}r\|t\|dr
    =\frac{\asymptthirdder}{6\sqrt{n}}\|t\|.
\end{align*}
We can now write
\begin{align}
    &\|A_1\|\\
    \leq&\left\|\frac{\int_{\|t\|<\delta\sqrt{n}}\exp\left[-\frac{1}{2}t^T\left(-\frac{1}{n}\partial_{22}\loglikelihood_2(\mle) - 2R_n(t)\right)t\right] \partial_{1}\left[\loglikelihood_2(\cdot,n^{-1/2}t+\mle_2)-\loglikelihood_2(\cdot,\mle_2)\right]_{\cdot=\mle_1}\pi_2(n^{-1/2}t+\mle_2)dt   }{\int_{\|t\|<\delta\sqrt{n}}\exp\left[-\frac{1}{2}t^T\left(-\frac{1}{n}\partial_{22}\loglikelihood_2(\mle) - 2R_n(t)\right)t\right]\pi_2(n^{-1/2}t+\mle_2)dt}\right\|\notag.
\end{align}\normalsize
Then we note that, in the integrand within the numerator, using the mean value theorem,
\begin{align*}
&\partial_1\loglikelihood_2(\mle_1,n^{-1/2}t+\mle_2)-\partial_{1}\loglikelihood_2(\mle_1,\mle_2)\\
=&\frac{t^T}{\sqrt{n}}\partial_{12}\loglikelihood_2(\mle_1,\mle_2)+\frac{t^T}{\sqrt{n}}\int_0^1\left[\partial_{12}\loglikelihood_2(\mle_1,\mle_2+n^{-1/2}rt)-\partial_{12}\loglikelihood_2(\mle_1,\mle_2)\right]dr,
\end{align*}
and the norm of the last term on the right-hand side is less than 
\begin{align*}
    & \left\|\frac{t}{\sqrt{n}}\right\|\,\,\int_{0}^1\left\|\partial_{12}\loglikelihood_2(\mle_1,\mle_2+n^{-1/2}rt)-\partial_{12}\loglikelihood_2(\mle_1,\mle_2)\right\|_{op}dr\\
    \leq &
    \sup_{s\neq 0,\|s\|\leq\delta}\frac{\left\|\partial_{12}\loglikelihood_2(\mle_1,\mle_2+s)-\partial_{12}\loglikelihood_2(\mle_1,\mle_2)\right\|_{op}}{\|s\|}\; \cdot \;
    \left\|\frac{t}{\sqrt{n}}\right\|^2\int_0^1rdr \leq M_4(X_1,X_2) \|t\|^2 / 2,
\end{align*}
by \cref{assump:der_bounds_asymptotic}.
Thus, we obtain
\begin{align}
    \|A_1\| \leq&\left\|\frac{\int_{\|t\|<\delta\sqrt{n}}\exp\left[-\frac{1}{2}t^T\left(-\frac{1}{n}\partial_{22}\loglikelihood_2(\mle) - 2R_n(t)\right)t\right]\frac{t^T}{\sqrt{n}}\partial_{12}\loglikelihood_2(\mle_1,\mle_2)\,\pi_2(n^{-1/2}t+\mle_2)dt  }{\int_{\|t\|<\delta\sqrt{n}}\exp\left[-\frac{1}{2}t^T\left(-\frac{1}{n}\partial_{22}\loglikelihood_2(\mle) - 2R_n(t)\right)t\right]\pi_2(n^{-1/2}t+\mle_2)dt}\right\|\notag\\
& +\frac{M_4(X_1,X_2)}{2}\cdot\frac{\int_{\|t\|<\delta\sqrt{n}}\exp\left[-\frac{1}{2}t^T\left(-\frac{1}{n}\partial_{22}\loglikelihood_2(\mle) - 2R_n(t)\right)t\right]\left\|t\right\|^2\,\pi_2(n^{-1/2}t+\mle_2)dt  }{\int_{\|t\|<\delta\sqrt{n}}\exp\left[-\frac{1}{2}t^T\left(-\frac{1}{n}\partial_{22}\loglikelihood_2(\mle) - 2R_n(t)\right)t\right]\pi_2(n^{-1/2}t+\mle_2)dt}\notag\\
=:& \|A_{11}\|+\|A_{12}\|,\label{eq:a_1_split}
\end{align}

\subsubsection*{Controlling $A_{12}$.}
We note that, for large enough $n$, using the lower and upper bounds on $\pi_2$ within the ball of radius $\delta$ around $\mle_2$,
and the bound on $R_n(t)$ in \eqref{eq:def_r_n},
\begin{align}
    \frac{1}{\sqrt{n}}\|A_{12}\| 
    \leq & \frac{\asymptdermixed}{2\sqrt{n}}\cdot M^2 \cdot\frac{\int_{\|t\|<\delta\sqrt{n}}\exp\left[-\frac{1}{2}t^T\left(-\frac{1}{n}\partial_{22}\loglikelihood_2(\mle)-\frac{\delta}{3}\asymptthirdder I_{d_2}\right)t\right]\|t\|^2dt  }{\int_{\|t\|<\delta\sqrt{n}}\exp\left[-\frac{1}{2}t^T\left(-\frac{1}{n}\partial_{22}\loglikelihood_2(\mle)+\frac{\delta}{3}\asymptthirdder I_{d_2}\right)t\right]dt}\notag\\
    &\xrightarrow[n\to\infty]{\mathbbm{P}}0,\label{eq:control_A_12}
\end{align}
since both integrals converge to constants, using the assumption that $-\frac{1}{n}\partial_{22}\loglikelihood_2(\mle)-\frac{\delta}{3}\asymptthirdder I_{d_2}$
ultimately dominates $\kappa I_{d_2}$ in the numerator, and for the denominator we note that $-n^{-1} \partial_{22} L_2(\mle)$ is positive definite with probability going to one by \cref{assump:positive_definite_asymptotic} and the uniform law of large numbers.

\subsubsection*{Controlling $A_{11}$.}
For $A_{11}$, in the numerator we subtract and add $\pi_2(\mle_2)$ to $\pi_2(\mle_2+tn^{-1/2})$, and use the triangle inequality.
This results in $\|A_{11}\|\leq \|A_{111}\|+\|A_{112}\|$, where $\|A_{112}\|$ has $\pi_2(\mle_2)$ instead of $\pi_2(\mle_2+tn^{-1/2})$ in the numerator. 
Next note that
\[
\int_{\|t\|<\delta\sqrt{n}}\exp\left[-\frac{1}{2}t^T\left(-\frac{1}{n}\partial_{22}\loglikelihood_2(\mle)\right)t\right] t \; dt=0,
\]
as it is the integral of an odd function. Recall the inequality $|e^x-1|\leq |x|e^{|x|}$, and note that, by \cref{eq:def_r_n}, for $\|t\|\leq \delta\sqrt{n}$,
\begin{align*}
    \left|\exp[t^TR_n(t)t]-1\right|\leq &\|R_n(t)\|_{op}\|t\|^2\exp\left[\|R_n(t)\|_{op}\|t\|^2\right]\\
    \leq& \frac{\asymptthirdder\|t\|^{3}}{6n^{1/2}}\exp\left[\frac{\delta\asymptthirdder}{6}\|t\|^2\right].
\end{align*}
Hence, on an event with probability going to one,
\begin{align*}
    &\frac{1}{\sqrt{n}}\|A_{112}\|\\
    =& \frac{1}{n}\left\|\frac{\partial_{12}\loglikelihood_2(\mle_1,\mle_2) \pi_2(\mle_2) \cdot \int_{\|t\|<\delta\sqrt{n}}\exp\left[-\frac{1}{2}t^T\left(-\frac{1}{n}\partial_{22}\loglikelihood_2(\mle)\right)t\right]\left(\exp\left[t^TR_n(t)t\right] - 1\right)t\,dt  }{\int_{\|t\|<\delta\sqrt{n}}\exp\left[-\frac{1}{2}t^T\left(-\frac{1}{n}\partial_{22}\loglikelihood_2(\mle) - 2R_n(t)\right)t\right]\pi_2(n^{-1/2}t+\mle_2)dt} \right\|\\
    \leq &\frac{\asymptthirdder|\pi_2(\mle_2)|\|\partial_{12}\loglikelihood_2(\mle_1,\mle_2)\|_{op}\int_{\|t\|<\delta\sqrt{n}}\exp\left[-\frac{1}{2}t^T\left(-\frac{1}{n}\partial_{22}\loglikelihood_2(\mle)-\frac{\delta\asymptthirdder}{3}I_{d_2}\right)t\right]\|t\|^4dt }{6n^{3/2}\int_{\|t\|<\delta\sqrt{n}}\exp\left[-\frac{1}{2}t^T\left(-\frac{1}{n}\partial_{22}\loglikelihood_2(\mle)+\frac{\delta\asymptthirdder}{3}I_{d_2}\right)t\right]\pi_2(n^{-1/2}t+\mle_2)dt}\\
    \leq &\frac{\asymptthirdder\asymptdermixed M^2}{6n^{1/2}}\cdot\frac{\int_{\|t\|<\delta\sqrt{n}}\exp\left[-\frac{1}{2}t^T\left(-\frac{1}{n}\partial_{22}\loglikelihood_2(\mle)-\frac{\delta\asymptthirdder}{3}I_{d_2}\right)t\right]\|t\|^4dt }{\int_{\|t\|<\delta\sqrt{n}}\exp\left[-\frac{1}{2}t^T\left(-\frac{1}{n}\partial_{22}\loglikelihood_2(\mle)+\frac{\delta\asymptthirdder}{3}I_{d_2}\right)t\right]dt}\\
    &\xrightarrow[n\to\infty]{\mathbbm{P}}0,
\end{align*}
similarly to the control of $A_{12}$ above.
By \cref{assump:priors_asymptotic} and the mean value theorem, for some $\tilde{t}$, such that $\|\tilde{t}\|\leq\delta\sqrt{n}$,
$\pi_2(n^{-1/2}t+\mle_2)-\pi_2(\mle_2) = \pi_2'(n^{-1/2}\tilde{t}+\mle_2)n^{-1/2}t$, and using \cref{eq:def_r_n},
\begin{align*}
    \frac{1}{\sqrt{n}}\|A_{111}\|
    \leq & \frac{\asymptdermixed\sup_{\|s\|\leq\delta}\pi_2'(\mle_2+s)}{n^{1/2}\inf_{\|s\|\leq\delta}\pi_2(\mle_2+s)}\\
    &\cdot\frac{\int_{\|t\|<\delta\sqrt{n}}\exp\left[-\frac{1}{2}t^T\left(-\frac{1}{n}\partial_{22}\loglikelihood_2(\mle)-\frac{\delta\asymptthirdder}{3}I_{d_2}\right)t\right]\|t\|^2\,\, dt  }{\int_{\|t\|<\delta\sqrt{n}}\exp\left[-\frac{1}{2}t^T\left(-\frac{1}{n}\partial_{22}\loglikelihood_2(\mle)+\frac{\delta\asymptthirdder}{3}I_{d_2}\right)t\right]dt}\xrightarrow[n\to\infty]{\mathbbm{P}}0,
\end{align*}
as in the control of $A_{12}$ above.
\subsubsection*{Controlling $A_2$}
It remains to control the term
\begin{align*}
    \frac{1}{\sqrt{n}}\|A_2\| 
    = & \frac{1}{\sqrt{n}\int\exp\left[\loglikelihood_2(\mle_1,\mle_2+n^{-1/2}t)-\loglikelihood_2(\mle_1,\mle_2)\right]\pi_2(\mle_2+n^{-1/2}t)dt}\\
    &\cdot\Bigg\|\int_{\|t\|>\delta\sqrt{n}}\exp\left[\loglikelihood_2(\mle_1,\mle_2+n^{-1/2}t)-\loglikelihood_2(\mle_1,\mle_2)\right]\\
    &\hspace{4cm}\cdot\partial_{1}\left[\loglikelihood_2(\cdot,\mle_2+n^{-1/2}t)-\loglikelihood_2(\cdot,\mle_2)\right]_{\cdot=\mle_1} \pi_2(\mle_2+n^{-1/2}t)dt   \Bigg\|.
\end{align*}
We appeal to the second part of \cref{assump:tails_asymptotic}, restrict the integral in the denominator to $\|t\|\leq \delta\sqrt{n}$,
and use \eqref{eq:def_r_n} in the integrand in the denominator, to obtain
\begin{align}
    \frac{1}{\sqrt{n}}\|A_2\| 
    \leq & \frac{1}{\sqrt{n}} e^{-n\gamma_2}\frac{\int_{\|t\|>\delta\sqrt{n}}\left[\left\|\partial_{1}\loglikelihood_2(\mle_1,n^{-1/2}t+\mle_2)\right\|+\left\|\partial_{1}\loglikelihood_2(\mle_1,\mle_2)\right\|\right]\pi_2(n^{-1/2}t+\mle_2)dt }{\int_{\|t\|<\delta\sqrt{n}}\exp\left[-\frac{1}{2}t^T\left(-\frac{1}{n}\partial_{22}\loglikelihood_2(\mle)+\frac{\delta\asymptthirdder}{3}I_{d_2}\right)t\right]\pi_2(n^{-1/2}t+\mle_2)dt}.\label{eq:bound_on_a2}
\end{align}
For the first term, completing the integral in the numerator and using  \cref{assump:integrability_asymptotic} and Jensen's inequality, we obtain
\[
\sqrt{n}e^{-n\gamma_2}\int_{\|t\|>\delta\sqrt{n}}\frac{1}{n}\left\|\partial_{1}\loglikelihood_2(\mle_1,n^{-1/2}t+\mle_2)\right\|\pi_2(n^{-1/2}t+\mle_2)dt \leq e^{-n\gamma_2} n^{(m+d_2-1)/2} M_5(X_1,X_2)^{1/2}.
\]
For the second term, we use \cref{assump:der_bounds_asymptotic}. This leads to 
\begin{align*}
\frac{1}{\sqrt{n}}\|A_2\|    \leq & e^{-n\gamma_2} \frac{1}{\inf_{\|s\|\leq\delta}\pi_2(\mle_2+s)}\cdot\frac{\asymptintegrals^{1/2} n^{(m+d_2-1)/2} + \asymptdermixed n^{(d_2-1)/2}  }{\int_{\|t\|<\delta\sqrt{n}}\exp\left[-\frac{1}{2}t^T\left(-\frac{1}{n}\partial_{22}\loglikelihood_2(\mle)+\frac{\delta\asymptthirdder}{3}I_{d_2}\right)t\right]dt},
\end{align*}
and therefore 
$\frac{1}{\sqrt{n}}\|A_2\|\xrightarrow{\mathbbm{P}}0$ as $n\to\infty$.

\end{proof}

\subsubsection{Limit of $H_n$}

\begin{lemma}\label{lem:h_n_asymptotic_proof}
Under the assumptions of \cref{thm:bvm_cut}, as $n\to\infty$,
\begin{align}
&\frac{1}{n}\left\|\partial_{11}\loglikelihood_2(\mle_1,\mle_2|X_1,X_2)-\partial_{11}F(\mle_1|X_1,X_2)\right.\notag\\
&\left.- \partial_{12}\loglikelihood_2(\mle|X_1,X_2)\left(\partial_2\partial_2\loglikelihood_2(\mle|X_1,X_2)\right)^{-1}\partial_{12}\loglikelihood_2(\mle|X_1,X_2)^T  \right\|_{op}\xrightarrow{\mathbbm{P}}0.\label{eq:h_n_asymptotic_proof}
\end{align}
\end{lemma}

\begin{proof}
We use the same notation as in the proof of \cref{lem:l_prime_n_asymptotic_proof}. Let us take $\delta_2>0$, small enough that
\begin{itemize}
\item simultaneously \cref{assump:third_der_asymptotic,assump:der_bounds_asymptotic} are both satisfied; 
\item there exists a fixed $M>0$, independent of $n$, such that
\[
\lim_{n\to\infty}\mathbbm{P}\left[\forall \|s\|\leq\delta_2,\, \frac{1}{M}\leq \pi_2(\mle_2+s)\leq M\text{ and }\|\pi_2'(\mle_2+s)\|\leq M\right]=1,
\]
which is possible by \cref{assump:priors_asymptotic} and the assumption that $\mle_2\to\truepar_2$ in probability;
\item there exists a fixed $\kappa>0$, independent of $n$, such that
\[
\lim_{n\to\infty}\mathbbm{P}\left[-\frac{1}{n}\partial_{22}\loglikelihood_2(\mle)-\frac{\delta_2}{3}\asymptthirdder I_{d_2}\succcurlyeq\kappa I_{d_2}\right]=1,
\]
which is possible by \cref{assump:third_der_asymptotic,assump:positive_definite_asymptotic,assump:finite_hessian_asymptotic}, the assumption that $\mle\xrightarrow{\mathbbm{P}}\truepar$, and the uniform law of large numbers.
\end{itemize}
Moreover, let us take $\gamma_2>0$, such that
\begin{align*}
&\lim_{n\to\infty}\mathbbm{P}\left[\underset{\|\theta_2-\mle_2\|>\delta_2}{\sup}\frac{1}{n}\left(\loglikelihood_2(\mle_1,\theta_2)-\loglikelihood_2(\mle_1,\mle_2)\right)\leq - \gamma_2\right]=1,
\end{align*}
which is possible by \cref{assump:tails_asymptotic}.
Now, differentiating under the integral sign, and writing $v^{\otimes 2}=vv^T$ for a vector $v$,
\begin{align}
    \partial_{11} F(\mle_1) &= 
    \frac{\int e^{\loglikelihood_2(\mle_1,\theta_2)} \partial_{11} \loglikelihood_2(\mle_1,\theta_2)\pi_2(d\theta_2)}{F(\mle_1)}
    + \frac{\int e^{\loglikelihood_2(\mle_1,\theta_2)} (\partial_1 \loglikelihood_2(\mle_1,\theta_2))^{\otimes 2}\pi_2(d\theta_2)}{F(\mle_1)}\notag\\
    &- \frac{\left(\int e^{\loglikelihood_2(\mle_1,\theta_2)} \partial_1 \loglikelihood_2(\mle_1,\theta_2)\pi_2(d\theta_2)\right)^{\otimes 2}}{F(\mle_1)^2}.\nonumber
\end{align}
We define three terms which, divided by $n$, go to zero, thus establishing Lemma~\ref{lem:h_n_asymptotic_proof}:
\begin{align}
    \bar{A}_1 &= \partial_{11} L_2(\mle) - \frac{\int e^{\loglikelihood_2(\mle_1,\theta_2)} \partial_{11} \loglikelihood_2(\mle_1,\theta_2)\pi_2(d\theta_2)}{F(\mle_1)},\nonumber\\
    \bar{A}_2 &= \frac{\int e^{\loglikelihood_2(\mle_1,\theta_2)} (\partial_1 \loglikelihood_2(\mle_1,\theta_2) - \partial_1 \loglikelihood_2(\mle))^{\otimes 2}\pi_2(d\theta_2)}{F(\mle_1)} +  \partial_{12}\loglikelihood_2(\mle)\left(\partial_2\partial_2\loglikelihood_2(\mle)\right)^{-1}\partial_{12}\loglikelihood_2(\mle)^T\nonumber\\
    \bar{A}_3 &= \left[\frac{\int e^{\loglikelihood_2(\mle_1,\theta_2)}\left(\partial_1\loglikelihood_2(\mle_1,\theta_2)-\partial_1\loglikelihood_2(\mle_1,\mle_2)\right)\pi_2(d\theta_2)}{F(\mle_1)}\right] \notag\\
    &\hspace{1cm}\cdot\left[\frac{\int e^{\loglikelihood_2(\mle_1,\theta_2)}\left(\partial_1\loglikelihood_2(\mle_1,\theta_2)-\partial_1\loglikelihood_2(\mle_1,\mle_2)\right)\pi_2(d\theta_2)}{F(\mle_1)}\right]^T,\label{eq:cov_comparison_bound_split}
\end{align}
and it can be checked that $\bar{A}_1-\bar{A}_2+\bar{A}_3$ is equal to the expression within the norm in \eqref{eq:h_n_asymptotic_proof}.

\paragraph{Controlling term $\bar{A}_1$. }
We start from the expression
\begin{align}
	&\frac{1}{n}\|\bar{A}_1\|_{op}\notag\\
    &=\frac{1}{n}\Bigg\|\frac{\int e^{\loglikelihood_2(\mle_1,\mle_2+t/\sqrt{n})-\loglikelihood_2(\mle_1,\mle_2)}\left[\partial_{11}\loglikelihood_2(\mle_1,\mle_2+t/\sqrt{n})-\partial_{11}\loglikelihood_2(\mle_1,\mle_2)\right]\pi_2(\mle_2+t/\sqrt{n})dt}{\int e^{\loglikelihood_2(\mle_1,\mle_2+t\sqrt{n})-\loglikelihood_2(\mle_1,\mle_2)}\pi_2(\mle_2+t/\sqrt{n})dt}\Bigg\|_{op},\label{eq:a_1_bar_starting_expression}
\end{align}
and we can deal with the denominator as in the control of $A_{12}$ in \eqref{eq:control_A_12}: we restrict the integral to $\|t\|\leq \delta_2\sqrt{n}$, use \cref{eq:def_r_n} 
to bound $\loglikelihood_2(\mle_1,\mle_2+t\sqrt{n})-\loglikelihood_2(\mle_1,\mle_2)$, use the bound on the prior density $\pi_2$ around $\mle_2$ and finally obtain that the denominator is of order $1$ in probability as $n\to\infty$.
For the numerator, we split the integral into two parts, $\|t\|\leq \delta_2\sqrt{n}$ and $\|t\|>\delta_2\sqrt{n}$.
On the one hand, with probability going to $1$ as $n\to\infty$, by \cref{assump:der_bounds_asymptotic} and the bound on the prior density $\pi_2$ around $\mle_2$, we have
\begin{align}
	&\frac{1}{n}\Bigg\|\int_{\|t\|\leq \delta_2\sqrt{n}} e^{\loglikelihood_2(\mle_1,\mle_2+t/\sqrt{n})-\loglikelihood_2(\mle_1,\mle_2)}\left[\partial_{11}\loglikelihood_2(\mle_1,\mle_2+t/\sqrt{n})-\partial_{11}\loglikelihood_2(\mle_1,\mle_2)\right]\pi_2(\mle_2+t/\sqrt{n})dt \Bigg\|_{op}\label{eq:a_1_bar_num1}\\
    &\leq n^{-1/2} \cdot M_4(X_1,X_2) \cdot M \cdot \int_{\|t\|\leq \delta_2\sqrt{n}} e^{\loglikelihood_2(\mle_1,\mle_2+t/\sqrt{n})-\loglikelihood_2(\mle_1,\mle_2)} \|t\| dt ,\notag
\end{align}
and we use \cref{eq:def_r_n} to bound the norm of the integral by a quantity of order $1$ in probability as $n\to\infty$. Thus the above expression goes to zero. On the other hand,
\begin{align}
	&\frac{1}{n}\Bigg\|\int_{\|t\|> \delta_2\sqrt{n}} e^{\loglikelihood_2(\mle_1,\mle_2+t/\sqrt{n})-\loglikelihood_2(\mle_1,\mle_2)}\left[\partial_{11}\loglikelihood_2(\mle_1,\mle_2+t/\sqrt{n})-\partial_{11}\loglikelihood_2(\mle_1,\mle_2)\right]\pi_2(\mle_2+t/\sqrt{n})dt \Bigg\|_{op}\label{eq:a_1_bar_num2}\\
    &\leq \frac{1}{n} e^{-n \gamma_2} n^{d_2/2}  \cdot \Bigg\|\int_{\|\theta_2-\mle_2\|> \delta_2} \left[\partial_{11}\loglikelihood_2(\mle_1,\theta_2)-\partial_{11}\loglikelihood_2(\mle_1,\mle_2)\right]\pi_2(d\theta_2) \Bigg\|_{op},\notag
\end{align}
and the norm of the integral is less than the integral of the norm which can be bounded by a polynomial in $n$, via \cref{assump:integrability_asymptotic}. The resulting expression goes to zero as $n\to\infty$ thanks to the exponential term $e^{-n\gamma_2}$. Thus $n^{-1}\|\bar{A}_1\|_{op}\xrightarrow{\mathbbm{P}}0$ as $n\to\infty$.

\paragraph{Controlling term $\bar{A}_3$. }
The term $n^{-1} \bar{A}_3$ goes to zero in probability as it is smaller than $(n^{-1/2} \|L'(\mle)\|)^2$ and $n^{-1/2} \|L'(\mle)\|$ goes to zero by \cref{lem:l_prime_n_asymptotic_proof}.

\paragraph{Controlling term $\bar{A}_2$.}
We start by splitting the integal within $\bar{A}_{2}$ and applying the triangle inequality to bound 
$n^{-1}\|\bar{A}_2\|_{op}$ by $n^{-1}\|\bar{A}_2\|_{op} \leq n^{-1}\|\bar{A}_{21}\|_{op} + n^{-1}\|\bar{A}_{22}\|_{op}$, with 
\begin{align}
    \|\bar{A}_{21}\|_{op} &= \Bigg\| \frac{\int_{\|\theta_2-\mle_2\| \leq \delta_2} e^{\loglikelihood_2(\mle_1,\theta_2)} (\partial_1 \loglikelihood_2(\mle_1,\theta_2) - \partial_1 \loglikelihood_2(\mle))^{\otimes 2}\pi_2(d\theta_2)}{F(\mle_1)}\notag\\
    &\hspace{1cm}+  \partial_{12}\loglikelihood_2(\mle)\left(\partial_2\partial_2\loglikelihood_2(\mle)\right)^{-1}\partial_{12}\loglikelihood_2(\mle)^T \Bigg\|_{op},\notag\\
    \|\bar{A}_{22}\|_{op} &= \Bigg\|\frac{\int_{\|\theta_2-\mle_2\| > \delta_2} e^{\loglikelihood_2(\mle_1,\theta_2)} (\partial_1 \loglikelihood_2(\mle_1,\theta_2) - \partial_1 \loglikelihood_2(\mle))^{\otimes 2}\pi_2(d\theta_2)}{F(\mle_1)} \Bigg\|_{op}.\label{eq:a_2_bar_split}
\end{align}

\paragraph*{Controlling $\bar{A}_{22}$.}
We apply a change of variable $\theta_2 \leftrightarrow t = \sqrt{n}(\theta_2 - \mle_2)$, divide numerator and denominator by $\exp(\loglikelihood_2(\mle))$, 
and deal with the denominator as previously (see the control of the term $\bar{A}_1$).
Thus we focus on the resulting numerator:
\begin{align*}
	&\int_{\|t\|>\delta_2\sqrt{n}} e^{\loglikelihood_2(\mle_1,\mle_2+t/\sqrt{n}) - \loglikelihood_2(\mle)}\left(\partial_1\loglikelihood_2(\mle_1,\mle_2+t/\sqrt{n})-\partial_1\loglikelihood_2(\mle_1,\mle_2)\right)^{\otimes 2}\pi_2(\mle_2+t/\sqrt{n})dt,
\end{align*}
and we see that we can bound the exponential term by $\exp(-n\gamma_2)$, as in the control of $\bar{A}_{1}$, and employ \cref{assump:der_bounds_asymptotic,assump:integrability_asymptotic} to bound the other terms by a polynomial in $n$. Thus $n^{-1}\|\bar{A}_{22}\|_{op}$ goes to zero as $n\to\infty$.

\paragraph*{Controlling $\bar{A}_{21}$.}
Note that, for all $\|t\|\leq \delta_2\sqrt{n}$, there is some $\overline{R}_n(t)$ in $\mathbbm{R}^{d_1\times d_2}$, such that, via the mean value theorem,
\begin{align}\label{eq:def_r_n_bar}
	&\partial_1\loglikelihood_2(\mle_1,\mle_2+t/\sqrt{n})-\partial_1\loglikelihood_2(\mle_1,\mle_2)=\partial_{12}\loglikelihood_2(\mle_1,\mle_2)t/\sqrt{n}+ \overline{R}_n(t) \, t,
    \end{align}
    with $\|\overline{R}_n(t) \, t \|_{op}\leq \frac{\asymptdermixed}{2} \|t\|^2.$
We then expand $\left(\partial_1\loglikelihood_2(\mle_1,\mle_2+t/\sqrt{n})-\partial_1\loglikelihood_2(\mle_1,\mle_2)\right)^{\otimes 2}$ as
\begin{align*}
    &\frac{ (\partial_{12}\loglikelihood_2(\mle_1,\mle_2)t)^{\otimes 2}}{n} +  \frac{\left[(\partial_{12}\loglikelihood_2(\mle_1,\mle_2)t)(\overline{R}_n(t)t)^T + (\overline{R}_n(t)t)(\partial_{12}\loglikelihood_2(\mle_1,\mle_2)t)^T\right]}{\sqrt{n}} + (\overline{R}_n(t) \, t)^{\otimes 2}.
\end{align*}
We obtain
\begin{align}
   & \|\bar{A}_{21}\|_{op} \\
    \leq& \Bigg\| \frac{\int_{\|t\| \leq \delta_2\sqrt{n}} e^{\loglikelihood_2(\mle_1,\mle_2+t/\sqrt{n})} \frac{(\partial_{12} \loglikelihood_2(\mle) t )^{\otimes 2}}{n}\pi_2(\mle_2+t/\sqrt{n})dt}{\int e^{\loglikelihood_2(\mle_1,\mle_2+t/\sqrt{n})}\pi_2(\mle_2+t/\sqrt{n})dt} +  \partial_{12}\loglikelihood_2(\mle)\left(\partial_2\partial_2\loglikelihood_2(\mle)\right)^{-1}\partial_{12}\loglikelihood_2(\mle)^T \Bigg\|_{op}\notag\\
    &+ \frac{1}{\int e^{\loglikelihood_2(\mle_1,\mle_2+t/\sqrt{n})}\pi_2(\mle_2+t/\sqrt{n})dt} \notag\\
    &\cdot\Bigg\|\hspace{-1mm}\int_{\|t\| \leq \delta_2\sqrt{n}}\hspace{-1mm} e^{\loglikelihood_2(\mle_1,\mle_2+t/\sqrt{n})}\Bigg\{ \hspace{-1mm}\frac{\left[(\partial_{12}\loglikelihood_2(\mle_1,\mle_2)t)(\overline{R}_n(t)t)^T + (\overline{R}_n(t)t)(\partial_{12}\loglikelihood_2(\mle_1,\mle_2)t)^T\right]}{\sqrt{n}}\notag \\
    & \hspace{8.5cm}+ (\overline{R}_n(t) \, t)^{\otimes 2}\Bigg\}
    \pi_2(\mle_2+t/\sqrt{n})dt\Bigg\|_{op}\notag\\
    &= \|\bar{A}_{211}\|_{op} + \|\bar{A}_{212}\|_{op}.\label{eq:a_21_bar_split}
\end{align}
We first deal with $\bar{A}_{212}$. We multiply and divide by $\exp(\loglikelihood_2(\mle))$,
and use the Taylor expansion of $\loglikelihood_2$ in \eqref{eq:def_r_n}. For the denominator we restrict the integral to $\|t\|\leq \delta_2\sqrt{n}$
and use a reasoning similar to the one in the control of $\bar{A}_1$ to show that it is of order $1$ in probability as $n\to\infty$.
For the numerator, we use \eqref{eq:def_r_n} to write
\begin{align*}
    &\left\|\int_{\|t\| \leq \delta_2\sqrt{n}} \hspace{-1mm}\exp\left( \hspace{-1mm}-\frac{1}{2}t^T \hspace{-1mm} \left(-\frac{1}{n}\partial_{22} L_2(\mle) - 2R_n(t)\right) \hspace{-1mm} t \hspace{-1mm}\right)
    \frac{(\partial_{12}\loglikelihood_2(\mle_1,\mle_2)t)(\overline{R}_n(t)t)^T}{\sqrt{n}} \pi_2(\mle_2+t/\sqrt{n})dt\right\|_{op}\\
    &\leq 
    \frac{M}{\sqrt{n}}\int_{\|t\| \leq \delta_2\sqrt{n}} \exp\left(-\frac{1}{2}t^T \left(-\frac{1}{n}\partial_{22} L_2(\mle) + \frac{\delta_2 M_1(X_1,X_2)}{3} I_{d_2}\right) t\right)
    \frac{M_4(X_1,X_2)^2}{2} \|t\|^3dt,
\end{align*}
using \cref{assump:der_bounds_asymptotic} and \eqref{eq:def_r_n_bar}. The resulting expression is of order $n^{-1/2}$ in probability as $n\to\infty$.
The exact same reasoning applies to the term with  $(\overline{R}_n(t)t) (\partial_{12}\loglikelihood_2(\mle_1,\mle_2)t)^T$. The last term is that with $(\overline{R}_n(t) \, t)^{\otimes 2}$, and we bound its operator norm by $M_4(X_1,X_2)^2/4 \|t\|^4$ using \eqref{eq:def_r_n_bar}, and we obtain a term of order $1$ in probability as $n\to\infty$. Thus, $n^{-1}\|\bar{A}_{212}\|_{op}$ goes to zero in probability as $n\to\infty$. It remains to control $\|\bar{A}_{211}\|_{op}$.
We first write the term $\bar{A}_{211}$ as 
\begin{align}
 &   \frac{\int_{\|t\| \leq \delta_2\sqrt{n}} e^{\loglikelihood_2(\mle_1,\mle_2+t/\sqrt{n})} n^{-1}(\partial_{12} \loglikelihood_2(\mle) t )^{\otimes 2}\pi_2(\mle_2+t/\sqrt{n})dt}{\int e^{\loglikelihood_2(\mle_1,\mle_2+t/\sqrt{n})}\pi_2(\mle_2+t/\sqrt{n})dt} 
    \hspace{-1mm}+  \hspace{-1mm}\partial_{12}\loglikelihood_2(\mle)\hspace{-1mm}\left(\partial_2\partial_2\loglikelihood_2(\mle)\right)^{-1}\hspace{-1mm}\partial_{12}\loglikelihood_2(\mle)^T \notag\\
  &  = \frac{\int_{\|t\| \leq \delta_2\sqrt{n}} e^{\loglikelihood_2(\mle_1,\mle_2+t/\sqrt{n})} n^{-1} \partial_{12} \loglikelihood_2(\mle)\left(t t^T + \left(\frac{1}{n}\partial_{22}L_2(\mle)\right)^{-1} \right) \partial_{12} \loglikelihood_2(\mle)^T \pi_2(\mle_2+t/\sqrt{n})dt}{\int e^{\loglikelihood_2(\mle_1,\mle_2+t/\sqrt{n})}\pi_2(\mle_2+t/\sqrt{n})dt} \notag
   \\
    &+   \frac{\int_{\|t\| > \delta_2\sqrt{n}} e^{\loglikelihood_2(\mle_1,\mle_2+t/\sqrt{n})} \pi_2(\mle_2+t/\sqrt{n})dt}{\int e^{\loglikelihood_2(\mle_1,\mle_2+t/\sqrt{n})}\pi_2(\mle_2+t/\sqrt{n})dt}  \left\{\partial_{12}\loglikelihood_2(\mle)\left(\partial_2\partial_2\loglikelihood_2(\mle)\right)^{-1}\partial_{12}\loglikelihood_2(\mle)^T\right\}\notag\\
    &=: \bar{A}_{2111} + \bar{A}_{2112}.\label{eq:a_211_split}
\end{align}
Using the triangle inequality, we need to control the norm of these two terms. The norm of the second term, $\|\bar{A}_{2112}\|_{op}$, goes to zero in probability as $n\to\infty$ using \cref{assump:tails_asymptotic},
and the convergence of the term in curly brackets to a constant matrix.

For the first term $\bar{A}_{2111}$, the crux is the identity $\frac{\int \exp(-\frac{1}{2}x^TAx)x x^T dx}{\int \exp(-\frac{1}{2}x^T Ax) dx} =A^{-1}$ for a positive definite matrix $A$,
where here $A = -\frac{1}{n}\partial_{22}\loglikelihood_2(\mle)$.
To keep the equations legible, we thus focus on the part of $\bar{A}_{2111}$ that reads:
\begin{align}
(*) = \frac{\int_{\|t\| \leq \delta_2\sqrt{n}} e^{\loglikelihood_2(\mle_1,\mle_2+t/\sqrt{n})}
\left(t t^T + \left(\frac{1}{n}\partial_{22}L_2(\mle)\right)^{-1} \right) \pi_2(\mle_2+t/\sqrt{n})dt}{\int e^{\loglikelihood_2(\mle_1,\mle_2+t/\sqrt{n})}\pi_2(\mle_2+t/\sqrt{n})dt}. \label{eq:a_2111_star}
\end{align}
If the norm of that term goes to zero, then $n^{-1} \|\bar{A}_{211}\|_{op}$ goes to zero because $(n^{-1}\partial_{12} L_2(\mle))$ goes to a constant in probability as $n\to\infty$. Restricting the integral in the denominator, adding and subtracting $\pi_2(\mle_2)$ to the integrand in the numerator,  and using the triangle inequality, the norm of $(*)$ is bounded by
\begin{align}
    & \Bigg\|
\frac{\int_{\|t\| \leq \delta_2\sqrt{n}} e^{\loglikelihood_2(\mle_1,\mle_2+t/\sqrt{n})}
\left(t t^T + \left(\frac{1}{n}\partial_{22}L_2(\mle)\right)^{-1} \right) \pi_2(\mle_2)dt}{
\int_{\|t\| \leq \delta_2\sqrt{n}} e^{\loglikelihood_2(\mle_1,\mle_2+t/\sqrt{n})}\pi_2(\mle_2+t/\sqrt{n})dt}
\Bigg\|_{op}\nonumber\\
&+ \Bigg\|
\frac{\int_{\|t\| \leq \delta_2\sqrt{n}} e^{\loglikelihood_2(\mle_1,\mle_2+t/\sqrt{n})}
\left(t t^T + \left(\frac{1}{n}\partial_{22}L_2(\mle)\right)^{-1} \right) \{\pi_2(\mle_2+t/\sqrt{n})-\pi_2(\mle_2)\}dt}{
\int_{\|t\| \leq \delta_2\sqrt{n}} e^{\loglikelihood_2(\mle_1,\mle_2+t/\sqrt{n})}\pi_2(\mle_2+t/\sqrt{n})dt}
\Bigg\|_{op}.\label{eq:a_bar_2111_priordensitydiff}
\end{align}
Upon dividing numerators and denominator by $\exp(L_2(\mle))$, we proceed as before to control the resulting denominator, e.g. in the control of $\bar{A}_1$. We focus on
the first numerator,
\begin{align*}
&\pi_2(\mle_2)    \Bigg\|
\int_{\|t\| \leq \delta_2\sqrt{n}} e^{\loglikelihood_2(\mle_1,\mle_2+t/\sqrt{n})-\loglikelihood_2(\mle)}
\left(t t^T + \left(\frac{1}{n}\partial_{22}L_2(\mle)\right)^{-1} \right) dt
\Bigg\|_{op}\\
&\leq M \Bigg\|
\int_{\|t\| \leq \delta_2\sqrt{n}} e^{-\frac{1}{2}t^T (-\frac{1}{n}\partial_{22}L_2(\mle))t} e^{t ^T R_n(t)t}
\left(t t^T + \left(\frac{1}{n}\partial_{22}L_2(\mle)\right)^{-1} \right) dt
\Bigg\|_{op},
\end{align*}
where we have used \eqref{eq:def_r_n}. Then we replace $\exp(tR_n(t) t)$ by 
$1 + (\exp(tR_n(t) t) - 1)$. We note that 
\begin{align*}
&\Bigg\|
\int_{\|t\| \leq \delta_2\sqrt{n}} e^{-\frac{1}{2}t^T (-\frac{1}{n}\partial_{22}L_2(\mle))t} 
\left(t t^T + \left(\frac{1}{n}\partial_{22}L_2(\mle)\right)^{-1} \right) dt
\Bigg\|_{op} \xrightarrow[n\to\infty]{\mathbb{P}} 0,
\end{align*}
because $\int e^{-x^T A x/ 2} xx^T dx=A^{-1} \int e^{-x^T A x / 2}dx$, for any positive definite $A$, and because the domain of the integral above converges to the entire space.
Regarding the term with $\exp(t^T R_n(t) t) - 1$, we use the bound $|e^a - 1| \leq |a| e^{|a|}$, 
which implies by \eqref{eq:def_r_n} that 
\begin{align*}
    \left|\exp(t^T R_n(t) t) - 1\right| \leq \frac{\|t\|^3 M_1(X_1,X_2)}{6\sqrt{n}} \exp\left(\frac{\delta_2 M_1(X_1,X_2)}{6}\|t\|^2\right),
\end{align*}
and therefore
\begin{align*}
&\Bigg\|
\int_{\|t\| \leq \delta_2\sqrt{n}} e^{-\frac{1}{2}t^T (-\frac{1}{n}\partial_{22}L_2(\mle))t} 
\left(e^{t^TR_n(t) t} - 1\right)\left(t t^T + \left(\frac{1}{n}\partial_{22}L_2(\mle)\right)^{-1} \right) dt
\Bigg\|_{op}\\
&\leq 
\frac{1}{\sqrt{n}}
\int_{\|t\| \leq \delta_2\sqrt{n}} e^{-\frac{1}{2}t^T (-\frac{1}{n}\partial_{22}L_2(\mle) - \frac{\delta_2 M_1(X_1,X_2)}{3})t} 
\frac{\|t\|^3 M_1(X_1,X_2)}{6} \left(t t^T + \left(\frac{1}{n}\partial_{22}L_2(\mle)\right)^{-1} \right) dt\\
&\xrightarrow[n\to\infty]{\mathbbm{P}}0
,
\end{align*}
thanks to the $n^{-1/2}$ term in front of the norm.
For the second numerator, in \eqref{eq:a_bar_2111_priordensitydiff},
we use the same reasoning as in the control of $A_{111}$ in the proof of \cref{lem:l_prime_n_asymptotic_proof}:
$\pi_2(\mle_2+t/\sqrt{n})-\pi_2(\mle_2)=\pi^\prime_2(\mle_2 + \tilde{t}/\sqrt{n}) t / \sqrt{n}$
and $\sup_{\|s\|\leq \delta_2} \pi^\prime(\mle_2 + s)\leq M$ on an event of probability going to one. On these events, the numerator in \eqref{eq:a_bar_2111_priordensitydiff} is upper-bounded by
\begin{align*}
    & \frac{M}{\sqrt{n}} \Bigg\| \int_{t\leq \delta_2 \sqrt{n}} e^{-\frac{1}{2}t^T (-\frac{1}{n}\partial_{22} L_2(\mle) - \frac{\delta_2 M_1(X_1,X_2)}{3}) t}\left(t t^T + \left(\frac{1}{n}\partial_{22}L_2(\mle)\right)^{-1} \right) t\, dt \Bigg\|_{op},
\end{align*}
which goes to zero in probability as $n\to\infty$. This concludes the proof that $(*)$ has a norm going to zero,
and thus $n^{-1}\|\bar{A}_{211}\|_{op}$ goes to zero in probability as $n\to\infty$, and this finally concludes the proof of \cref{lem:h_n_asymptotic_proof}.\end{proof}

\subsection{Verifying the assumptions\label{suppl:verifying_assumptions}}

\begin{example}
    The assumptions of \cref{thm:bvm_cut} do not require the model to be well-specified. Take the biased data example, where
    \[
    f_1(x_1|\theta_1)=\frac{1}{(2\pi)^{d_1/2}}e^{-\|x_1-\theta_1\|^2/2},\qquad f_2(x_2|\theta_1,\theta_2,x_1)=\frac{1}{(2\pi)^{d_1/2}}e^{-\|x_2-\theta_1-\theta_2\|^2/2}
    \]
    and the priors $\pi_1$ and $\pi_2$ are centred Gaussian. Let us not assume anything about the data generating process, except that the pairs $\{(X_{1,j},X_{2,j})\}_{j=1}^{\infty}$ are i.i.d. with a finite mean. We see that
    $\mle_1=\overline{X}_1$, $\mle_2=\overline{X}_2-\mle_1$,
    and, using Slutsky's theorem,
    \[
    \theta_1^*=\mathbbm{E}[X_{1,1}],\qquad\theta_2^*=\mathbbm{E}[X_{1,2}]-\mathbbm{E}[X_{1,1}]\qquad\text{and}\qquad (\mle_1,\mle_2)\xrightarrow[n\to\infty]{\mathbbm{P}}(\theta_1^*,\theta_2^*).
    \]
    Now, for Gaussian priors, \cref{assump:priors_asymptotic} is satisfied. As for \cref{assump:tails_asymptotic}, if $\delta_2>\delta_1$ then
    \begin{align*}
&\underset{\|\theta_2-\mle_2\|>\delta_2}{\sup_{\|\theta_1-\mle_1\|\leq\delta_1}}\frac{\loglikelihood_2(\theta_1,\theta_2|X_1,X_2)-\loglikelihood_2(\mle_1,\mle_2|X_1,X_2)}{n}\\
\leq &\sup_{\|\theta_1+\theta_2-\mle_1-\mle_2\|>\delta_2-\delta_1}\frac{\sum_{i=1}^{n}\|x_{2,i}-\mle_1-\mle_2\|^2-\sum_{i=1}^{n}\|x_{2,i}-\theta_1-\theta_2\|^2}{2n}\\
=&\sup_{\|\theta_1+\theta_2-\mle_1-\mle_2\|>\delta_2-\delta_1}\sum_{i=1}^{n}\frac{\left<\theta_1+\theta_2-\mle_1-\mle_2,2x_{2,i}-\theta_1-\theta_2-\mle_1-\mle_2\right>}{2n}\\
=&\sup_{\|\theta_1+\theta_2-\mle_1-\mle_2\|>\delta_2-\delta_1}\sum_{i=1}^{n}\frac{\left<\theta_1+\theta_2-\mle_1-\mle_2,x_{2,i}-\theta_1-\theta_2\right>}{2n}\\
=&\sup_{\|\theta_1+\theta_2-\mle_1-\mle_2\|>\delta_2-\delta_1}\frac{\left<\theta_1+\theta_2-\mle_1-\mle_2,\mle_1+\mle_2-\theta_1-\theta_2\right>}{2}\\
\leq & -\frac{1}{2}(\delta_2-\delta_1)^2=-\gamma_2
\end{align*}
and
\begin{align*}
\sup_{\|\theta_1-\mle_1\|>\delta_1}\frac{\loglikelihood_1(\theta_1,\theta_2|X_1)-\loglikelihood_1(\mle_1,\mle_2|X_1)}{n}=&\sup_{\|\theta_1-\mle_1\|>\delta_1}\sum_{i=1}^n\frac{\left<\theta_1-\mle_1,2x_{1,i}-\theta_1-\mle_1\right>}{2n}\\
=&\sup_{\|\theta_1-\mle_1\|>\delta_1}\frac{\left<\theta_1-\mle_1,\mle_1-\theta_1\right>}{2}
\leq -\frac{\delta_1^2}{2}=-\gamma_1.
\end{align*}
Hence, \cref{assump:tails_asymptotic} is satisfied.
Furthermore, for $s\neq 0$, with $\|s\|\leq \delta$,
\begin{align}
    \frac{|\loglikelihood_2(\mle+s|X_1,X_2)-\loglikelihood_2(\mle|X_1,X_2)|}{n\|s\|}=\left|\frac{\left<-s_1-s_2,s_1-s_2\right>}{2\|s\|}\right|=\frac{\|s_1+s_2\|^2}{2\|s\|}\leq& \frac{\|s_1\|^2+\|s_2\|^2}{\|s\|}\notag\\
    =&\|s\|\leq \delta.\label{eq:ex_assump_der}
\end{align}
Hence, \cref{assump:third_der_asymptotic} is satisfied for any $\delta>0$ and with $M_1(X_1,X_2)=\delta$.
Moreover, \cref{assump:finite_hessian_asymptotic} is satisfied with $M_2=M_3=1$, \cref{assump:positive_definite_asymptotic} is satisfied with $J_1(\theta_1)=J_2(\theta_1,\theta_2)=I_{d_1}$ and \cref{assump:der_bounds_asymptotic} is satisfied for any $\delta>0$ and with $M_4(X_1,X_2)=\max\{1,\delta/2\}$ (by an argument similar to \cref{eq:ex_assump_der}). Finally, to verify \cref{assump:integrability_asymptotic}, note that
\begin{align*}
    \frac{1}{n^2}\int\left\|\partial_1L_2(\mle_1,\theta_2|X_1,X_2)\right\|^2\pi_2(\theta_2)d\theta_2=& \frac{1}{n^2}\int\left\|\sum_{i=1}^n\left(X_{2,i}-\mle_1-\theta_2\right)\right\|^2\pi_2(\theta_2)d\theta_2\\
    =&\int\left\|\mle_2-\theta_2\right\|^2\pi_2(\theta_2)d\theta_2\leq 2\|\mle_2\|^2+2\Tr[\Sigma_2],
\end{align*}
where $\Sigma_2$ is the covariance of $\pi_2$. Hence, \cref{assump:integrability_asymptotic} is satisfied with \\ $M_5(X_1,X_2)=\max\left\{1,2\|\mle_2\|^2,2\Tr[\Sigma_2]\right\}$.
\end{example}

\section{Laplace approximation and its error}

This appendix is devoted to the proofs
of the results in Section~\ref{section:nonasymptotic_control}, namely
\cref{thm:non_asymptotic_simplified} and \cref{prop:biased_data}.
In \cref{sec:more_detailed_non_asymptotic},
we state regularity assumptions on the model and 
a non-asymptotic bound on the total variation (TV) distance
between the cut posterior and a Normal approximation, 
which we call the main bound (\cref{thm:without_correction1}).
The bound involves the term $\loglikelihood'(\mle)$, which is
not explicit, and 
the Normal approximation in  \cref{thm:without_correction1}
has covariance matrix equal to the inverse of minus the Hessian of the log-density of the cut posterior,
which is intractable because of the integral in the feedback term. Thus this covariance is different from the tractable matrix $J_{n}(\mle)$ in the proposed Laplace approximation in \cref{thm:non_asymptotic_simplified}.
Thus, we state technical results controlling 
$\loglikelihood'(\mle)$ (\cref{lem:gradient_bound_new}) as well as the difference between the covariance matrices (\cref{lem:covariance_comparison_new}).
Section~\ref{sec:more_detailed_non_asymptotic} ends with what we call our final bound, \cref{thm:non_asymptotic_final}, where the TV distance between the cut posterior and the Laplace approximation is bounded by explicit terms.

In \cref{sec:proof_of_theorem2}, we state the assumptions for \cref{thm:non_asymptotic_simplified}, and provide its proof, using \cref{thm:without_correction1}, \cref{lem:gradient_bound_new} and \cref{lem:covariance_comparison_new}. 
In \cref{sec:proof_of_main_bound} we prove \cref{thm:without_correction1},
and in \cref{sec:proof_technical_lemmas} we prove the technical lemmas.
In \cref{sec:proof_example_biased}, we prove \cref{prop:biased_data},
which is an application of Theorems~\ref{thm:without_correction1}-\ref{thm:non_asymptotic_final} to a concrete example, for which the bounds can be made explicit functions of the dimension of the parameter space.

\subsection{Detailed, non-asymptotic bounds on the error}\label{sec:more_detailed_non_asymptotic}

In this section, we shall derive a general non-asymptotic bound on the error of the Laplace approximation, in 
\cref{thm:non_asymptotic_final}. First, we present our assumptions,
and then we state an intermediate result, \cref{thm:without_correction1},
which we call the main bound. It controls the error between the cut posteriors
and a centered Normal distribution with covariance matrix given by the inverse of minus the Hessian of the cut posterior log-density.

\subsubsection{Assumptions}\label{sec:non_asymptotic_assumptions}
Recall the notation given in \cref{eq:cutloglikelihood,eq:loglikelihoods}. We introduce the following notation:
\begin{align}\label{eq:logposterior}
\logposterior(\theta_1,\theta_2)=\loglikelihood(\theta_1,\theta_2)+\log\pi_1(\theta_1)+\log\pi_2(\theta_2),
\end{align}
and, recalling the definition of $J_n(\hat{\theta})$ in \cref{eq:j_n_simplified},
we let
\begin{align}
	\fisherinformation(\mle)=-\frac{\loglikelihood''(\mle)}{n};\qquad \postfisherinformation(\mle)=-\frac{\logposterior''(\mle)}{n};\qquad \jtilde=-\frac{1}{n}\partial_{22}\loglikelihood_2(\mle);\qquad\jbreve=-\frac{\loglikelihood_1''(\mle_1)}{n}.\label{eq:j_bar_j_hat}
\end{align}
In what follows, $B_{\eta}(x)$ denotes the open ball of radius $\eta>0$ around $x$. 
We require that there exist $\delta_1,\delta_2,\delta_3, \epsilon_1,\epsilon_2>0$ such that the following assumptions hold.
\begin{assumption}\label{assump:m_estimator1}
    The two-step M-estimator (2SM) $\mle=(\mle_1,\mle_2)$ in \cref{eq:2sM} is unique. 
\end{assumption}
\begin{assumption}\label{assump:fisher_inform1}
    All of $\postfisherinformation(\mle),\fisherinformation(\mle),\jtilde,\jbreve$ in \cref{eq:j_bar_j_hat} are strictly positive definite.
\end{assumption}
\begin{remark}
    In order to verify that the posterior Fisher information matrix $\postfisherinformation(\mle)$ is strictly positive definite and its eigenvalues are lower bounded, we may use \cref{lem:covariance_comparison_new} below, together with Weyl's inequality. Indeed, all we need to do is verify that the matrix $J_n(\mle)$ of \cref{eq:j_n_simplified}
    is positive definite with lower-bounded eigenvalues, which can be done numerically. If $n$ is large enough then $\postfisherinformation(\mle)$ is automatically positive definite with eigenvalues bounded away from zero. The positive-definiteness of $\fisherinformation(\mle)$ can be verified in an analogous way, using \cref{lem:covariance_comparison_new} below.
\end{remark}
\begin{assumption}\label{assump:third_der1}
	The generalized log-cut-posterior $\logposterior$ is two times differentiable inside $\{\param:\|\param-\mle\|\leq \delta_1\vee\delta_2\} $. There exists  $\thirdder>0$ such that:
	\begin{align*}
		{\sup_{s\neq 0,\,\|s_{1:d_1}\|\leq \delta_1,\|s_{(1+d_1):d_2}\|\leq \delta_2}}\frac{\left\|\postfisherinformation(\mle)^{-1/2}\left(\logposterior''(\mle+s)-\logposterior''(\mle)\right)\postfisherinformation(\mle)^{-1/2}\right\|_{op}}{n\left\|s\right\|}\leq \thirdder.
	\end{align*}
    Moreover,
    $\sqrt{\delta_1^2+\delta_2^2}\, \thirdder<1$.
    
\end{assumption}

\begin{assumption}\label{assump:third_der_modules1}
There exist $\thirdderone,\thirddertwo>0$, such that
    \begin{align*}
		&\sup_{t\neq 0,\,\|t\|\leq \delta_1}({n\left\|t\right\|})^{-1}{\left\|\left(\loglikelihood_1''(\mle_1+t)-\loglikelihood_1''(\mle_1)\right)\right\|_{op}}\leq \thirdderone;\\
    &\sup_{s\neq 0,\,\|s\|\leq \delta_2}({n\left\|s\right\|})^{-1}{\left\|\left(\partial_{22}\loglikelihood_2(\mle_1,\mle_2+s)-\partial_{22}\loglikelihood_2(\mle_1,\mle_2)\right)\right\|_{op}}\leq \thirddertwo.
    \end{align*}
\end{assumption}

\begin{assumption}\label{assump:tails1}
For some $\epsilon_1,\epsilon_2>0$, 
\begin{align*}
    \sup_{\|\theta-\mle_1\|>\delta_1}\frac{\loglikelihood_1(\theta)-\loglikelihood_1(\mle_1)}{n}\leq -\epsilon_1;\qquad {\sup_{\|\theta_1-\mle_1\|\leq\delta_1,\|\theta_2-\mle_2\|>\delta_2}}\frac{\loglikelihood_2(\theta_1,\theta_2)-\loglikelihood_2(\mle_1,\mle_2)}{n}\leq -\epsilon_2.
    \end{align*}
    
\end{assumption}

\begin{assumption}\label{assump:derivative_bounds_new1}
There exist $\firstder,\secondderone,\thirddermixed,\thirddermixedone,\secondder,\seconddermixed>0$ such that 
$\delta_1\firstder+\delta_3\secondder<\epsilon_2$ and 
    \begin{align*}
    &\sup_{\|s_1\|<\delta_1,\|s_2\|<\delta_3}\frac{\left\|\partial_1\loglikelihood_2(\mle_1+s_1,\mle_2+s_2)\right\|}{n}\leq\firstder;\quad \sup_{\|s_1\|<\delta_1,\|s_2\|<\delta_3}\frac{\left\|\partial_2\loglikelihood_2(\mle_1+s_1,\mle_2+s_2)\right\|}{n}\leq\secondder;\quad\\
    &  \sup_{\|s\|<\delta_2}n^{-1}{\left\|\partial_{11}\loglikelihood_2(\mle_1,\mle_2+s)\right\|_{op}}\leq \secondderone;\quad \sup_{\|s\|<\delta_2} n^{-1}{\left\|\partial_{12}\loglikelihood_2(\mle_1,\mle_2+s)\right\|_{op}}\leq \seconddermixed;\\
    &\sup_{s\neq 0,\,\|s\|\leq \delta_2} (n\|s\|)^{-1}{\left\|\partial_{11}\loglikelihood_2(\mle_1,\mle_2+s)-\partial_{11}\loglikelihood_2(\mle_1,\mle_2)\right\|_{op}}\leq\thirddermixedone;\\
    &\sup_{s\neq 0,\|s\|\leq \delta_2} ({n\|s\|})^{-1}{\left\|\partial_{12}\loglikelihood_2(\mle_1,\mle_2+s)-\partial_{12}\loglikelihood_2(\mle_1,\mle_2)\right\|_{op}}\leq\thirddermixed.
    \end{align*}
\end{assumption}


\begin{assumption}\label{assump:der_integrability_new}
There exist $\firstdersquaredintegral, \secondderintegral>0$, such that
\[
\frac{1}{n^2}\int_{\|\theta-\mle_2\|>\delta_2}\left\|\partial_1L_2(\mle_1,\theta)\right\|^2\pi_2(\theta)d\theta\leq\firstdersquaredintegral;\quad \frac{1}{n}\int_{\|\theta-\mle_2\|>\delta_2}\left\|\partial_{11}\loglikelihood_2(\mle_1,\theta)\right\|_{op}\pi_2(\theta)d\theta\leq\secondderintegral.
\]
 Moreover, for some $\delta>0$, the following integrals exist:
\[
\int\sup_{\|\theta_1-\mle_1\|<\delta}\left\|\partial_{11}L_2(\theta_1,\theta)\right\|_{op}\pi_2(\theta)d\theta\qquad\text{and}\qquad \int\sup_{\|\theta_1-\mle_1\|<\delta}\left\|\partial_1L_2(\theta_1,\theta)\right\|^2\pi_2(\theta)d\theta.
\]

\end{assumption}

\begin{assumption}\label{assump:priors11}
	There exist $\priorboundone, \priorboundtwo, \priorboundthree, \priorboundfour, \priorboundfive$ such that
	\begin{align*}
		&{}{\sup_{\|\param_1-\mle_1\|\leq\delta_1,\|\param_2-\mle_2\|\leq\delta_2}}\left|({\prior_1(\param_1)\prior_2(\param_2)})^{-1}\right|\leq \priorboundone;\quad \sup_{\|\param_1-\mle_1\|\leq\delta_1}\left|{\prior_1(\param_1)}^{-1}\right|\leq \priorboundtwo;\quad 
	\sup_{\|\param_2-\mle_2\|\leq\delta_2}\left|{\prior_2(\param_2)}^{-1}\right|\leq \priorboundthree;
	\\
        &{\sup_{\|\param_1-\mle_1\|\leq\delta_1,\|\param_2-\mle_2\|\leq\delta_2}}\left|\prior_1(\param_1)\prior_2(\param_2)\right|\leq \priorboundfour;\qquad
        {\sup_{\|\param_1-\mle_1\|\leq\delta_1,\|\param_2-\mle_2\|\leq\delta_2}}\left| (\log(\prior_1(\param_1)\prior_2(\param_2)))'\right|\leq \priorboundfive.
	\end{align*}
\end{assumption}

\begin{assumption}\label{assump:priors21}
	There exist $\priorboundsix, \priorboundseven, \priorboundthirdder>0$ such that
	\begin{align*}
       & \sup_{\|\param_2-\mle_2\|\leq\delta_2}\left|\prior_2(\param_2)\right|\leq \priorboundsix;\quad\sup_{\|\theta_2-\hat{\theta}_2\| \leq \delta_2}\| \pi_2'(\theta_2)\|\leq\priorboundseven;\\
    &\sup_{t\neq 0, \|t\|\leq \delta_2}(\|t\|)^{-1}{\|(\log\pi_2)''(\mle_2+t)-(\log\pi_2)''(\mle_2)\|}\leq \priorboundthirdder.
	\end{align*}
\end{assumption}

\subsubsection{Main bound}\label{sec:non_asymptotic_main_bound}
In order to state our first main bound, \cref{thm:without_correction1}, and other non-asymptotic results below, we need some additional notation:
\begin{align*}
        &\postminevaluemle=\left\|\postfisherinformation(\mle)^{-1}\right\|_{op}^{-1};\qquad\decayhatbar :={}
    \exp\left[-\frac{1}{2}
        \left((\delta_1\wedge \delta_2)\sqrt{n}-        \sqrt{\Tr(\postfisherinformation(\mle)^{-1})}\right)^2\postminevaluemle
        \right];\\
    &\jhatplus:=\jbreve+\frac{\delta_1\thirdderone}{3}I_{d_1};\\
    &\minevalmleplus=\left\|\jhatplus^{-1}\right\|_{op}^{-1};  \qquad \decayhatplus :=
    \exp\left[-\frac{1}{2}
        \left(\delta_1\sqrt{n}-
        \sqrt{\Tr\left[\left(\jhatplus\right)^{-1}\right]}\right)^2\minevalmleplus
        \right].
\end{align*}
We also write $d:=d_1+d_2$, and $\mathcal{N}$ for the Normal distribution.

\begin{thm}\label{thm:without_correction1}
 Suppose that \cref{assump:m_estimator1,assump:fisher_inform1,assump:third_der1,assump:tails1,assump:priors11} hold. Moreover, assume that 
\begin{align}\label{eq:n_size_main}
n>\max\left\{\frac{\Tr(\postfisherinformation(\mle)^{-1})}{(\delta_1\wedge\delta_2)^2},\,\frac{\Tr(\jhatplus^{-1})}{(\delta_1)^2}\right\}.
\end{align}
Then
\begin{align*}
    &\mathrm{TV}\left(
        \tilde{\pi}_{\rm cut}\, ,\,
            \mathcal{N}(0,\postfisherinformation(\mle)^{-1})
    \right)
    \leq (C_1+C_2)n^{-1/2}+C_3n^{d_1/2}e^{-n\epsilon_1}+C_4n^{d_2/2}e^{-n(\epsilon_2-\delta_1\firstder-\delta_3\secondder)}+2\decayhatbar,
\end{align*}
for
\begin{align*}
&C_1=\frac{\|\postfisherinformation(\mle)^{-1/2}L'(\mle)\|}{\sqrt{\left(1-\sqrt{\delta_1^2+\delta_2^2}\thirdder\right)}};\quad C_2=\frac{1}{\sqrt{\left(1-\sqrt{\delta_1^2+\delta_2^2}\thirdder\right)}}\cdot\frac{\thirdder\sqrt{d^2+2d}}{2\sqrt{1-\decayhatbar}}+\frac{\priorboundfive}{2\sqrt{\left(1-\sqrt{\delta_1^2+\delta_2^2}\thirdder\right)}},\\
&C_3=\frac{2\priorboundtwo\det{\jhatplus}^{1/2}}{(2\pi)^{d_1/2}(1-\decayhatplus)};\quad C_4=\frac{2\priorboundthree\Gamma(d_2/2+1)}{\left(\delta_3\sqrt{\pi}\right)^{d_2}},
\end{align*}
where $\Gamma$ is Euler's Gamma function. 
\end{thm}
\cref{thm:without_correction1} is proved in \cref{sec:proof_of_main_bound}.

\subsubsection{Additional lemma: controlling $\|\loglikelihood'(\mle)\|$}\label{sec:non_asymptotic_gradient}
In order to control $\|\loglikelihood'(\mle)\|$, which appears in $C_1$ above, we provide \cref{lem:gradient_bound_new}. We define
\begin{align}
\label{eq:def:jbartwo}
&\jbartwo:=\jtilde-\frac{1}{n}(\log\pi_2)''(\mle_2);\qquad
\jbarminus:=\jbartwo-\delta_2\left(\frac{\thirddertwo}{3}+\frac{\priorboundthirdder}{3n}\right) I_{d_2};\\
&
\jbarplus:=\jbartwo+\delta_2\left(\frac{\thirddertwo}{3}+\frac{\priorboundthirdder}{3n}\right)I_{d_2};\qquad \minevalbarplus=\left\|\jbarplus^{-1}\right\|_{op}^{-1}; \nonumber\\
        &\pminus:=\left\|\jbarminus^{-1}(\log\pi_2)'(\mle_2)\right\|;\qquad\expplus:=\exp\left(\frac{1}{2n}(\log\pi_2)'(\mle_2)^T\jbarplus^{-1}(\log\pi_2)'(\mle_2)\right);\nonumber\\
        &\expminus:=\exp\left(\frac{1}{2n}(\log\pi_2)'(\mle_2)^T\jbarminus^{-1}(\log\pi_2)'(\mle_2)\right);\nonumber\\
        &\decaybarplus:=
    \exp\left[-\frac{1}{2}
        \left(\delta_2\sqrt{n}-\sqrt{\Tr\left[\left(\jbarplus\right)^{-1}\right]}-\frac{\|\jbarplus^{-1}(\log\pi_2)'(\mle_2)\|}{\sqrt{n}}\right)^2\minevalbarplus
        \right].\nonumber
\end{align}

We have the following result:
\begin{lemma}\label{lem:gradient_bound_new}
    Suppose \cref{assump:priors11,assump:priors21,assump:third_der_modules1,assump:der_integrability_new,assump:tails1,assump:derivative_bounds_new1} hold and retain the notation introduced therein.  Assume that $\jbarminus$ is positive definite. Suppose that
    \begin{align}\label{eq:n_size_grad_bound}
n >
\left(
({2\delta_2})^{-1}\left(\sqrt{\Tr\!\left[\left(\jbarplus\right)^{-1}\right]}
+
\sqrt{ \Tr\!\left[\left(\jbarplus\right)^{-1}\right] + 4 \delta_2 \,\left\|\jbarplus^{-1}(\log\pi_2)'(\hat\theta_2)\right\|}
\right)
\right)^2.
    \end{align}
    Then
    \begin{align*}
\|L'(\mle)\|\leq D_1'+D_2'+D_3'+D_4'n^{d_2/2+1}e^{-n\epsilon_2},
\end{align*}
where
\begin{align*}
&D_1'=\frac{\thirddermixed\left(\Tr[\jbarminus^{-1}]+\frac{1}{n}\left(\pminus\right)^2\right)\det{\jbarplus}^{1/2}\expminus}{\det{\jbarminus}^{1/2}\left(1-\decaybarplus\right)\,\expplus};\\
&D_2':=\frac{3\seconddermixed\det{\jbarplus}^{1/2}(\thirddertwo+\priorboundthirdder/n)\,\Tr[\jbarminus^{-1}]^2 \left(\expminus\right)^2}{\det{\jbarminus}^{1/2}(1-\decaybarplus)\expplus};\\
&D_3':=\frac{3\seconddermixed\det{\jbarplus}^{1/2} \|(\log\pi_2)'(\mle_2)\|\,\Tr[\jbarminus^{-1}]\left(\expminus\right)^2}{\det{\jbarminus}^{1/2}(1-\decaybarplus)\expplus};\quad D_4'=\frac{\det{\jbarplus}^{1/2}\left(\sqrt{\firstdersquaredintegral}+\firstder\right)}{(2\pi)^{d_2/2}\left(1-\decaybarplus\right)\,\expplus}.
\end{align*}
\end{lemma}
\Cref{lem:gradient_bound_new} is proved in \cref{sec:proof_gradient_bound}.

\subsubsection{Additional lemma: controlling $\|\postfisherinformation(\mle)-J_{n}(\mle)\|_{op}$}\label{sec:non_asymptotic_covariance}
Now, we present results to upper-bound the distance between the covariance matrix appearing in \cref{thm:without_correction1} and the tractable matrix $J_n$ given in \cref{eq:j_n_simplified}, which can be used in practice. We introduce an additional piece of notation, with $\jbartwo$ as in \cref{eq:def:jbartwo}:
\begin{align*}
&
\minevalpost=\left\|\jbartwo^{-1}\right\|_{op}^{-1},\quad\decaybartwo:=
    \exp\left[-\frac{1}{2}
        \left(\delta_2\sqrt{n}-\sqrt{\Tr\left[\left(\jbartwo\right)^{-1}\right]}\right)^2\minevalpost
        \right].
\end{align*}
\begin{lemma}\label{lem:covariance_comparison_new}
Retain the assumptions of \cref{lem:gradient_bound_new} and assume that 
\begin{align}\label{eq:n_size_cov_comparison}
n>(\delta_2)^{-2}{\Tr\left[\left(\jbartwo\right)^{-1}\right]}{}.
\end{align}
    Then, we have
    \begin{align*}
\left\|\postfisherinformation(\mle)-J_{n}(\mle)\right\|_{op} \leq& \frac{1}{n}(D_1'+D_2'+D_3'+D_4'n^{d_2/2+1}e^{-n\epsilon_2})^2 + \frac{1}{\sqrt{n}}(E_1'+E_2'n^{(d_2+3)/2}e^{-n\epsilon_2}+E_3'+E_4'+E_5'),
\end{align*}
where $\postfisherinformation(\mle)$ is defined in \cref{eq:j_bar_j_hat} and $J_{n}(\mle)$ defined in \cref{eq:j_n_simplified},
$D_1',\dots,D_4'$ are given in \cref{lem:gradient_bound_new} and
\begin{align*}
    &E_1'=\frac{\thirddermixedone\left(\sqrt{\Tr[\jbarminus^{-1}]}+\frac{1}{\sqrt{n}}\pminus\right)\det{\jbarplus}^{1/2}\expminus}{\det{\jbarminus}^{1/2}(1-\decaybarplus)\,\expplus};\\
    &E_2'=\frac{\priorboundthree\det{\jbarplus}^{1/2}}{(2\pi)^{d_2/2}(1-\decaybarplus)\,\expplus}\left(2\firstdersquaredintegral+2\left(\firstder\right)^2+\frac{\secondderintegral+\secondderone}{n}+\frac{(\seconddermixed)^2}{n\minevalpost}\right); \\
    &E_3'= \frac{8\thirddermixed\det{\jbarplus}^{1/2}\expminus}{\det{\jbarminus}^{1/2}(1-\decaybarplus)\,\expplus}\cdot\Bigg[\seconddermixed\Bigg(2\Tr\left[\jbarminus^{-1}\right]^{3/2}+\frac{\left(\pminus\right)^{3}}{n^{3/2}}\Bigg)\\
    &\hspace{7cm}+\thirddermixed\Bigg(\frac{3\Tr\left[\jbarminus^{-1}\right]^{2}}{\sqrt{n}}+\frac{\left(\pminus\right)^{4}}{n^{5/2}}\Bigg)\Bigg];\\
    &E_4'= \frac{\sqrt{8}(\seconddermixed)^2\det{\jbarplus}^{1/2}\,\left(\expminus\right)^2}{\det{\jbarminus}^{1/2}(1-\decaybarplus)\,\expplus}\Bigg\{\left(\thirddertwo+\frac{\priorboundthirdder}{n}\right)\hspace{-1mm}\Bigg[\Tr\left[\jbarminus^{-1}\right]^{5/2}\\
    &\hspace{1cm}+\frac{\Tr\left[\jbarminus^{-1}\right]^{3/2}}{\minevalpost}\Bigg]+\|(\log\pi_2)'(\mle_2)\|\Bigg[\Tr\left[\jbarminus^{-1}\right]^{3/2}+\frac{\Tr\left[\jbarminus^{-1}\right]^{1/2}}{\minevalpost}\Bigg]\Bigg\};\\
    &E_5'= \frac{\sqrt{n}(\seconddermixed)^2\det{\jbarplus}^{1/2}\left(\sqrt{3\decaybartwo}\,\Tr\left[\jbartwo^{-1}\right]+\decaybartwo\left(\minevalpost\right)^{-1}\right)}{\det{\jbartwo}^{1/2}(1-\decaybarplus)\,\expplus}.
\end{align*}
\end{lemma}

\begin{remark}
    The complexity of the bound in \cref{lem:gradient_bound_new}, as well as in \cref{lem:covariance_comparison_new}, is due to our efforts to handle the dependence on dimension carefully.
\end{remark}
\Cref{lem:covariance_comparison_new} is proved in \cref{sec:proof_of_covaraince_comparison}.
As a result of \cref{lem:covariance_comparison_new}, we have the following useful comparison of the two Gaussian measures of interest, which we prove in \cref{sec:proof_gaussian_comparison}.
\begin{innercustomthm}\label{prop:gaussian_comparison}
    Under the assumptions of \cref{lem:covariance_comparison_new}, we have
    \begin{align*}
        &\mathrm{TV}\left(
            \mathcal{N}(0,\postfisherinformation(\mle)^{-1}), \mathcal{N}(0,J_{n}(\mle)^{-1})
    \right)\\
    \leq& \frac{3}{2}\sqrt{d_1}\min\left\{\|J_{n}(\mle)^{-1}\|_{op},\|\postfisherinformation(\mle)^{-1}\|_{op}\right\}\\
    &\cdot\left\{\frac{1}{n}(D_1'+D_2'+D_3'+D_4'n^{d_2/2+1}e^{-n\epsilon_2})^2+\frac{1}{\sqrt{n}}(E_1'+E_2'n^{(d_2+3)/2}e^{-n\epsilon_2}+E_3'+E_4'+E_5')\right\},
    \end{align*}
    where $D_1',\dots,D_4'$ are given in \cref{lem:gradient_bound_new} and $E_1',\dots,E_5'$ are given in \cref{lem:covariance_comparison_new}.
\end{innercustomthm}

\subsubsection{Final bound}

Now, we are ready to present our final bound.

\begin{thm}\label{thm:non_asymptotic_final}
 Retain the notation of \cref{sec:non_asymptotic_assumptions,sec:non_asymptotic_main_bound,sec:non_asymptotic_gradient,sec:non_asymptotic_covariance}. Suppose that \cref{assump:m_estimator1,assump:fisher_inform1,assump:third_der1,assump:tails1,assump:priors11,assump:priors11,assump:priors21,assump:third_der_modules1,assump:der_integrability_new,assump:tails1,assump:derivative_bounds_new1} hold. Moreover, assume that 
\begin{align*}
n>\max\left\{\frac{\Tr(\postfisherinformation(\mle)^{-1})}{(\delta_1\wedge\delta_2)^2},\,\frac{\Tr(\jhatplus^{-1})}{(\delta_1)^2},\frac{\Tr(\jbartwo^{-1})}{(\delta_2)^2},\frac{\Tr(\jbarplus^{-1}) + \delta_2 \,\left\|\jbarplus^{-1}(\log\pi_2)'(\hat\theta_2)\right\|}{(\delta_2)^2}\right\}.
\end{align*}
Then
\begin{align*}
    &\mathrm{TV}\left(
        \tilde{\pi}_{\rm cut}\, ,\,
            \mathcal{N}(0,J_n(\mle)^{-1})
    \right)\\
    \leq&\frac{\left\|\postfisherinformation(\mle)^{-1/2}\right\|_{op}\left(D_1'+D_2'+D_3'+D_4'n^{d_2/2+1}e^{-n\epsilon_2}\right)}{\sqrt{\left(1-\sqrt{\delta_1^2+\delta_2^2}\thirdder\right)}\,\sqrt{n}}+\frac{C_2}{\sqrt{n}}+C_3n^{d_1/2}e^{-n\epsilon_1}\\
    &+C_4n^{d_2/2}e^{-n(\epsilon_2-\delta_1\firstder-\delta_3\secondder)}+2\decayhatbar+\frac{3}{2}\sqrt{d_1}\min\left\{\|J_{n}(\mle)^{-1}\|_{op},\|\postfisherinformation(\mle)^{-1}\|_{op}\right\}\\
    &\cdot\Bigg\{\frac{1}{n}(D_1'+D_2'+D_3'+D_4'n^{d_2/2+1}e^{-n\epsilon_2})^2+\frac{1}{\sqrt{n}}(E_1'+E_2'n^{(d_2+3)/2}e^{-n\epsilon_2}+E_3'+E_4'+E_5')\Bigg\},
\end{align*}
where $C_2,C_3,C_4$ are given in \cref{thm:without_correction1}, $D_1',D_2',D_3',D_4'$ in \cref{lem:gradient_bound_new} and $E_1',E_2',E_3',E_4',E_5'$ in \cref{lem:covariance_comparison_new}.

\end{thm}
\begin{proof}
    \Cref{thm:non_asymptotic_final} follows directly from \cref{thm:without_correction1,lem:gradient_bound_new,lem:covariance_comparison_new,prop:gaussian_comparison}.
\end{proof}

\subsection{Assumptions and proof of \Cref{thm:non_asymptotic_simplified}\label{sec:proof_of_theorem2}}

\subsubsection{Assumptions of \Cref{thm:non_asymptotic_simplified}}

Recall the definitions of \cref{eq:logposterior,eq:j_bar_j_hat} and let $\jbartwo$ be as in \cref{eq:def:jbartwo}.
We assume that there exist $\delta_1,\delta_2,\delta_3,m,\lambda>0$, such that
\begin{assumption}\label{assump:m_estimator_simplified}
    The two-step estimator of interest $\mle=(\mle_1,\mle_2)$ is unique. 
\end{assumption}
\begin{assumption}\label{assump:fisher_inform_simplified}
   All of $J_n(\mle),  \jtilde, \jbreve, \postfisherinformation(\mle),\jbartwo$ are strictly positive definite and their smallest eigenvalues are strictly lower bounded by $\lambda>0$.
\end{assumption}

\begin{assumption}\label{assump:third_der_simplified}
	The generalized log-posterior $\logposterior$, as well as $\loglikelihood_1$ and $\loglikelihood_2$ are two times differentiable inside $\{\param:\|\param-\mle\|\leq \delta_1\vee\delta_2\}$. Moreover,
	\begin{align*}
    &\underset{\|s_{(1+d_1):d_2}\|\leq \delta_2}{\sup_{s\neq 0,\,\|s_{1:d_1}\|\leq \delta_1}}\frac{\left\|\logposterior''(\mle+s)-\logposterior''(\mle)\right\|_{op}}{n\left\|s\right\|};\sup_{t\neq 0,\,\|t\|\leq \delta_1}\frac{\left\|\loglikelihood_1''(\mle_1+t)-\loglikelihood_1''(\mle_1)\right\|_{op}}{n\left\|t\right\|};
    \\
    &\sup_{s\neq 0,\,\|s\|\leq \delta_2}\frac{\left\|\partial_{22}\loglikelihood_2(\mle_1,\mle_2+s)-\partial_{22}\loglikelihood_2(\mle_1,\mle_2)\right\|_{op}}{n\left\|s\right\|};\\
    &\sup_{\|s_1\|<\delta_1,\|s_2\|<\delta_3}\frac{\left\|\partial_1\loglikelihood_2(\mle_1+s_1,\mle_2+s_2)\right\|}{n};\qquad \sup_{\|s_1\|<\delta_1,\|s_2\|<\delta_3}\frac{\left\|\partial_2\loglikelihood_2(\mle_1+s_1,\mle_2+s_2)\right\|}{n}
	\end{align*}
    are all upper bounded by some $K>0$.
\end{assumption}

\begin{assumption}\label{assump:der_integrability_new_simplified}
For some $\delta>0$, the following integrals exist:
\[
\int\sup_{\|\theta_1-\mle_1\|<\delta}\left\|\partial_{11}L_2(\theta_1,\theta_2)\right\|_{op}\pi_2(\theta_2)d\theta_2\qquad\text{and}\qquad \int\sup_{\|\theta_1-\mle_1\|<\delta}\left\|\partial_1L_2(\theta_1,\theta_2)\right\|^2\pi_2(\theta_2)d\theta_2.\]
\end{assumption}

\begin{assumption}\label{assump:third_der_modules_simplified}
The following quantities: 
    \begin{align*}
    &\sup_{\|s\|<\delta_2}\frac{\left\|\partial_{12}\loglikelihood_2(\mle_1,\mle_2+s)\right\|_{op}}{n};\qquad \sup_{\|s\|<\delta_2}\frac{\left\|\partial_{11}\loglikelihood_2(\mle_1,\mle_2+s)\right\|_{op}}{n};\\
    &\sup_{s\neq 0,\|s\|\leq \delta_2}\frac{\left\|\partial_{12}\loglikelihood_2(\mle_1,\mle_2+s)-\partial_{12}\loglikelihood_2(\mle_1,\mle_2)\right\|_{op}}{n\|s\|};\,\sup_{s\neq 0,\|s\|\leq \delta_2}\frac{\left\|\partial_{11}\loglikelihood_2(\mle_1,\mle_2+s)-\partial_{11}\loglikelihood_2(\mle_1,\mle_2)\right\|_{op}}{n\|s\|};\\
    &\frac{1}{n^m}\int_{\|\theta-\mle_2\|>\delta_2}\left\|\partial_1L_2(\mle_1,\theta_2)\right\|^2\pi_2(\theta_2)d\theta_2;\qquad\frac{1}{n^m}\int_{\|\theta-\mle_2\|>\delta_2}\left\|\partial_{11}\loglikelihood_2(\mle_1,\theta_2)\right\|_{op}\pi_2(\theta_2)d\theta_2,
    \end{align*}
   all exist and are all upper bounded by some $K'>0$.
\end{assumption}

\begin{assumption}\label{assump:priors11_simplified}
	For $i\in\{1,2\}$, the following quantities:
	\begin{align*}
		&\sup_{\|\param_i-\mle_i\|\leq\delta_i}\left|{\prior_i(\param_i)}^{-1}\right|;\qquad
        \sup_{\|\param_i-\mle_i\|\leq\delta_i}\left|\prior_i(\param_i)\right|;\qquad\sup_{\|\param_i-\mle_i\|\leq\delta_i}\left|\partial\prior_i(\param_i)\right|;\\
        &\sup_{t\neq 0, \|t\|\leq \delta_2} {\|t\|}^{-1}{\|(\log\pi_2)''(\mle_2+t)-(\log\pi_2)''(\mle_2)\|_{op}};\qquad \left\|(\log\pi_2)''(\mle_2)\right\|_{op},
	\end{align*}
    all exist and are all upper bounded by some 
    $K''>0$.
\end{assumption}

\begin{assumption}\label{assump:tails_simplified}
For any fixed $\tilde{\delta}_1>0$, there exists fixed $\epsilon_1>0$ such that
\begin{align*}
    \mathbbm{P}\left[\sup_{\|\theta-\mle_1\|>\tilde{\delta}_1}\frac{\loglikelihood_1(\theta)-\loglikelihood_1(\mle_1)}{n}\leq -\epsilon_1\right]\geq p_1(\tilde{\delta}_1,n).
\end{align*}
Moreover, for any fixed $\bar{\delta}_2>0$, there exist fixed $\bar{\delta}_1,\epsilon_2>0$ such that
\begin{align}   \label{eq:assump_tails_simpl} 
\mathbbm{P}\left[{}{\sup_{\|\theta_1-\mle_1\|\leq\bar{\delta}_1,\|\theta_2-\mle_2\|>\bar{\delta}_2}}\frac{\loglikelihood_2(\theta_1,\theta_2)-\loglikelihood_2(\mle_1,\mle_2)}{n}\leq -\epsilon_2\right]\geq p_2(\bar{\delta}_2,n).
    \end{align}
\end{assumption}

\subsubsection{Proof of \Cref{thm:non_asymptotic_simplified}}\label{sec:proof_of_non_asymptotic_simplified}
\Cref{thm:non_asymptotic_simplified} follows from \Cref{thm:without_correction1,lem:gradient_bound_new,lem:covariance_comparison_new,prop:gaussian_comparison}. 
Under the setup and assumptions of \cref{thm:non_asymptotic_simplified}, with probability at least $\tilde{p}(n_2)$, \cref{assump:m_estimator1} is satisfied and for some fixed and deterministic $\delta_1,\delta_2,\delta_3,m>0$, the constants and quantities of \cref{assump:priors21,assump:third_der1,assump:priors11,assump:third_der_modules1,assump:der_integrability_new,assump:derivative_bounds_new1} are such that:
\begin{align}
&\bullet\quad\thirdder,\thirdderone,\thirddertwo,\firstder,\secondderone \text{are upper bounded by some fixed, deterministic } K;\label{eq:part_2_1}\\
&\bullet\quad\thirddermixed,\thirddermixedone,\seconddermixed,\firstdersquaredintegral n^{-m}, \secondderintegral n^{-m} \text{ are upper bounded by some fixed, deterministic }K';\label{eq:part_2_2}\\
&\bullet\quad\priorboundone, \priorboundtwo, \priorboundthree, \priorboundfour, \priorboundfive,\priorboundsix, \priorboundseven, \priorboundthirdder,\|(\log\pi_2)''(\mle_1)\|_{op} \text{ are upper bounded}\notag\\
&\hspace{1cm}\text{by some fixed, deterministic }K''\label{eq:part_2_3};\\
&\bullet\quad J_{n}(\mle),\jtilde,\jbreve\succ\lambda I_{d_2}.\label{eq:part_2_31}
\end{align}
Choose $0<\bar{\delta}_2<\min\{\delta_2,{\lambda}{K}^{-1}\}$ and let $\bar{\delta}_1,\epsilon_2$ satisfy \cref{eq:assump_tails_simpl}. Moreover, choose
\[0<\tilde{\delta}_1<\min\{\bar{\delta}_1,\delta_1,{\epsilon_2}{K}^{-1},\sqrt{({\lambda}{K}^{-1})^2-(\bar{\delta}_2)^2}\}.\] 
Then, $\tilde{\delta}_1K<\epsilon_2$ so, for sufficiently small $\delta_3>0$,
\[
\firstder\leq K\text{ and }\secondder\leq K\qquad\Longrightarrow\qquad\tilde{\delta}_1\firstder+\delta_3\secondder<\epsilon_2,
\]
as in \cref{assump:derivative_bounds_new1}. Moreover, by \cref{assump:tails_simplified}, with probability at least $p_1(\tilde{\delta}_1,n_2)+p_1(\bar{\delta}_2,n_2)-1$, 
\begin{align}\label{eq:part_2_4}
\sup_{\|\theta-\mle_1\|>\tilde{\delta}_1}\frac{\loglikelihood_1(\theta)-\loglikelihood_1(\mle_1)}{n}\leq -\epsilon_1\qquad\text{and}\qquad\underset{\|\theta_2-\mle_2\|>\bar{\delta}_2}{\sup_{\|\theta_1-\mle_1\|\leq\tilde{\delta}_1}}\frac{\loglikelihood_2(\theta_1,\theta_2)-\loglikelihood_2(\mle_1,\mle_2)}{n}\leq -\epsilon_2,
\end{align}
for some $\epsilon_1>0$, as in \cref{assump:tails1}. 

Furthermore, recall that $\bar{\delta}_2<\frac{\lambda}{K}$. Hence, if \cref{eq:part_2_1,eq:part_2_2,eq:part_2_3,eq:part_2_31,eq:part_2_4} hold and \cref{assump:m_estimator1} is satisfied then, for sufficiently large $n$, 
\begin{align*}
&\jtilde-\frac{\bar{\delta}_2(\thirddertwo+\priorboundthirdder/n)+3\|(\log\pi_2)''(\mle_2)\|_{op}/n}{3}I_{d_2};\\
&\jtilde-\frac{1}{n}(\log\pi_2)''(\mle_2)-2\bar{\delta}_2\left(\frac{\thirddertwo}{3}+\frac{\priorboundthirdder}{3n}\right) I_{d_2}
\end{align*}
are both positive definite with their smallest eigenvalues bounded away from zero, as in the assumptions of \cref{lem:covariance_comparison_new,lem:gradient_bound_new}. Since $\bar{\delta}_2<\delta_2$, the bound in \cref{lem:covariance_comparison_new} is then of order $n^{-1/2}$ and it follows that for sufficiently large $n$, 
$
\postfisherinformation(\mle)\succ\lambda I_{d_2}$ and $
\jhat\succ\lambda I_{d_2}$, making \cref{assump:fisher_inform1} satisfied.
Moreover, recall that $\tilde{\delta}_1<\sqrt{({\lambda}{K}^{-1})^2-(\bar{\delta}_2)^2}$ and so  if \cref{eq:part_2_1} holds and \cref{assump:m_estimator1,assump:fisher_inform1} are satisfied with $
\postfisherinformation(\mle)\succ\lambda I_{d_2}$ then 
$\sqrt{\tilde{\delta}_1^2+\bar{\delta}_2^2}\, \thirdder<1$,
as in \cref{assump:third_der1}.

Finally, recall again that $\bar{\delta}_2<\delta_2$ and $\tilde{\delta}_1<\delta_1$. The above analysis leads us to the conclusion that, with probability at least $p(n_1,n_2,d_1,d_2)$ the bound in \cref{prop:gaussian_comparison} is of order $n^{-1/2}$, the bound in \cref{lem:gradient_bound_new} is of order one and the bound in \cref{thm:without_correction1} is hence of order $n^{-1/2}$, all of which together yields \cref{eq:tv_bound_simplified}.

\subsection{Proof of \cref{thm:without_correction1}}\label{sec:proof_of_main_bound}

\subsubsection{Introduction to the proof}

We start with the following lemma on centered Normal distributions. 

	\begin{lemma}\label{lemma1}
		Let $Z\sim\mathcal{N}(0,\Sigma)$. Then, for any $t>0$,
		\begin{align}
			&\mathbbm{P}\left[\left\|Z\right\|-\sqrt{\Tr(\Sigma)}\geq \sqrt{2\left\|\Sigma\right\|_{op}t}\right]\leq e^{-t};\label{concentration1}\\
			&\mathbbm{P}\left[\left\|Z\right\|^2-\Tr(\Sigma)\geq 2\sqrt{\Tr\left(\Sigma^2\right)t}+2\left\|\Sigma\right\|_{op}t\right]\leq e^{-t}.\label{concentration2}
		\end{align}
	\end{lemma}
	\begin{proof}
		The proof follows an argument similar to the one of \cite[page 135]{vershynin}. Specifically, \cref{concentration2} comes from \cite[Proposition 1]{hsu}. In order to prove \cref{concentration1}, we note that
		\begin{align*}
			\mathbbm{P}\left[\left\|Z\right\|-\sqrt{\Tr(\Sigma)}\geq \sqrt{2\left\|\Sigma\right\|_{op}t}\right]=&\mathbbm{P}\left[\left\|Z\right\|^2-\Tr(\Sigma)\geq 2\left\|\Sigma\right\|_{op} t + 2\sqrt{2\Tr\left(\Sigma\right)\left\|\Sigma\right\|_{op}t}\right]\\
			\leq & \mathbbm{P}\left[\left\|Z\right\|^2-\Tr(\Sigma)\geq 2\sqrt{\Tr\left(\Sigma^2\right)t}+2\left\|\Sigma\right\|_{op}t\right]\leq e^{-t}.
		\end{align*}
	\end{proof}
    Next, we introduce the notation
    \begin{align*}
        \normone{s}:=\left\|\left(\postfisherinformation(\mle)^{-1/2}s\right)_{1:d_1}\right\|;\qquad \normtwo{s}:=\left\|\left(\postfisherinformation(\mle)^{-1/2}s\right)_{d_1+1:d_2}\right\|.
    \end{align*}
	We let $g=\mathbbm{1}_A$ be the indicator of a measurable set $A$, 
     and for $\theta \sim \pi_{\text{cut}}$, consider
	\begin{align}
    D_g^{M}:=&\left|\mathbbm{E}\left[g\left(\sqrt{n}\postfisherinformation(\mle)^{1/2}(\theta-\mle)\right)\right]-\mathbbm{E}_{Z\sim\mathcal{N}(0,I_d)}\left[g(Z)\right]\right|.\label{dgmle}
	\end{align}
    Finally, we let
    $\posterior$
    denote the density of the cut posterior.
	\subsubsection{Initial decomposition of distance $D_g^{M}$}
	Now, let
	\begin{align}
	&h_g^{M}(u):=g(u)-\frac{n^{-d/2}\left|\text{det}\left(\postfisherinformation(\mle)^{-1/2}\right)\right|}{C^{M}_n} \innerint{t}g(t)\posterior(n^{-1/2}\postfisherinformation(\mle)^{-1/2} t+\mle) dt,\nonumber\\
	 \text{for}\quad &C_n^{M}:=n^{-d/2}\left|\text{det}\left(\postfisherinformation(\mle)^{-1/2}\right)\right|\innerint{u}\posterior(n^{-1/2} \postfisherinformation(\mle)^{-1/2}u+\mle)du.\nonumber 
	\end{align}
	Note that, for
	\begin{align}\label{fn_mle}
	F_n^{M}:=\innerint{u}\frac{e^{-u^Tu/2}}{(2\pi)^{d/2}}du,
	\end{align}
	we have
	\begin{align}
		D_g^{M}=& \left|\int_{\mathbbm{R}^d}h_g^{M}(u)\frac{e^{-u^Tu/2}}{(2\pi)^{d/2}}du\right.\notag\\
		&\left.-n^{-d/2} \left|\text{det}\left(\postfisherinformation(\mle)^{-1/2}\right)\right|\int_{\normone{u}>\delta_1\sqrt{n_1}\text{ or }\normtwo{u}>\delta_2\sqrt{n}} h_g^{M}(u) \posterior(n^{-1/2}\postfisherinformation(\mle)^{-1/2} u+\mle)du\right|\notag\\
		\leq &\frac{1}{F_n^{M}}\left|\innerint{u} h_g^{M}(u)\frac{e^{-u^Tu/2}}{(2\pi)^{d/2}}du\right|\notag\\
		&+\left|\int_{\normone{u}>\delta_1\sqrt{n_1}\text{ or }\normtwo{u}>\delta_2\sqrt{n}}h_g^{M}(u)\left[\frac{e^{-u^Tu/2}}{(2\pi)^{d/2}}-\frac{\left|\text{det}\left(\postfisherinformation(\mle)^{-1/2}\right)\right|}{n^{d/2}}\,\posterior(n^{-1/2} u+\mle)\right] du\right|\notag\\
		=&\left|\innerint{u}g(u)\frac{e^{-u^Tu/2}}{F_n^{M}(2\pi)^{d/2}}du\right.\notag\\
		&\left.-\frac{n^{-d/2}\left|\text{det}\left(\postfisherinformation(\mle)^{-1/2}\right)\right|}{C_n^{M}} \innerint{u} g(u)\posterior(n^{-1/2}\postfisherinformation(\mle)^{-1/2} u+\mle) du\right|\notag\\
        &+\Bigg|\int_{\normone{u}>\delta_1\sqrt{n_1}\text{ or }\normtwo{u}>\delta_2\sqrt{n}}h^{M}_g(u)\notag\\
        &\hspace{2cm}\cdot\left[\frac{e^{-u^Tu/2}}{(2\pi)^{d/2}}-n^{-d/2}\left|\text{det}\left(\postfisherinformation(\mle)^{-1/2}\right)\right|\,\posterior(n^{-1/2} \postfisherinformation(\mle)^{-1/2}u+\mle)\right] du\Bigg|\notag\\
		=:&I_1^{M}+I_2^{M}.\label{i1i2mle}
	\end{align}

\subsubsection{Controlling term $I_1^{M}$}

\subsubsection*{Strategy.}

In order to control $I_1^{M}$, we employ the log-Sobolev inequality.
Define the Kullback-Leibler divergence (or relative entropy) between two measures $\nu$ and $\mu$ such that $\nu\ll\mu$ as:
\begin{align}\label{def:kl_divergence}
\text{KL}(\nu\|\mu)=\int\log\left(\frac{d\nu}{d\mu}(x)\right)\nu(dx).
\end{align}
Define also the Fisher divergence as
\begin{align}\label{def:fisher_divergence}
	\text{Fisher}(\nu\|\mu)=\int\left\|\left(\log\left(\frac{d\nu}{d\mu}\right)\right)'(x)\right\|^2\nu(dx).
\end{align}
The following definition plays a central role in our proof.

\begin{definition}[Log-Sobolev inequality (LSI)]\label{def:log-sobolev}
Let $\mu$ be a probability measure on $\Omega\subset\mathbbm{R}^d$ and let $\alpha>0$. We say that $\mu$ satisfies the log-Sobolev inequality with constant $\alpha$, or LSI($\alpha$) for short, if 
\begin{align}\label{general_lsi}
	\int f^2\log f^2 d\mu - \int f^2d\mu \log\left(\int f^2 d\mu\right)\leq \frac{2}{\alpha}\int\left\| f'\right\|^2d\mu,\quad\text{for all }f:\Omega\to\mathbbm{R}^+.
\end{align}
Equivalently, $\mu$ satisfies LSI($\alpha$) if, for any probability measure $\nu$ on $\Omega$ such that $\mu\ll\nu$ and $\nu\ll\mu$,
\begin{align}\label{measure_lsi}
	\text{KL}(\nu\|\mu)\leq \frac{1}{2\alpha}\text{Fisher}(\nu\|\mu).
\end{align}
\end{definition}

We use the \textit{Bakry-\'Emery} criterion \citep{bakry-emery}. \cref{prop_lsi} is a version for measures on $\mathbbm{R}^d$ truncated to convex sets \cite[Theorem 2.1]{kolesnikov}, \cite[Theorem A1]{Schlichting:2019}.

\begin{innercustomthm}[Bakry-\'Emery criterion]\label{prop_lsi}
Let $\Omega\subset\mathbbm{R}^d$ be convex, let $H:\Omega\to\mathbbm{R}$ and consider a probability measure $\mu(dx)\propto e^{-H(x)}\mathbbm{1}_\Omega(x)dx$. Assume that $H''(x)\succeq \alpha I_{d\times d}$, for some $\alpha>0$. Then $\mu$ satisfies LSI($\alpha$) (see Definition \ref{def:log-sobolev}).
\end{innercustomthm}

To control $I_1^{M}$, we control the Fisher divergence between the rescaled cut posterior and the Normal inside a ball around the 2SM estimator. Then, we establish LSI inside that ball for the rescaled cut posterior and thus control the KL divergence. Pinsker's inequality, e.g. {\cite[Theorem 2.16]{massart}},
finally provides control of the total variation distance.

\subsubsection*{Final argument.}


For $t$ such that $\normone{t}\leq\sqrt{n}\delta_1$, $\normtwo{t}\leq\sqrt{n}\delta_2$, we have
\begin{align*}
	-n^{-1}\postfisherinformation(\mle)^{-1/2}\logposterior''(\mle+n^{-1/2}\postfisherinformation^{-1/2}t)\postfisherinformation(\mle)^{-1/2}
    \succeq&I_d-\sqrt{\delta_1^2+\delta_2^2}\, \thirdder I_{d\times d}\\
    \succeq& \left(1- \sqrt{\delta_1^2+\delta_2^2}\, \thirdder\right)I_{d\times d}.
\end{align*}
This means that, inside the convex set 
$B_0=\{t\in\mathbbm{R}^d\,:\,\normone{t}\leq\sqrt{n}\delta_1, \normtwo{t}\leq\sqrt{n}\delta_2 \}$,
the measure whose density, up to a normalizing constant, is given by 
$$t\mapsto e^{\logposterior(\mle+n^{-1/2}\postfisherinformation(\mle)^{-1/2}t)}\mathbbm{1}_{\{\normone{t}\leq\sqrt{n}\delta_1, \normtwo{t}\leq\sqrt{n}\delta_2 \}}(t),$$
is $ (1- \sqrt{\delta_1^2+\delta_2^2}\, \thirdder)$-strongly log-concave (see e.g. \cite{saumard}). 
Using the {Bakry-\'Emery criterion}, given by \cref{prop_lsi}, 
$[\sqrt{n}\postfisherinformation(\mle)^{1/2}(\theta-\mle)]_{B_{0}}$, with $\theta\sim \pi_{\text{cut}}$, satisfies the {log-Sobolev inequality} 
LSI$(1- \sqrt{\delta_1^2+\delta_2^2}\, \thirdder)$
(see Definition \ref{def:log-sobolev}). Recall $F_n^{M}$ in \cref{fn_mle}.
By combining the log-Sobolev inequality with {Pinsker's inequality}, we obtain that, for all functions $g$ which are indicators of measurable sets,
\begin{align}
	I_1^{M}\leq& \mathrm{TV}\left(\left[\mathcal{L}\left(\sqrt{n}\postfisherinformation(\mle)^{1/2}\left(\theta-\mle\right)\right)\right]_{B_{0}},\left[\mathcal{N}\left(0,I_d\right)\right]_{B_{0}}\right)\notag\\
	\underset{\text{inequality}}{\stackrel{\text{Pinsker's}}\leq}&\sqrt{\frac{1}{2}\text{KL}\left(\left.\left[\mathcal{N}(0,I_d)\right]_{B_{0}}\,\right\|\,\left[\mathcal{L}\left(\sqrt{n}\postfisherinformation(\mle)^{1/2}\left(\theta-\mle\right)\right)\right]_{B_{0}}\right)}\notag\\
	\underset{\text{inequality}}{\stackrel{\text{log-Sobolev}}\leq} &\frac{1}{2\sqrt{1- \sqrt{\delta_1^2+\delta_2^2}\thirdder}} \sqrt{\text{Fisher}\left(\left.\left[\mathcal{N}(0,I_d)\right]_{B_{0}}\,\right\|\,\left[\mathcal{L}\left(\sqrt{n}\postfisherinformation(\mle)^{1/2}\left(\theta-\mle\right)\right)\right]_{B_{0}}\right)} \notag\\
	&\hspace{-1cm}\leq \frac{1}{2\sqrt{1-\sqrt{\delta_1^2+\delta_2^2}\thirdder}}\sqrt{\int_{u\in B_0}\frac{e^{-\left\|u\right\|^2/2}}{F_n^{M}(2\pi)^{d/2}}\left\|u+\frac{\postfisherinformation(\mle)^{-1/2}\loglikelihood'(n^{-1/2}\postfisherinformation(\mle)^{-1/2}u+\mle)}{\sqrt{n}}\right\|^2du}\notag\\
	&+\frac{1}{2\sqrt{n\left(1-\sqrt{\delta_1^2+\delta_2^2}\thirdder\right)}}\cdot\sqrt{\int_{u\in B_0}\frac{e^{-\left\|u\right\|^2/2}}{F_n^{M}(2\pi)^{d/2}}\left\|\frac{\prior'(n^{-1/2}\fisherinformation(\mle)^{-1/2}u+\mle)}{\prior(n^{-1/2}\fisherinformation(\mle)^{-1/2}u+\mle)}\right\|^2du}\notag\\
	\underset{\text{theorem}}{\stackrel{\text{Taylor's}}\leq} &\frac{1}{2\sqrt{n\left(1-\sqrt{\delta_1^2+\delta_2^2}\thirdder\right)}}\sqrt{\int_{u\in B_0}\frac{e^{-\left\|u\right\|^2/2}}{F_n^{M}(2\pi)^{d/2}}\left(\|\fisherinformation(\mle)^{-1/2}L'(\mle)\|+\frac{\thirdder}{2}\left\|u\right\|^2\right)^2du}\notag\\
    &+\frac{\priorboundfive}{2\sqrt{n\left(1-\sqrt{\delta_1^2+\delta_2^2}\thirdder\right)}}\notag\\
    &\hspace{-1.2cm}\leq\hspace{-1mm} \frac{1}{\sqrt{n\left(1-\sqrt{\delta_1^2+\delta_2^2}\thirdder\right)}}\hspace{-1mm}\left(\hspace{-1mm}\|\fisherinformation(\mle)^{-1/2}L'(\mle)\|+\frac{\thirdder\sqrt{d^2+2d}}{2\sqrt{1-\decayhatbar}}\right)+\frac{\priorboundfive}{2\sqrt{n\left(1-\sqrt{\delta_1^2+\delta_2^2}\thirdder\right)}},
    \label{i1_mle_tv} 
\end{align}
which gives rise to the first summand in the bound in \cref{thm:without_correction1}. The last inequality in \cref{i1_mle_tv} follows from Lemma \ref{lemma1} and the fact that, for 
$U\sim\mathcal{N}(0,I_{d\times d})$,
$\mathbbm{E}\|U\|^4\leq d^2+2d$.

\subsubsection{Controlling term $I_2^{M}$ }

Since $|g|\leq 1$, we have
\begin{align*}
		I_2^{M} &\leq\int_{B_0^C}|g(u)|\left|n^{-d/2} \text{det}\left(\postfisherinformation(\mle)^{-1/2}\right)\posterior(n^{-1/2}\postfisherinformation(\mle)^{-1/2} u+\mle)-\frac{e^{-u^Tu/2}}{(2\pi)^{d/2}}\right|du\\
		&+ \frac{n^{-d/2}\det{\postfisherinformation(\mle)^{-1/2}}}{C_n^{M}}\innerint{t}|g(t)|\posterior(n^{-1/2}\postfisherinformation(\mle)^{-1/2}t+\hat{\theta}_n)dt\\
		&\cdot\int_{B_0^C}\left|n^{-d/2} \text{det}\left(\postfisherinformation(\mle)^{-1/2}\right)\posterior(n^{-1/2}\postfisherinformation(\mle)^{-1/2} u+\mle)-\frac{e^{-u^Tu/2}}{(2\pi)^{d/2}}\right|du\\
        \leq&\int_{\|s_1\|>\delta_1\sqrt{n}\text{ or }\|s_2\|>\delta_2\sqrt{n}}\left|n^{-d/2} \posterior(n^{-1/2}s+\mle)-\det{\postfisherinformation(\mle)}^{1/2}\frac{e^{-(s^T\postfisherinformation(\mle)s/2}}{(2\pi)^{d/2}}\right|ds\\
		&+ \frac{n^{-d/2}\det{\postfisherinformation(\mle)^{-1/2}}}{C_n^{M}}\innerint{t}\posterior(n^{-1/2}\postfisherinformation(\mle)^{-1/2}t+\hat{\theta}_n)dt\\
		&\cdot\int_{\|s_1\|>\delta_1\sqrt{n}\text{ or }\|s_2\|>\delta_2\sqrt{n}}\left|n^{-d/2} \posterior(n^{-1/2}s+\mle)-\det{\postfisherinformation(\mle)}^{1/2}\frac{e^{-s^T\postfisherinformation(\mle)s/2}}{(2\pi)^{d/2}}\right|ds\\
		\leq & 2\int_{\|s_1\|>\delta_1\sqrt{n}\text{ or }\|s_2\|>\delta_2\sqrt{n}}\left|n^{-d/2} \posterior(n^{-1/2}s+\mle)-\det{\postfisherinformation(\mle)}^{1/2}\frac{e^{-s^T\postfisherinformation(\mle)s/2}}{(2\pi)^{d/2}}\right|ds.
	\end{align*}
We shall bound this term by decomposing it into three terms. We get
\begin{align*}
I_{2}^{M}
\leq&2\int_{\mathbbm{R}^{d_2}}\int_{\|s_1\|>\delta_1\sqrt{n}}n^{-d/2} \posterior(n^{-1/2}s+(\mle_1,\mle_2))ds_2ds_1\\
&+2\int_{\|s_2\|>\delta_2\sqrt{n}}\int_{\|s_1\|\leq\delta_1\sqrt{n}}n^{-d/2} \posterior(n^{-1/2}(s_1,s_2)+(\mle_1,\mle_2))ds_2ds_1\\
&+2\int_{\|s_1\|>\delta_1\sqrt{n}\text{ or }\|s_2\|>\delta_2\sqrt{n}}\det{\postfisherinformation(\mle)}^{1/2}\frac{e^{-s^T\postfisherinformation(\mle)s/2}}{(2\pi)^{d/2}}ds\\
=:&I_{2,1}^{M}+I_{2,2}^{M}+I_{2,3}^{M}.
\end{align*}

\paragraph{Step 1}
First, we look at
\begin{align*}
&\int_{\mathbbm{R}^{d_2}}\int_{\|t_1\|>\delta_1\sqrt{n}}\frac{\exp\left[L_1\left(\mle_1+t_1n^{-1/2}\right)\right]\exp\left[L_2\left(\hat{\theta}_1+t_1n^{-1/2},\hat{\theta}_2+t_2n^{-1/2}\right)\right]}{\int\exp\left(L_2\left(\hat{\theta}_1+t_1/\sqrt{n},t\right)\right)\pi_2(t)dt}\\
&\hspace{9cm}\cdot \pi_1(\hat{\theta}_1+t_1/\sqrt{n})\pi_2(\hat{\theta}_2+t_2/\sqrt{n})dt_1dt_2\\
=&\int_{\|t_1\|>\delta_1\sqrt{n}}\hspace{-1mm}\frac{n^{d_2/2}\exp\left[L_1(\hat{\theta}_1+t_1/\sqrt{n})\right]\int_{\mathbbm{R}^{d_2}}\exp\left[L_2\left(\hat{\theta}_1+t_1/\sqrt{n},s\right)\right]\pi_2(s)ds}{\int_{\mathbbm{R}^{d_2}}\exp\left(L_2\left(\hat{\theta}_1+t_1/\sqrt{n},t\right)\right)\pi_2(t)dt}\pi_1(\hat{\theta}_1+t_1/\sqrt{n})\,dt_1\\
=&n^{d_2/2}\int_{\|t_1\|>\delta_1\sqrt{n}}\exp\left[L_1(\hat{\theta}_1+t_1/\sqrt{n})\right]\pi_1(\hat{\theta}_1+t_1/\sqrt{n})dt_1.
\end{align*}
Moreover,
\begin{align*}
&\int\int\frac{\exp\left[L_1(\hat{\theta}_1+s_1/\sqrt{n})\right]\exp\left[L_2\left(\hat{\theta}_1+s_1/\sqrt{n},\hat{\theta}_2+s_2/\sqrt{n}\right)\right]}{\int\exp\left(L_2\left(\hat{\theta}_1+s_1/\sqrt{n},t\right)\right)\pi_2(t)dt}\\
    &\hspace{7cm}\cdot \pi_1(\hat{\theta}_1+s_1/\sqrt{n})\pi_2(\hat{\theta}_2+s_2/\sqrt{n})ds_1ds_2\\
=&n^{d_2/2}\int\frac{\exp\left[L_1(\hat{\theta}_1+s_1/\sqrt{n})\right]\int\exp\left[L_2\left(\hat{\theta}_1+s_1/\sqrt{n},s\right)\right]\pi_2(s)ds}{\int\exp\left(L_2\left(\hat{\theta}_1+s_1/\sqrt{n},t\right)\right)\pi_2(t)dt} \pi_1(\hat{\theta}_1+s_1/\sqrt{n})ds_1\\
=&n^{d_2/2}\int\exp\left[L_1(\hat{\theta}_1+s_1/\sqrt{n})\right]\pi_1(\hat{\theta}_1+s_1/\sqrt{n})ds_1.
\end{align*}
Therefore,
\begin{align*}
I_{2,1}^{M}:=&\frac{2\int_{\|t_1\|>\delta_1\sqrt{n}}\exp\left[L_1(\hat{\theta}_1+t_1/\sqrt{n})\right]\pi_1(\hat{\theta}_1+t_1/\sqrt{n})dt_1}{\int\exp\left[L_1(\hat{\theta}_1+s_1/\sqrt{n})\right]\pi_1(\hat{\theta}_1+s_1/\sqrt{n})ds_1}\\
=&\frac{2\int_{\|t_1\|>\delta_1\sqrt{n}}\exp\left[L_1(\hat{\theta}_1+t_1/\sqrt{n})-\loglikelihood_1(\mle_1)\right]\pi_1(\hat{\theta}_1+t_1/\sqrt{n})dt_1}{\int\exp\left[L_1(\hat{\theta}_1+s_1/\sqrt{n})-\loglikelihood_1(\mle_1)\right]\pi_1(\hat{\theta}_1+s_1/\sqrt{n})ds_1}\\
\leq & \frac{2\priorboundtwo e^{-n\epsilon_1}\int_{\|t_1\|>\delta_1\sqrt{n}}\pi_1(\hat{\theta}_1+t_1/\sqrt{n})dt_1}{\int_{\|t_1\|\leq \delta_1\sqrt{n}}\exp\left(-s_1^T\left(-\frac{1}{n}\loglikelihood_1''(\mle)+\delta_1\thirdderone/3\right)s_1/2\right)ds_1}\\
\leq &\frac{2\priorboundtwo\det{\jhatplus}^{1/2}n^{d_1/2}e^{-n\epsilon_1}}{(2\pi)^{d/2}\decayhatplus},
\end{align*}
giving rise to the second summand in the bound of Theorem \ref{thm:without_correction1}.

\paragraph{Step 2.}
Now, 
\begin{align*}
&\int_{\|t_2\|>\delta_2\sqrt{n}}\int_{\|t_1\|<\delta_1\sqrt{n}}\frac{\exp\left[L_1(\hat{\theta}_1+t_1/\sqrt{n})\right]\exp\left[L_2\left(\hat{\theta}_1+t_1/\sqrt{n},\hat{\theta}_2+t_2/\sqrt{n}\right)\right]}{\int\exp\left(L_2\left(\hat{\theta}_1+t_1/\sqrt{n},t\right)\right)\pi_2(t)dt}\\
&\hspace{9cm}\cdot \pi_1(\hat{\theta}_1+t_1/\sqrt{n})\pi_2(\hat{\theta}_2+t_2/\sqrt{n})dt_1dt_2\\  
=& \int_{\|t_2\|>\delta_2\sqrt{n}}\int_{\|t_1\|<\delta_1\sqrt{n}}\frac{\exp\left[L_1(\hat{\theta}_1+t_1/\sqrt{n})\right]\exp\left[L_2\left(\hat{\theta}_1+t_1/\sqrt{n},\hat{\theta}_2+t_2/\sqrt{n}\right)-L_2(\hat{\theta}_1,\hat{\theta}_2)\right]}{\int\exp\left(L_2\left(\hat{\theta}_1+t_1/\sqrt{n},t\right)-L_2(\hat{\theta}_1,\hat{\theta}_2)\right)\pi_2(t)dt}\\
&\hspace{9cm}\cdot \pi_1(\hat{\theta}_1+t_1/\sqrt{n})\pi_2(\hat{\theta}_2+t_2/\sqrt{n})dt_1dt_2\\ 
\leq& \int_{\|t_2\|>\delta_2\sqrt{n}}\int_{\|t_1\|<\delta_1\sqrt{n}}\frac{e^{-n\epsilon_2}\exp\left[L_1(\hat{\theta}_1+t_1/\sqrt{n})\right]}{\int\exp\left(L_2\left(\hat{\theta}_1+t_1/\sqrt{n},t\right)-L_2(\hat{\theta}_1,\hat{\theta}_2)\right)\pi_2(t)dt}\\
&\hspace{9cm}\cdot \pi_1(\hat{\theta}_1+t_1/\sqrt{n})\pi_2(\hat{\theta}_2+t_2/\sqrt{n})dt_1dt_2\\ 
\leq& \int_{\|t_1\|<\delta_1\sqrt{n}}\frac{n^{d_2/2}e^{-n\epsilon_2}\exp\left[L_1(\hat{\theta}_1+t_1/\sqrt{n})\right]}{\int_{\|t-\hat{\theta}_2\|\leq\delta_3}\exp\left(L_2\left(\hat{\theta}_1+t_1/\sqrt{n},t\right)-L_2(\hat{\theta}_1,\hat{\theta}_2)\right)\pi_2(t)dt} \cdot\pi_1(\hat{\theta}_1+t_1/\sqrt{n})dt_1\\ 
\leq& \int_{\|t_1\|<\delta_1\sqrt{n}}\frac{n^{d_2}e^{-n\epsilon_2}\exp\left[L_1(\hat{\theta}_1+t_1/\sqrt{n})\right]\cdot\pi_1(\hat{\theta}_1+t_1/\sqrt{n})}{\int_{\|t\|\leq\delta_3\sqrt{n}}\exp\left(L_2\left(\hat{\theta}_1+t_1/\sqrt{n},\hat{\theta}_2+t/\sqrt{n}\right)-L_2(\hat{\theta}_1,\hat{\theta}_2)\right)\pi_2(\hat{\theta}_2+t/\sqrt{n})dt} dt_1.
\end{align*}
Now, we note that, for $\|t\|<\delta_3\sqrt{n}$ and $\|t_1\|<\delta_1\sqrt{n}$,
\begin{align*}
    &\left|L_2\left(\hat{\theta}_1+t_1/\sqrt{n},\hat{\theta}_2+t/\sqrt{n}\right)-L_2(\hat{\theta}_1,\hat{\theta}_2)\right|\\
    \leq & \left|L_2\left(\hat{\theta}_1+t_1/\sqrt{n},\hat{\theta}_2+t/\sqrt{n}\right)-L_2(\hat{\theta}_1+t_1/\sqrt{n},\hat{\theta}_2)\right|+\left|L_2(\hat{\theta}_1+t_1/\sqrt{n},\hat{\theta}_2)-L_2(\hat{\theta}_1,\hat{\theta}_2)\right|\\
    \leq &\sup_{\|s_1\|<\delta_1\sqrt{n},\|s\|<\delta_3\sqrt{n}}\left\|\partial_2L_2(\mle_1+s_1/\sqrt{n},\mle_2+s/\sqrt{n})\right\|\delta_3 + \sup_{\|s_1\|<\delta_1\sqrt{n}}\left\|\partial_1L_2(\mle_1+s_1/\sqrt{n},\mle_2)\right\|\delta_1\\
    \leq &n(\delta_1\firstder+\delta_3\secondder).
\end{align*}
We choose $\delta_3,\delta_1$ so small that $\delta_1\firstder+\delta_3\secondder<\epsilon_2$. In that case, we have the following:
\begin{align*}
    I_{2,2}^{M}&=\int_{\|t_2\|>\delta_2\sqrt{n}}\int_{\|t_1\|<\delta_1\sqrt{n}}\frac{2\exp\left[L_1(\hat{\theta}_1+t_1/\sqrt{n})\right]\exp\left[L_2\left(\hat{\theta}_1+t_1/\sqrt{n},\hat{\theta}_2+t_2/\sqrt{n}\right)\right]}{\int\exp\left(L_2\left(\hat{\theta}_1+t_1/\sqrt{n},t\right)\right)\pi_2(t)dt}\\
&\hspace{8cm}\cdot \pi_1(\hat{\theta}_1+t_1/\sqrt{n})\pi_2(\hat{\theta}_2+t_2/\sqrt{n})dt_1dt_2\\  
&\cdot\frac{1}{n^{d_2/2}\int\exp\left[L_1(\hat{\theta}_1+s_1/\sqrt{n})\right]\pi_1(\hat{\theta}_1+s_1/\sqrt{n})ds_1}\\
\leq & \frac{2n^{d_2/2}e^{-n(\epsilon_2-\delta_1\firstder-\delta_3\secondder)}\priorboundthree}{\text{Vol}(V_d(\delta_3))}\cdot\frac{\int_{\|t_1\|<\delta_1\sqrt{n}}\exp\left[L_1(\hat{\theta}_1+t_1/\sqrt{n})\right]\pi_1(\hat{\theta}_1+t_1/\sqrt{n})dt_1}{\int\exp\left[L_1(\hat{\theta}_1+s_1/\sqrt{n})\right]\pi_1(\hat{\theta}_1+s_1/\sqrt{n})ds_1}\\
\leq & \frac{2n^{d_2/2}e^{-n(\epsilon_2-\delta_1\firstder-\delta_3\secondder)}\priorboundthree}{\text{Vol}(V_{d_2}(\delta_3))},
\end{align*}
which gives rise to the third summand in the bound of Theorem \ref{thm:without_correction1}.

\paragraph{Step 3.}
Note that
\begin{align*}
    I_{2,3}^{M}=&2\int_{\|s_1\|>\delta_1\sqrt{n}\text{ or }\|s_2\|>\delta_2\sqrt{n}}\det{\postfisherinformation(\mle)}^{1/2}\frac{e^{-s^T\postfisherinformation(\mle)s/2}}{(2\pi)^{d/2}}ds\\
    \leq & 2\mathbbm{P}\left[\|\mathcal{N}(0,\postfisherinformation(\mle)^{-1}\|>(\delta_1\wedge \delta_2)\sqrt{n}\right]
    \leq 2\decayhatbar,
\end{align*}
by Lemma \ref{lemma1}, which gives rise to the last summand in the bound in Theorem \ref{thm:without_correction1}.

\subsection{Proof of technical lemmas\label{sec:proof_technical_lemmas}}

\subsubsection{Proof of \cref{lem:gradient_bound_new}}\label{sec:proof_gradient_bound}
We split $L'(\mle)$ into $A_1$ and $A_2$ as in \cref{eq:l_prime_asymp_dec} in the proof of \cref{lem:l_prime_n_asymptotic_proof}. Moreover, we split the bound on $\|A_1\|$ into the same two parts $A_{11}$ and $A_{12}$ as in \cref{eq:a_1_split}, except that we take a step back and do not plug in $M_4$ in place of
\[\sup_{s\neq 0,\|s\|\leq\delta}\left\|\partial_{12}\loglikelihood_2(\mle_1,\mle_2+s)-\partial_{12}\loglikelihood_2(\mle_1,\mle_2)\right\|_{op}.\]

\paragraph{Controlling $A_{12}$.}
We obtain
\begin{align*}
    \|A_{12}\|\leq & \sup_{s\neq 0,\|s\|\leq\delta}\frac{\left\|\partial_{12}\loglikelihood_2(\mle_1,\mle_2+s)-\partial_{12}\loglikelihood_2(\mle_1,\mle_2)\right\|_{op}}{2n\|s\|}\\
    &\hspace{-1cm}\cdot\frac{\int_{\|t\|<\delta_2\sqrt{n}}\exp\left[-\frac{1}{2}t^T\left(-\frac{1}{n}\partial_{22}\loglikelihood_2(\mle)-\frac{(\log\pi_2)''(\mle_2)}{n}-2\tilde{R}_n(t)\right)t+\frac{1}{\sqrt{n}}\left[(\log\pi_2)'(\mle)\right]^Tt\right]\|t\|^2dt  }{\int_{\|t\|<\delta_2\sqrt{n}}\exp\left[-\frac{1}{2}t^T\left(-\frac{1}{n}\partial_{22}\loglikelihood_2(\mle)-\frac{(\log\pi_2)''(\mle_2)}{n}-2\tilde{R}_n(t)\right)t+\frac{1}{\sqrt{n}}\left[(\log\pi_2)'(\mle)\right]^Tt\right]dt},
    \end{align*}
where $\tilde{R}_n$ arises from the following Taylor expansion, for $\|t\|<\delta_2\sqrt{n}$:
\begin{align}
    &\loglikelihood_2(\mle_1,n^{-1/2}t+\mle_2)+(\log\pi_2)(n^{-1/2}t+\mle_2) -\loglikelihood_2(\mle_1,\mle_2)-(\log\pi_2)(\mle_2)\notag\\
    =&\frac{1}{\sqrt{n}}\left[(\log\pi_2)'(\mle)\right]^Tt+\frac{1}{2n}t^T[\partial_{22}\loglikelihood_2(\mle_1,\mle_2)+(\log\pi_2)''(\mle_2)]t+t^T\tilde{R}_n(t)t\label{eq:r_n_tilde_def},
\end{align}
and
\begin{align}
\|\tilde{R}_n(t)\|_{op}\leq\frac{(\thirddertwo+\priorboundthirdder/n)\|t\|}{6n^{1/2}}\leq \frac{\delta_2(\thirddertwo+\priorboundthirdder/n)}{6}.\label{eq:r_n_tilde_bound}
\end{align}
It follows that
    \begin{align}
    &\|A_{12}\|\notag\\
    \leq & \frac{\thirddermixed\hspace{-1mm}\int_{\|t\|<\delta_2\sqrt{n}}\exp\hspace{-1mm}\left[-\frac{1}{2}t^T\left(-\frac{1}{n}\partial_{22}\loglikelihood_2(\mle)-\frac{\delta_2}{3}\thirddertwo-\frac{(\log\pi_2)''(\mle_2)}{n}-\frac{\delta_2}{3n}\priorboundthirdder\right)t+\frac{1}{\sqrt{n}}(\log\pi_2)'(\mle_2)t\right]\hspace{-1mm}\|t\|^2dt  }{2\int_{\|t\|<\delta_2\sqrt{n}}\exp\left[-\frac{1}{2}t^T\left(-\frac{1}{n}\partial_{22}\loglikelihood_2(\mle)+\frac{\delta_2}{3}\thirddertwo-\frac{(\log\pi_2)''(\mle_2)}{n}+\frac{\delta_2}{3n}\priorboundthirdder\right)t+\frac{1}{\sqrt{n}}(\log\pi_2)'(\mle_2)t\right]dt}\notag\\
    \leq & \frac{ \thirddermixed\left(\Tr[\jbarminus^{-1}]+\frac{1}{n}\left\|\jbarminus^{-1}(\log\pi_2)'(\mle_2)\right\|^2\right)\det{\jbarplus}^{1/2}}{\det{\jbarminus}^{1/2}(1-\decaybarplus)\exp\left(\frac{1}{2n}(\log\pi_2)'(\mle_2)^T\jbarplus^{-1}(\log\pi_2)'(\mle_2)\right)}\notag\\
    &\hspace{5cm}\cdot \exp\left(\frac{1}{2n}(\log\pi_2)'(\mle_2)^T\jbarminus^{-1}(\log\pi_2)'(\mle_2)\right),\label{eq:a_12_bound}
\end{align}
by Lemma \ref{lemma1}, which gives rise to term $D_1'$. 

\paragraph{Controlling $A_{11}$.}
Moreover, we note that
\begin{align*}
    &\|A_{11}\|\\
    \leq&\frac{\pi_2(\mle_2) }{\sqrt{n}\int_{\|t\|<\delta_2\sqrt{n}}\exp\left[-\frac{1}{2}t^T\left(-\frac{1}{n}\partial_{22}\loglikelihood_2(\mle)-\frac{(\log\pi_2)''(\mle_2)}{n}-2\tilde{R}_n(t)\right)t+\frac{\left[(\log\pi_2)'(\mle)\right]^Tt}{\sqrt{n}}\right]dt}\\
    &\cdot \Bigg\|\int_{\|t\|<\delta_2\sqrt{n}}\exp\left[-\frac{1}{2}t^T\left(-\frac{1}{n}\partial_{22}\loglikelihood_2(\mle)-\frac{1}{n}(\log\pi_2)''(\mle_2)\right)t\right]\\
    &\hspace{5cm}\cdot \exp\left[t^T\tilde{R}_n(t)t+\frac{\left[(\log\pi_2)'(\mle_2)\right]^Tt}{\sqrt{n}}\right]t^T\partial_{12}\loglikelihood_2(\mle_1,\mle_2) \,dt\Bigg\| \\
    \leq&\frac{1}{\sqrt{n}\int_{\|t\|<\delta_2\sqrt{n}}\exp\left[-\frac{1}{2}t^T\left(-\frac{1}{n}\partial_{22}\loglikelihood_2(\mle)+\frac{\delta_2}{3}\thirddertwo-\frac{(\log\pi_2)''(\mle_2)}{n}+\frac{\delta_2}{3n}\priorboundthirdder\right)t+\frac{\left[(\log\pi_2)'(\mle_2)\right]^Tt}{\sqrt{n}}\right]dt}\\
    &\cdot \int_{\|t\|<\delta_2\sqrt{n}}\exp\left[-\frac{1}{2}t^T\left(-\frac{1}{n}\partial_{22}\loglikelihood_2(\mle)-\frac{1}{n}(\log\pi_2)''(\mle_2)\right)t\right]\\
    &\hspace{3cm}\cdot \left|1-\exp\left[t^T\tilde{R}_n(t)t+\frac{[(\log\pi_2)'(\mle)]^T}{\sqrt{n}}t\right]\right| \left\|\partial_{12}\loglikelihood_2(\mle_1,\mle_2) t\right\|\,dt.
\end{align*}
Now, recall the inequality $|e^a-1|\leq |a|e^{|a|}$ and note that for $\|t\|\leq \delta_2\sqrt{n}$,
\begin{align*}
    &\left|\exp\left[t^T\tilde{R}_n(t)t+\frac{[(\log\pi_2)'(\mle_2)]^T}{\sqrt{n}}t\right]-1\right|\\
    \leq& \left[\frac{(\thirddertwo+\priorboundthirdder/n)\|t\|^3}{6\sqrt{n}}+\frac{|[(\log\pi_2)'(\mle_2)]^Tt|}{\sqrt{n}}\right]\exp\left[\frac{\delta_2(\thirddertwo+\priorboundthirdder/n)}{6}\|t\|^2+\frac{\left|[(\log\pi_2)'(\mle_2)]^Tt\right|}{\sqrt{n}}\right].
\end{align*}
Hence,
\begin{align*}
\|A_{11}\|\leq &\sqrt{2}\seconddermixed\det{\jbarplus}^{1/2}\\
& \hspace{-1.5cm}\cdot\frac{\int_{\|t\|<\delta_2\sqrt{n}}\exp\left[-\frac{1}{2}t^T\jbarminus t+\frac{\left|[(\log\pi_2)'(\mle_2)]^Tt\right|}{\sqrt{n}}\right]\left[\frac{1}{6}(\thirddertwo+\priorboundthirdder/n) \|t\|^4+\|(\log\pi_2)'(\mle_2)\|\|t\|^2\right]\,dt}{\int_{\|t\|<\delta_2\sqrt{n}}\exp\left[-\frac{1}{2}t^T\jbarplus t+\frac{\left[(\log\pi_2)'(\mle_2)\right]^Tt}{\sqrt{n}}\right]dt}.
\end{align*}
Now, for a Gaussian random vector $X$ with mean zero and covariance $\Sigma$, and any fixed $a,p$,
\[
\mathbbm{E}\left[\|X\|^pe^{|a^TX|}\right]\leq \sqrt{\mathbbm{E}[\|X\|^{2p}]\mathbbm{E}\left[e^{2|a^TX|}\right]},\qquad\text{and}\qquad \mathbbm{E}\left[e^{2|a^TX|}\right]\leq 2e^{2a^T\Sigma a},
\]
so that we obtain
\begin{align}
\|A_{11}\|
\leq & \frac{3\seconddermixed\det{\jbarplus}^{1/2}\left\{(\thirddertwo+\priorboundthirdder/n)\,\Tr[\jbarminus^{-1}]^2 + \|(\log\pi_2)'(\mle_2)\|\,\Tr[\jbarminus^{-1}]\right\}\left(\expminus\right)^2}{\det{\jbarminus}^{1/2}(1-\decaybarplus)\exp\left(\frac{1}{2n}(\log\pi_2)'(\mle_2)^T\jbarplus^{-1}(\log\pi_2)'(\mle_2)\right)},\label{eq:a_11_bound}
\end{align}
which gives rise to terms $D_2'$ and $D_3'.$

\paragraph{Controlling $A_2$.}
Finally, similarly to \cref{eq:bound_on_a2}, we note that
\begin{align*}
    &\|A_2\|\\
    \leq & ne^{-n\epsilon_2}\frac{\det{\jbarplus}^{1/2}\int_{\|t\|>\delta_2\sqrt{n}}\frac{1}{n}\left[\left\|\partial_{1}\loglikelihood_2(\mle_1,n^{-1/2}t+\mle_2)\right\|+\left\|\partial_{1}\loglikelihood_2(\mle_1,\mle_2)\right\|\right]\hspace{-1mm}\pi_2(n^{-1/2}t+\mle_2)dt }{(2\pi)^{d_2/2}(1-\decaybarplus)\exp\left(\frac{1}{2n}(\log\pi_2)'(\mle_2)^T\jbarplus^{-1}(\log\pi_2)'(\mle_2)\right)}\\
    \leq&n^{d_2/2+1}e^{-n\epsilon_2}\frac{\det{\jtildeplus}^{1/2}\left(\sqrt{\firstdersquaredintegral}+\firstder\right)}{(2\pi)^{d_2/2}(1-\decaybarplus)\exp\left(\frac{1}{2n}(\log\pi_2)'(\mle_2)^T\jbarplus^{-1}(\log\pi_2)'(\mle_2)\right)},
\end{align*}
which gives rise to term $D_4'$.\qed

\subsubsection{Proof of \cref{lem:covariance_comparison_new}\label{sec:proof_of_covcomparison}} \label{sec:proof_of_covaraince_comparison}
As a reminder,
$\postfisherinformation(\mle)
=J_{n}(\mle)-\frac{1}{n}D$,
where
\[
D=\begin{bmatrix}
	\partial_{11}\loglikelihood_2(\mle)-\partial_{11}\log\int e^{\loglikelihood_2(\mle_1,\theta)}\pi_2(\theta)d\theta-\partial_{12}\loglikelihood_2(\mle)\left(\partial_2\partial_2\logposterior_2(\mle)\right)^{-1}\partial_{12}\loglikelihood_2(\mle)^T&0\\
	0&0
\end{bmatrix}.
\]
We note that
\[
\|D\|_{op}=\left\|\partial_{11}\loglikelihood_2(\mle)-\partial_{11}\log\int e^{\loglikelihood_2(\mle_1,\theta)}\pi_2(\theta)d\theta-\partial_{12}\loglikelihood_2(\mle)\left(\partial_2\partial_2\logposterior_2(\mle)\right)^{-1}\partial_{12}\loglikelihood_2(\mle)^T\right\|_{op}.
\]
 As in \cref{eq:cov_comparison_bound_split}, in the proof of \cref{lem:h_n_asymptotic_proof}, we split the bound on $\|D\|_{op}$ into three parts: $\bar{A}_1,\bar{A}_2,\bar{A}_3$.

 \paragraph{Controlling $\bar{A}_3$.}
 By \cref{lem:gradient_bound_new,eq:l_prime_asymp_dec},
\begin{align*}
\|\bar{A}_3\|_{op}=
	&\left\| \left[\frac{\int e^{\loglikelihood_2(\mle_1,\theta_2)}\left(\partial_1\loglikelihood_2(\mle_1,\theta_2)-\partial_1\loglikelihood_2(\mle_1,\mle_2)\right)\pi_2(\theta_2)d\theta_2}{\int e^{\loglikelihood_2(\mle_1,\theta_2)}\pi_2(\theta_2)d\theta_2}\right]\right.\\
    &\hspace{2cm}\cdot\left.\left[\frac{\int e^{\loglikelihood_2(\mle_1,\theta_2)}\left(\partial_1\loglikelihood_2(\mle_1,\theta_2)-\partial_1\loglikelihood_2(\mle_1,\mle_2)\right)\pi_2(\theta_2)d\theta_2}{\int e^{\loglikelihood_2(\mle_1,\theta_2)}\pi_2(\theta_2)d\theta_2}\right]^T\right\|_{op}\\
\leq &(D_1'+D_2'+D_3'+D_4'n^{d_2/2+1}e^{-n\epsilon_2})^2,
\end{align*}
for $D_1',D_2',D_3',D_4'$ given in \cref{lem:gradient_bound_new}.

\paragraph{Controlling $\bar{A}_1$.}
In order to control $\|\bar{A}_1\|_{op}$ we start with the expression as in \cref{eq:a_1_bar_starting_expression}. We divide the numerator and denominator of this expression by $\pi(\mle_2)$. Then, in order to control the denominator, we restrict the integral to $\|t\|\leq \delta_2\sqrt{n}$ and then lower bound it analogously to \cref{eq:a_12_bound} or \cref{eq:a_11_bound}, by
\begin{align}\label{eq:denom_lower_bound}
(2\pi)^{d_2/2}\det{\jbarplus}^{-1/2}(1-\decaybarplus)\expplus.
\end{align}
In order to control the numerator, we split it into the integral over the inner region $\|t\|\leq\delta_2\sqrt{n}$ and the integral over the outer region $\|t\|>\delta_2\sqrt{n}$, similarly to \cref{eq:a_1_bar_num1,eq:a_1_bar_num2}. To control the former, we use \cref{assump:derivative_bounds_new1,eq:r_n_tilde_def,eq:r_n_tilde_bound} and upper bound it by
\begin{align*}
&\sqrt{n}\thirddermixedone\int_{\|t\|<\delta_2\sqrt{n}}\exp\bigg[-\frac{1}{2}t^T\left(-\frac{1}{n}\partial_{22}\loglikelihood_2(\mle)-\frac{\delta_2}{3}\thirddertwo-\frac{(\log\pi_2)''(\mle_2)}{n}-\frac{\delta_2}{3n}\priorboundthirdder\right)t\\
&\hspace{9cm}+\frac{1}{\sqrt{n}}(\log\pi_2)'(\mle_2)t\bigg]\|t\|dt \\
\leq &\sqrt{n}\thirddermixedone(2\pi)^{d_2/2}\det{\jbarminus}^{-1/2}\left(\sqrt{\Tr[\jbarminus^{-1}]}+\frac{1}{\sqrt{n}}\left\|\jbarminus^{-1}(\log\pi_2)'(\mle_2)\right\|\right)\expminus.
\end{align*}
Moreover, we upper bound the integral over the outer region by 
\[
\frac{n^{d_2/2}e^{-n\epsilon_2}}{\pi_2(\mle_2)}\int_{\|\theta-\mle_2\|>\delta_2}\left\|\partial_{11}\loglikelihood_2(\mle_1,\theta)-\partial_{11}\loglikelihood_2(\mle_1,\mle_2)\right\|_{op}\pi_2(\theta)d\theta\leq  n^{d_2/2}e^{-n\epsilon_2}\priorboundthree\left(\secondderintegral+\secondderone\right).
\]
As a consequence, we obtain that
\begin{align*}
    \|\bar{A}_1\|\leq & \frac{\sqrt{n}\thirddermixedone\left(\sqrt{\Tr[\jbarminus^{-1}]}+\frac{1}{\sqrt{n}}\left\|\jbarminus^{-1}(\log\pi_2)'(\mle_2)\right\|\right)\det{\jbarplus}^{1/2}\expminus}{\det{\jbarminus}^{1/2}(1-\decaybarplus)\,\expplus}\\
	&+\frac{n^{d_2/2+1}e^{-n\epsilon_2}\det{\jbarplus}^{1/2}}{(2\pi)^{d_2/2}(1-\decaybarplus)\,\expplus}\priorboundthree\left(\secondderintegral+\secondderone\right),
\end{align*}
giving rise to term $E_1'$ and contributing to term $E_2'$.

\paragraph{Controlling $\bar{A}_2$.}
We now need to control $\bar{A}_2$ and we split the bound into two parts: $\bar{A}_{21}$ and $\bar{A}_{22}$, as in \cref{eq:a_2_bar_split}. The only difference here, as compared to \cref{eq:a_2_bar_split}, is that in the context of \cref{lem:covariance_comparison_new}, due to our definition in \cref{eq:j_n_simplified}, we need to replace term 
\begin{align}\label{eq:first_term}
\partial_{12}\loglikelihood_2(\mle)\hspace{-1mm}\left(\partial_{22}\loglikelihood_2(\mle)\right)^{-1}\hspace{-1mm}\partial_{12}\loglikelihood_2(\mle)^T
\end{align}
with
\begin{align}\label{eq:second_term}
\partial_{12}\loglikelihood_2(\mle)\hspace{-1mm}\left(\partial_{22}\logposterior_2(\mle)\right)^{-1}\hspace{-1mm}\partial_{12}\loglikelihood_2(\mle)^T.
\end{align}
in the definition of $\bar{A}_{21}$.

\paragraph{Controlling $\bar{A}_{22}$.}
We control $\|\bar{A}_{22}\|_{op}$ in a way similar to the one we used to control the denominator and the outer integral in the numerator in definition of $\bar{A}_1$. We obtain
\begin{align}
    &\|\bar{A}_{22}\|_{op}\notag\\
	\leq & \frac{\left(\pi_2(\mle_2)\right)^{-1}n^{d_2/2}e^{-n\epsilon_2}\int_{\|\theta_2-\mle_2\|>\delta_2}\left\|\partial_1\loglikelihood_2(\mle_1,\theta_2)-\partial_1\loglikelihood_2(\mle_1,\mle_2)\right\|^2\pi_2(\theta_2)d\theta_2}{\int_{\|t\|<\delta_2\sqrt{n}}\exp\left[-\frac{1}{2}t^T\left(-\frac{1}{n}\partial_{22}\loglikelihood_2(\mle)+\frac{\delta_2}{3}\thirddertwo-\frac{(\log\pi_2)''(\mle_2)}{n}+\frac{\delta_2}{3n}\priorboundthirdder\right)t+\frac{1}{\sqrt{n}}(\log\pi_2)'(\mle_2)t\right]dt}\notag\\
    \leq & \frac{2n^{d_2/2+2}e^{-n\epsilon_2}\priorboundthree\det{\jbarplus}\left(\firstdersquaredintegral+\left(\firstder\right)^2\right)}{(2\pi)^{d_2/2}(1-\decaybarplus)\,\expplus},\label{eq:a_bar_22_bound}
\end{align}
contributing to term $E_2'$.

\paragraph{Controlling $\bar{A}_{21}$.}
Now, to bound $\bar{A}_{21}$ we split the bound into two parts: $\bar{A}_{211}$ and $\bar{A}_{212}$, as in \cref{eq:a_21_bar_split} and recall the definition of $\overline{R}_n(t)$, given in \cref{eq:def_r_n_bar} and the definition of and bound on $\tilde{R}_n(t)$ given by \cref{eq:r_n_tilde_def,eq:r_n_tilde_bound}. Moreover, we note that in the setup of \cref{lem:covariance_comparison_new}, we have
\begin{align*}
	\|\overline{R}_n(t)t\|_{op}\leq \frac{\thirddermixed}{2}\|t\|^2.
\end{align*}

\paragraph{Controlling $\bar{A}_{212}$.} In order to control $\|\bar{A}_{212}\|_{op}$, we again divide the numerator and the denominator by $e^{\loglikelihood_2(\mle)}\pi_2(\mle_2)$. We then lower bound the denominator by \cref{eq:denom_lower_bound}. Moreover, we upper bound the numerator by
\begin{align*}
    &\int_{\|t\|<\delta_2\sqrt{n}}\exp\left[-\frac{1}{2}t^T\left(-\frac{1}{n}\partial_{22}\loglikelihood_2(\mle)-\frac{\delta_2}{3}\thirddertwo-\frac{(\log\pi_2)''(\mle_2)}{n}-\frac{\delta_2}{3n}\priorboundthirdder\right)t+\frac{1}{\sqrt{n}}(\log\pi_2)'(\mle_2)t\right]\\
    &\cdot\Bigg[\frac{1}{\sqrt{n}}t^T\overline{R}_n(t)tt^T\partial_{12}\loglikelihood_2(\mle_1,\mle_2)+\frac{1}{\sqrt{n}}\partial_{12}\loglikelihood_2(\mle_1,\mle_2)tt^T\overline{R}_n(t)^Tt+t^T\overline{R}_n(t)tt^T\overline{R}_n(t)^Tt\Bigg]dt\\
    \leq & \thirddermixed\int_{\|t\|\leq\delta_2\sqrt{n}}e^{-t^T\jbarminus t/2+[(\log\pi_2)'(\mle_2)]^Tt/\sqrt{n}}\left[ 2\sqrt{n}\seconddermixed\|t\|^3+\thirddermixed\|t\|^4\right]dt.
\end{align*}
As a result, we obtain
\begin{align*}
	&\|\bar{A}_{212}\|_{op}\\
	\leq & \frac{\thirddermixed\det{\jbarplus}^{1/2}\expminus}{\det{\jbarminus}^{1/2}(1-\decaybarplus)\,\expplus}\cdot\Bigg[8\sqrt{n}\seconddermixed\left(2\Tr\left[\jbarminus^{-1}\right]^{3/2}\right.\\
    &\left.+\frac{1}{n^{3/2}}\left\|\jbarminus^{-1}(\log\pi_2)'(\mle_2)\right\|^{3}\right)+8\thirddermixed\left(3\Tr\left[\jbarminus^{-1}\right]^{2}+\frac{1}{n^{2}}\left\|\jbarminus^{-1}(\log\pi_2)'(\mle_2)\right\|^{4}\right)\Bigg],
\end{align*}
which gives rise to term $E_3'$.

\paragraph{Controlling $\bar{A}_{211}$.}
We now split $\bar{A}_{211}$ into $\bar{A}_{2111}$ and $\bar{A}_{2112}$, as in \cref{eq:a_211_split} yet taking into account that in the context of \cref{lem:covariance_comparison_new} the term \cref{eq:first_term} is replaced with \cref{eq:second_term}.

\paragraph{Controlling $\bar{A}_{2112}$.}
After dividing the numerator and the denominator of $\bar{A}_{2112}$ by $e^{\loglikelihood_2(\mle)}\pi_2(\mle_2)$ and restricting the integral in the denominator to $\|t\|\leq\delta_2\sqrt{n}$, we obtain the following bound
\begin{align*}
   \|\bar{A}_{2112}\|_{op}\leq \frac{n^{d_2/2+1}e^{-n\epsilon_2}\priorboundthree\det{\jbarplus}^{1/2}(\seconddermixed)^2}{(2\pi)^{d_2/2}(1-\decaybarplus)\,\expplus\minevalpost},
\end{align*}
in a way similar to the one that let us bound $\|\bar{A}_{22}\|_{op}$ in \cref{eq:a_bar_22_bound}. This bound contributes to $E_2'$.

\paragraph{Controlling $\bar{A}_{2111}$.}
We shall first focus on upper-bounding the norm of the quantity $(*)$ given by \cref{eq:a_2111_star}, with the term in \cref{eq:first_term} replaced by \cref{eq:second_term}. Recall the definition of and the bound on $\tilde{R}_n(t)$, given by \cref{eq:r_n_tilde_def,eq:r_n_tilde_bound}.
Upon dividing the numerator and the denominator by $e^{\loglikelihood_2(\mle)}\pi_2(\mle_2)$ and restricting the integral in the denominator to $\|t\|\leq\delta_1\sqrt{n}$, we note that
\begin{align*}
\|(*)\|_{op}
\leq &\frac{1}{\int_{\|t\|<\delta_2\sqrt{n}}\exp\left[-\frac{1}{2}t^T\left(-\frac{1}{n}\partial_{22}\loglikelihood_2(\mle)-\frac{(\log\pi_2)''(\mle_2)}{n}-2\tilde{R}_n(t)\right)t+\frac{1}{\sqrt{n}}\left[(\log\pi_2)'(\mle)\right]^Tt\right]dt}\\
&\hspace{-1cm}\cdot \left\|\int_{\|t\|<\delta_2\sqrt{n}}\exp\left[-\frac{1}{2}t^T\left(-\frac{1}{n}\partial_{22}\logposterior_2(\mle)-2\tilde{R}_n(t)\right)t+\frac{1}{\sqrt{n}}\left[(\log\pi_2)'(\mle)\right]^Tt\right]\right.\\
&\left.\hspace{9cm}\cdot\left(tt^T+\left(\frac{1}{n}\partial_{22}\logposterior_2(\mle)\right)^{-1}\right)dt  \right\|_{op}.
\end{align*}
Now, recall the inequality $|e^x-1|\leq |x|e^{|x|}$ and note that for $\|t\|\leq \delta_2\sqrt{n}$,
\begin{align*}
    &\left|\exp\left[t^T\tilde{R}_n(t)t+\frac{[(\log\pi_2)'(\mle_2)]^T}{\sqrt{n}}t\right]-1\right|\\
    \leq& \frac{1}{\sqrt{n}}\left[\frac{\thirddertwo+\priorboundthirdder/n}{6}\|t\|^3+\|(\log\pi_2)'\|\|t\|\right]\exp\left[\frac{\delta_2\left(\thirddertwo+\priorboundthirdder/n\right)}{6}\|t\|^2 +\frac{\left|[(\log\pi_2)'(\mle_2)]^Tt\right|}{\sqrt{n}}\right].
\end{align*}
Recall also that $\int e^{-x^TAx/2}xx^Tdx = A^{-1}\int e^{-x^TA/2}$ for any positive definite $A$. It follows that
\begin{align*}
   \| (*)\|_{op}\leq& \frac{\det{\jbarplus}^{1/2}}{(2\pi)^{d_2/2}(1-\decaybarplus)\expplus}\\
   & \cdot\Bigg\{\int_{\|t\|>\delta_2\sqrt{n}}\exp\left[-\frac{1}{2}t^T\left(-\frac{1}{n}\partial_{22}\logposterior_2(\mle)\right)t\right]\left\|tt^T+\left(\frac{1}{n}\partial_{22}\logposterior_2(\mle)\right)^{-1}\right\|_{op}dt \\
&+\frac{1}{\sqrt{n}}\int_{\|t\|\leq\delta_2\sqrt{n}}\exp\left[-\frac{1}{2}t^T\jbarminus t+\frac{|[(\log\pi_2)'(\mle_2)]^Tt|}{\sqrt{n}}\right]\left\|tt^T+\left(\frac{1}{n}\partial_{22}\logposterior_2(\mle)\right)^{-1}\right\|_{op}\\
&\hspace{3cm}\cdot\left[\frac{\thirddertwo+\priorboundthirdder/n}{6}\|t\|^3+\|(\log\pi_2)'(\mle_2)\|\|t\|\right]dt \Bigg\}.
\end{align*}
Hence, using the Cauchy-Schwarz inequality,
\begin{align*}
    &\|\bar{A}_{2111}\|_{op}\\
    \leq &\frac{n(\seconddermixed)^2\det{\jbarplus}^{1/2}\left(\sqrt{3\decaybartwo}\,\Tr\left[\jbartwo^{-1}\right]+\decaybartwo\left(\minevalpost\right)^{-1}\right)}{\det{\jbartwo}^{1/2}(1-\decaybarplus)\,\expplus}\\
    &+\frac{\sqrt{8}\sqrt{n}(\seconddermixed)^2\det{\jbarplus}^{1/2}\,\left(\expminus\right)^2}{\det{\jbarminus}^{1/2}(1-\decaybarplus)\,\expplus}\Bigg\{\left(\thirddertwo+\frac{\priorboundthirdder}{n}\right)\hspace{-1mm}\Bigg[\Tr\left[\jbarminus^{-1}\right]^{5/2}+\frac{\Tr\left[\jbarminus^{-1}\right]^{3/2}}{\minevalpost}\Bigg]\\
    &+\|(\log\pi_2)'(\mle_2)\|\Bigg[\Tr\left[\jbarminus^{-1}\right]^{3/2}+\frac{\Tr\left[\jbarminus^{-1}\right]^{1/2}}{\minevalpost}\Bigg]\Bigg\},
\end{align*}
by \cref{lemma1}, which gives rise to terms $E_4'$ and $E_5'$.

\subsubsection{Proof of Proposition \ref{prop:gaussian_comparison}}\label{sec:proof_gaussian_comparison}

By \cite[Theorem 1.1]{devroye2023},
\begin{align*}
	\mathrm{TV}\left(
	\mathcal{N}(0,\postfisherinformation(\mle)^{-1}), \mathcal{N}(0,J_{n}(\mle)^{-1})
	\right)\leq& \frac{3}{2}\sqrt{\sum_{i=1}^d\lambda_i^2}\;,
\end{align*}
where $\lambda_i$'s are the eigenvalues of $J_{n}(\mle)^{-1}\postfisherinformation(\mle)-I_d$. 
Note that
\begin{align*}
	&J_{n}(\mle)^{-1}\postfisherinformation(\mle)-I_d=J_{n}(\mle)^{-1}\left(\postfisherinformation(\mle)-J_{n}(\mle)\right)
    \\
    =&-\frac{1}{n}J_{n}(\mle)^{-1}\begin{bmatrix}
		\partial_{11}\loglikelihood_2(\mle)-\partial_{11}\log\int e^{\loglikelihood_2(\mle_1,\theta_2)}\pi_2(\theta_2)d\theta_2-\partial_{12}\loglikelihood_2(\mle)\left(\partial_2\partial_2\logposterior_2(\mle)\right)^{-1}\partial_{12}\loglikelihood_2(\mle)^T &0\\
		0&0
	\end{bmatrix}.
\end{align*}
We note that, since the non-zero (top-left) block of the matrix above is of size $d_1\times d_1$, we have
\[
\sqrt{\sum_{i=1}^d\lambda_i^2}\leq \sqrt{d_1}\left\|J_{n}(\mle)^{-1}\right\|_{op}\left\|\postfisherinformation(\mle)-J_{n}(\mle)\right\|_{op}.
\]
By symmetry, $\sqrt{\sum_{i=1}^d\lambda_i^2}$ is also bounded by the right-hand side term above,
with $\|J_{n}(\mle)^{-1}\|_{op}$ replaced by $\|\postfisherinformation(\mle)^{-1}\|_{op}$,
and $\|\postfisherinformation(\mle)-J_{n}(\mle)\|_{op}$ is upper-bounded in \cref{lem:covariance_comparison_new}.\qed

\subsection{More detail on \cref{ex:biased} and proof of \cref{prop:biased_data}}
\label{sec:proof_example_biased}

The bound in \cref{prop:biased_data} can be established by applying \cref{thm:without_correction1} directly and additionally controlling $\|\loglikelihood'(\mle)\|$and $\mathrm{TV}(
\mathcal{N}(0,\postfisherinformation(\mle)^{-1}), \mathcal{N}(0,J_{n}(\mle)^{-1})
)$, also directly. We do so in \cref{sec:two_step_biased_data,sec:concentration_two_step_mle,sec:checking_assumptions_biased_data,sec:trace_biased_data,sec:gradient_biased_data,sec:final_biased_data}. Additionally, we checked whether the bound would change if we used \cref{lem:gradient_bound_new,lem:covariance_comparison_new,prop:gaussian_comparison} to control $\|\loglikelihood'(\mle)\|$ and $\mathrm{TV}(
\mathcal{N}(0,\postfisherinformation(\mle)^{-1}), \mathcal{N}(0,J_{n}(\mle)^{-1})
)$ (rather than controlling those terms directly) i.e. applied the bound of \cref{thm:non_asymptotic_final}. In \cref{sec:grad_bound_biased_data,sec:cov_compaison_biased_data} we demonstrate that it would indeed change: the term $d_2n^{-1/2}$ in the bound in \cref{prop:biased_data} would be replaced by $d_2^{5/4}n^{-1/2}$. 

In \cref{sec:verify_assumptions_example} we show that \cref{ex:biased} satisfies the assumptions of \cref{thm:bvm_cut}.


\subsubsection{Two-step estimator: focus on \cref{assump:m_estimator1}}\label{sec:two_step_biased_data}
\Cref{assump:m_estimator1} is satisfied. Indeed, the second likelihood is strongly log-concave. The first pseudo-likelihood is strictly log-concave. To see this, note that
for any vector $v$ with $\|v\|=1$,
\begin{align}
    &v^T\loglikelihood_1''(\theta_1)v\notag\\
    =&\frac{\sqrt{1+\alpha^2}}{\alpha}\sum_{j=1}^{n_1}\left(1+\left(\frac{\|x_{1,j}-\theta_1\|}{\alpha}\right)^2\right)^{-1/2}\left[\frac{1}{\alpha^2}\left(1+\left(\frac{\|x_{1,j}-\theta_1\|}{\alpha}\right)^2\right)^{-1}|v^T(x_{1,j}-\theta_1)|^2-1   \right]\label{eq:l_1_hessian}
\end{align}
and
\begin{align}
    \frac{1}{\alpha^2}\left(1+\left(\frac{\|x_{1,j}-\theta_1\|}{\alpha}\right)^2\right)^{-1}|v^T(x_{1,j}-\theta_1)|^2=\frac{|v^T(x_{1,j}-\theta_1)|^2}{\alpha^2+\|x_{1,j}-\theta_1\|^2}\leq \frac{\|(x_{1,j}-\theta_1)\|^2}{\alpha^2+\|x_{1,j}-\theta_1\|^2}<1,\label{eq:l_1_hessian_bound}
\end{align}
which implies that $v^T\loglikelihood_1''(\theta_1)v<0$ as required.
Hence the 2SM estimator exists and is unique.

\subsubsection{Concentration of the two-step estimator}\label{sec:concentration_two_step_mle}
Let us now derive a concentration inequality for the 2SM estimator. Starting with the second module, where the data are viewed as i.i.d. from an isotropic sub-Gaussian distribution, we have
\begin{align*}
  \|\mle_2+\mle_1\| = \|\bar{x}_2\|\leq C\sqrt{\frac{d_2\vee\log n}{n}}
\end{align*}
for an absolute constant $C$, with probability at least $1-{c}/{n}$, for some absolute constant $c$. This follows for instance from \cite[Exercise 6.3.5]{vershynin}. Now, for the first module, our analysis is inspired by the proof of \cite[Theorem 1]{chakraborty2024}. Let $\eta=\frac{65(d_1\vee\log n)^2}{\alpha^3\sqrt{n_1}}$ and 
\begin{align*}
   f_u(t)=\frac{1}{n_1}\sum_{j=1}^{n_1}\sqrt{1+\left(\frac{\|x_{1,j}-t\eta u\|}{\alpha}\right)^2}.
\end{align*}
If
$
\inf_{\|u\|=1}f_u(1)>f_u(0)
$
then $\|\mle_1\|\leq\eta$, because $f_u$ is strictly convex for any $u$.
Furthermore, 
\begin{align*}
    f_u'(t) &= -\frac{\eta}{\alpha n_1}\sum_{j=1}^{n_1}\frac{(x_{1,j}-t\eta u)^Tu}{\left(\alpha^2+\|x_{1,j}-t\eta u\|^2\right)^{1/2}};\\
    f_u''(t)
    &=\frac{\eta^2}{\alpha n_1}\sum_{j=1}^{n_1}\frac{1}{\left(\alpha^2+\|x_{1,j}-t\eta u\|^2\right)^{1/2}}\left[\|u\|^2-\frac{((x_{1,j}-t\eta u)^Tu)^2}{\left(\alpha^2+\|x_{1,j}-t\eta u\|^2\right)}\right].
\end{align*}
Now, using Taylor's theorem, for $\|u\|=1$ and some $\tilde{t}\in(0,1)$,
\begin{align*}
    &f_u(1)-f_u(0)
    =f'_u(0)+\frac{1}{2}f_u''\left(\tilde{t}\right)\\
     \geq&-\frac{\eta}{\alpha n_1}\sum_{j=1}^{n_1}\frac{x_{1,j}^Tu}{\left(\alpha^2+\|x_{1,j}\|^2\right)^{1/2}}+\frac{\eta^2}{2\alpha n_1}\sum_{j=1}^{n_1}\frac{1}{\left(\alpha^2+\|x_{1,j}-\tilde{t}\eta u\|^2\right)^{1/2}}\left[\|u\|^2-\frac{\|x_{1,j}-\tilde{t}\eta u\|^2\|u\|^2}{\left(\alpha^2+\|x_{1,j}-\tilde{t}\eta u\|^2\right)}\right]\\
    \geq &-\frac{\eta}{\alpha n_1}\sum_{j=1}^{n_1}\frac{x_{1,j}^Tu}{\left(\alpha^2+\|x_{1,j}\|^2\right)^{1/2}}+\frac{\eta^2}{2\alpha n_1}\sum_{j=1}^{n_1}\frac{\alpha^2}{\left(\alpha^2+2\|x_{1,j}\|^2+2\eta^2\right)^{3/2}}.
\end{align*}
Now, if $x_{1,j}$'s are i.i.d. standard multivariate Gaussian and $\|u\|=1$ then $x_{1,j}^Tu$ are i.i.d. $\mathcal{N}(0,\|u\|^2)$, i.e. i.i.d. univariate standard Gaussian. By the Chernoff bound, we have
\begin{align*}
    \mathbbm{P}\left[\frac{1}{n_1}\sum_{j=1}^{n_1}x_{1,j}^Tu\geq \sqrt{\frac{\log n_1}{n_1}}\right]\leq \frac{1}{n_1}=\frac{1}{\vartheta n}.
\end{align*}
Also, by \cref{lemma1},
\begin{align}
\mathbbm{P}\left[\max_{j\in[n_1]}\|x_{1,j}\|<\sqrt{d_1}+2\sqrt{\log n}\right]=&\left(\mathbbm{P}\left[\|x_{1,1}\|<\sqrt{d_1}+2\sqrt{\log n}\right]\right)^{n_1}\notag\\
\geq& \left(1-\frac{1}{n^2}\right)^{n_1}=\left(1-\frac{1}{n^2}\right)^{\vartheta n}\geq 1-\frac{\vartheta}{n},\label{eq:max_x_1_high_prob}
\end{align}
where the last inequality follows from Bernoulli's inequality.
It follows that, with probability at least $1-\frac{\left(\vartheta+\vartheta^{-1}\right)}{n}$, and for large enough $n$,
\begin{align*}
    f_u(1)-f_u(0)
    \geq &-\frac{\eta\sqrt{\log n_1}}{\alpha^2\sqrt{n_1}}+\frac{\eta^2\alpha}{2\left(\alpha^2+2d_1+4\log n+4\sqrt{d_1\log n}+2\eta^2\right)^{3/2}}\\
    \geq&\eta\left(\frac{\alpha\eta}{64\left(d_1\vee\log n\right)^{3/2}}-\frac{\sqrt{d_1\vee\log n}}{\alpha^2\sqrt{n_1}}\right)>0.
\end{align*}
So, with probability at least $1-\frac{\left(\vartheta+\vartheta^{-1}\right)}{n}$, and for large enough $n$,
\[
\|\mle_1\|\leq \frac{65(d_1\vee\log n)^2}{\alpha^3\sqrt{n_1}}=\frac{65(d_1\vee\log n)^2}{\vartheta^{1/2}\alpha^3\sqrt{n}}.
\]
Furthermore, for any $a>0$,
$
(\alpha^2+a\|x_{1,j}\|^2)^{-3/2}\in\left[0,\frac{1}{\alpha^3}\right]
$
and hence, by Hoeffding's and Jensen's inequalities, for any $t>0$,
\begin{align*}
    \frac{1}{n_1}\sum_{j=1}^{n_1}\frac{1}{(\alpha^2+a\|x_{1,j}\|^2)^{3/2}}\geq &\frac{1}{n_1}\sum_{j=1}^{n_1}\mathbbm{E}\left[\frac{1}{(\alpha^2+a\|x_{1,j}\|^2)^{3/2}} \right]-\frac{t}{n_1}\\
    \geq& \frac{1}{(\alpha^2+a\mathbbm{E}\|x_{1,j}\|^2)^{3/2}}-\frac{t}{n_1}=\frac{1}{(\alpha^3+ad_2)^{3/2}}-\frac{t}{n_1},
\end{align*}
with probability at least $1-\exp\left[-\frac{2\alpha^6t^2}{n} \right]$. Letting
$
t=\frac{n_1}{2(\alpha^2+ad_2)^{3/2}},
$
we have that
\[
\frac{1}{n_1}\sum_{j=1}^{n_1}\frac{1}{(\alpha^2+a\|x_{1,j}\|^2)^{3/2}}\geq\frac{1}{2(\alpha^3+ad_2)^{3/2}},
\]
with probability at least
$
1-\exp\left[-\frac{\alpha^6n_1}{2(\alpha^2+ad_2)^3}\right]\geq 1-\frac{1}{n}
$.
\subsubsection{Checking \cref{assump:der_integrability_new,assump:priors11,assump:priors21,assump:derivative_bounds_new1,assump:tails1,assump:third_der_modules1,assump:third_der1,assump:fisher_inform1} one by one}\label{sec:checking_assumptions_biased_data}


We check the rest of the assumptions in the reverse order. In what follows, we assume that, for any absolute constant $a>0$, and for some absolute constant $C>0$,
\begin{align}
&\max_{j\in[n_1]}\|x_{1,j}\|<\sqrt{d_1}+2\sqrt{\log n};\qquad \|\mle_1\|\leq \frac{65(d_1\vee\log n)^2}{\vartheta^{1/2}\alpha^3\sqrt{n}};\qquad  \notag\\
&\|\mle_2+\mle_1\| = \|\bar{x}_2\|\leq C\sqrt{\frac{d_2\vee\log n}{n}};\qquad\frac{1}{n_1}\sum_{j=1}^{n_1}\frac{1}{(\alpha^2+a\|x_{1,j}\|^2)^{3/2}}\geq\frac{1}{2(\alpha^2+ad_2)^{3/2}}.\label{eq:two_mle_bounds}
\end{align}
As proved in \cref{sec:concentration_two_step_mle}, these bounds hold with probability at least $1-\frac{c}{n}$, for some absolute constant $c>0$. 
Moreover, we shall set
\begin{align}\label{eq:deltas_biased_data}
&\delta_1=\frac{\alpha^4}{180d_2^{3/2}};\qquad \delta_2=3\delta_1=\frac{\alpha^4}{60d_2^{3/2}};\\
&\epsilon_1= \frac{67\delta_1^2\vartheta\alpha^2\sqrt{1+\alpha^2}}{270d_2^{3/2}}=\frac{67\vartheta\alpha^{10}\sqrt{1+\alpha^2}}{8748000d_2^{9/2}};\qquad \epsilon_2=2\delta_1^2=\frac{\alpha^8}{16200d_2^3}.\label{eq:epsilons_biased_data}
\end{align}
All the inequalities appearing below hold for large enough $n$. Moreover, if, for some $a,b>0$, there exists a constant $\tilde{C}$, independent of $n$ and $d_2$, such that $a\leq \tilde{C}b$, then we write $a\lesssim b$.

\paragraph{Focus on \cref{assump:priors11,assump:priors21}}
The priors are Gaussian so \cref{assump:priors11,assump:priors21} are satisfied. In particular, we can set
\begin{align}
&\priorboundthirdder=0;\label{eq:priorboundthirdder}\\
&\priorboundsix = \left(\frac{50}{\pi}\right)^{d_2/2};\\
&\priorboundseven=100\left(\frac{50}{\pi}\right)^{d_2/2}(\|\mle_2\|+\delta_2)\lesssim \left(\frac{50}{\pi}\right)^{d_2/2}; \\
&\priorboundone = (2\pi)^{d_1/2}\left(\frac{\pi}{50}\right)^{d_2/2}e^{50(\|\mle_2\|+\delta_2)^2+(\|\mle_1\|+\delta_1)^2/2}\lesssim \left(\frac{\pi}{5}\right)^{d_2};\\
&\priorboundfour = \left(\frac{50}{\pi}\right)^{d_2/2}\left(\frac{1}{2\pi}\right)^{d_1/2}=\left(\frac{25}{\pi^2}\right)^{d_2/2};\\
&\priorboundfive = 100(\|\mle_1\|+\delta_1) + (\|\mle_2\|+\delta_2)\lesssim \frac{(d_2\vee\log n)^2}{\sqrt{n}}+\frac{1}{d_2^{3/2}};\label{eq:prior_five_bounded_data}\\
&\priorboundthree = \left(\frac{\pi}{50}\right)^{d_2/2}e^{50(\|\mle_2\|+\delta_2)^2}\leq \left(\frac{\pi}{50}\right)^{d_2/2}\exp\left[50\left(\frac{4d_2^2}{\alpha^3\sqrt{\vartheta\,n}}  +\delta_2\right)^2 \right]\lesssim \left(\frac{\pi}{50}\right)^{d_2/2};\label{eq:prior_three_biased_data}\\
&\priorboundtwo = (2\pi)^{d_1/2}e^{(\|\mle_1\|+\delta_1)^2/2}\leq(2\pi)^{d_1/2}\exp\left(\frac{1}{2}\left(\frac{65(d_2\vee \log n)^2}{\alpha^3\sqrt{\vartheta\,n}}+\delta_1\right)^2 \right)\lesssim (2\pi)^{d_2/2}.\label{eq:prior_two_biased_data}
\end{align}

\paragraph{Focus on \cref{assump:der_integrability_new}}
We note that, 
\begin{align}
&\frac{1}{n^2}\int\|\partial_1\loglikelihood_2(\mle_1,\theta_2)\|^2\pi_2(\theta_2)d\theta_2\leq  \frac{1}{n^2}\cdot\left(\frac{50}{\pi}\right)^{d_2/2}\int\left\|\sum_{i=1}^{n}(x_{2,i}-\mle_1-\theta_2)   \right\|^2e^{-50\|\theta_2\|^2}d\theta_2\notag\\
&\hspace{5cm}\leq \|\bar{x}_2-\mle_1\|^2+\frac{d_2}{100}\leq \frac{d_2}{50}=:\firstdersquaredintegral,\label{eq:firstdersquared_biased_data}\\
&\text{and we may take }\secondderintegral=1.\label{eq:secondderintegral_biased_data}
\end{align}
Also, 
\begin{align*}
&\int\sup_{\|\theta_1-\mle_1\|<\delta}\|\partial_{11}\loglikelihood_2(\theta_1,\theta_2)\|_{op}\pi_2(\theta_2)d\theta_2 = n\quad\text{and}\\
&\int\sup_{\|\theta_1-\mle_1\|<\delta}\|\partial_1\loglikelihood_2(\theta_1,\theta_2)\|_{op}^2\pi_2(\theta_2)d\theta_2\leq n^2\sup_{\|\theta_1-\mle_1\|<\delta}\|\bar{x}_2-\theta_1\|^2+\frac{d_2n^2}{100}<\infty.
\end{align*}

\paragraph{Focus on \cref{assump:derivative_bounds_new1}} 

We note that $\partial_{12}\loglikelihood_2(\theta_1,\theta_2)=-nI_{d_2}$
so we may take 
\begin{align}\label{eq:seconddermixed_biased_data}
\seconddermixed=1.
\end{align}
Moreover, note that
\begin{align}
    \sup_{\|s_1\|\leq\delta_1,\|s_2\|\leq\delta_3}\frac{\left\|\partial_1\loglikelihood_2(\mle_1+s_1,\mle_2+s_2)\right\|}{n}\leq &\|\bar{x}_2\|+\|\mle_1+\mle_2\|+\delta_1+\delta_3=2\|\bar{x}_2\|+\delta_1+\delta_3\notag\\
    \leq& 2C\sqrt{\frac{d_2\vee\log n}{n}}+\delta_1+\delta_3=:\firstder.\notag\\
    &\hspace{-7.5cm}\text{Hence}\,\,\firstder\lesssim d_2^{-3/2}\text{ and }\secondder=\firstder.\label{eq:firstder_biased_data}\\
    &\hspace{-7.5cm}\text{We set }\delta_3=\frac{6}{7}\cdot\frac{\varepsilon}{2C\sqrt{\frac{d_2\vee\log n}{n}}+\sqrt{\varepsilon_2}}-\delta_1\geq \frac{8\sqrt{\epsilon_2}}{7\sqrt{2}}-\delta_1=\frac{\delta_1}{7}.\label{eq:delta_3_biased_data}
\end{align}
Then,
\begin{align}
    &\delta_1\firstder+\delta_3\secondder\notag\\
    =&\frac{6}{7}\cdot\frac{\varepsilon_2}{2C\sqrt{\frac{d_2}{n}}+\sqrt{\varepsilon_2}}\left[2C\sqrt{\frac{d_2\vee\log n}{n}}+\frac{6}{7}\cdot\frac{\varepsilon_2}{2C\sqrt{\frac{d_2\vee\log n}{n}}+\sqrt{\varepsilon_2}}\right]\notag\\
    <&\frac{6}{7}\cdot\frac{\varepsilon_2}{2C\sqrt{\frac{d_2}{n}}+\sqrt{\varepsilon_2}}\left[2C\sqrt{\frac{d_2}{n}}+\frac{\varepsilon_2}{2C\sqrt{\frac{d_2}{n}}+\sqrt{\varepsilon_2}}\right]\notag\\
    \leq &\frac{6}{7}\cdot\frac{\varepsilon_2}{C\sqrt{\frac{d_2\vee\log n}{n}}+\sqrt{C^2\frac{d_2\vee\log n}{n}+\varepsilon_2}}\left[2\sqrt{C^2\frac{d_2\vee\log n}{n}}+\frac{\varepsilon_2}{C\sqrt{\frac{d_2\vee\log n}{n}}+\sqrt{C^2\frac{d_2\vee\log n}{n}+\varepsilon_2}}\right]\notag\\
    =&\frac{6}{7}\left[\sqrt{C^2\frac{d_2\vee\log n}{n}+\epsilon_2}-C\sqrt{\frac{d_2\vee\log n}{n}}\right]\left[\sqrt{C^2\frac{d_2\vee\log n}{n}+\epsilon_2}+C\sqrt{\frac{d_2\vee\log n}{n}}\right]\notag\\
    =&\frac{6}{7}\epsilon_2.\label{eq:bound_epsilon_firstder_biased_data}\\
    &\hspace{-0.5cm}\text{Moreover,  }\partial_{11}\loglikelihood_2(\theta_1,\theta_2)=-nI_{d_2},\text{ so we may take }\secondderone = 1.\label{eq:secondderone_biased_data}\\
    &\hspace{-0.5cm}\text{Furthermore, }\thirddermixed=0=\thirddermixedone.\label{eq:thirddermixed_biased_data}
\end{align}

\paragraph{Focus on \cref{assump:tails1}}
First, we see that, if $\delta_2>\delta_1$,
\begin{align}
&\underset{\|\theta_2-\mle_2\|>\delta_2}{\sup_{\|\theta_1-\mle_1\|\leq\delta_1}}\frac{\loglikelihood_2(\theta_1,\theta_2)-\loglikelihood_2(\mle_1,\mle_2)}{n}\notag\\
\leq &\sup_{\|\theta_1+\theta_2-\mle_1-\mle_2\|>\delta_2-\delta_1}\frac{\sum_{i=1}^{n}\|x_{2,i}-\mle_1-\mle_2\|^2-\sum_{i=1}^{n}\|x_{2,i}-\theta_1-\theta_2\|^2}{2n}\notag\\
=&\sup_{\|\theta_1+\theta_2-\mle_1-\mle_2\|>\delta_2-\delta_1}\sum_{i=1}^{n}\frac{\left<\theta_1+\theta_2-\mle_1-\mle_2,2x_{2,i}-\theta_1-\theta_2-\mle_1-\mle_2\right>}{2n}\notag\\
=&\sup_{\|\theta_1+\theta_2-\mle_1-\mle_2\|>\delta_2-\delta_1}\sum_{i=1}^{n}\frac{\left<\theta_1+\theta_2-\mle_1-\mle_2,x_{2,i}-\theta_1-\theta_2\right>}{2n}\notag\\
=&\sup_{\|\theta_1+\theta_2-\mle_1-\mle_2\|>\delta_2-\delta_1}\frac{\left<\theta_1+\theta_2-\mle_1-\mle_2,\mle_1+\mle_2-\theta_1-\theta_2\right>}{2}\notag\\
\leq & -\frac{1}{2}(\delta_2-\delta_1)^2=-\epsilon_2.\label{eq:tail_epsilon2}
\end{align}
Now, notice that
\begin{align*}
    &\loglikelihood_1'''(\theta_1)[u,v,w]\\
    =&\frac{\sqrt{1+\alpha^2}}{\alpha}\sum_{i=1}^{n_1}\Bigg[\frac{3}{\alpha^4}\left(1+\left(\frac{\|x_{1,j}-\theta_1\|}{\alpha}\right)^2\right)^{-5/2}    (u^T(x_{1,j}-\theta_1))(v^T(x_{1,j}-\theta_1))(w^T(x_{1,j}-\theta_1))\\
    &-\frac{1}{\alpha^2}\left(1+\left(\frac{\|x_{1,j}-\theta_1\|}{\alpha}\right)^2\right)^{-3/2}\left((u^T(x_{1,j}-\theta_1))v^Tw+(v^T(x_{1,j}-\theta_1))u^Tw+(w^T(x_{1,j}-\theta_1))v^Tu\right)\Bigg].
\end{align*}
Therefore,
\begin{align}
    \|\loglikelihood_1'''(\theta_1)\|_{op}
    \leq&3\sqrt{1+\alpha^2}\sum_{i=1}^{n_1}\Bigg[\frac{\|x_{1,j}-\theta_1\|^3}{\left(\alpha^2+\|x_{1,j}-\theta_1\|^2 \right)^{5/2}}+\frac{\|x_{1,j}-\theta_1\|}{\left(\alpha^2+\|x_{1,j}-\theta_1\|^2 \right)^{3/2}}   \Bigg]\notag\\
    \leq & 3\sqrt{1+\alpha^2}n_1\left(\frac{6}{25\alpha^2}\sqrt{\frac{3}{5}}+\frac{2}{3\sqrt{3}\alpha^2}\right) \leq \frac{2\sqrt{1+\alpha^2}}{\alpha^2}n_1.\label{eq:third_der_l1}
\end{align}
Recall also the expression for $v^T\loglikelihood_1''(\theta_1)v$, for any test vector $v$ with $\|v\|=1$, given in \cref{eq:l_1_hessian}
and note that, using \cref{eq:two_mle_bounds},
\begin{align}
    -\frac{1}{n}v^T\loglikelihood_1''(\mle_1)v\geq &\frac{\sqrt{1+\alpha^2}}{\alpha n}\sum_{j=1}^{n_1}\left(1+\left(\frac{\|x_{1,j}-\mle_1\|}{\alpha}\right)^2\right)^{-1/2}\frac{\alpha^2}{\alpha^2+\|x_{1,j}-\mle_1\|^2}\notag\\
    \geq& \frac{1}{n_1}\sum_{j=1}^{n}\vartheta \frac{\alpha^2\sqrt{1+\alpha^2}}{(\alpha^2+\frac{3}{2}\|x_{1,j}\|^2)^{3/2}}\notag\\
    \geq& \vartheta \frac{\alpha^2\sqrt{1+\alpha^2}}{2(\alpha^2+\frac{3}{2}d_2)^{3/2}}.
    \label{eq:huber_l1_upper_bound}
\end{align}
Therefore, for $\|\theta_1-\mle_1\|=\delta_1$, using a Taylor expansion and \cref{eq:third_der_l1},
\begin{align}
    \frac{\loglikelihood_1(\theta_1)-\loglikelihood_1(\mle_1)}{n}\leq &\frac{1}{2n}(\theta_1-\mle_1)^T\loglikelihood_1''(\mle_1)(\theta_1-\mle_1)+\frac{\sqrt{1+\alpha^2}n_1}{3\alpha^2n}\delta_1^3\notag\\
    \leq & -\frac{\vartheta}{2}\cdot\frac{1}{n_1}\sum_{j=1}^{n_1}\frac{\alpha^2\sqrt{1+\alpha^2}\delta_1^2}{(\alpha^2+\|x_{1,j}-\mle_1\|^2)^{3/2}}+\frac{\sqrt{1+\alpha^2}\vartheta}{3\alpha^2}\delta_1^3\notag\\
    \stackrel{\cref{eq:huber_l1_upper_bound}}\leq& -\frac{\vartheta}{2}\cdot\frac{1}{n_1}\sum_{j=1}^{n_1}\frac{\alpha^2\sqrt{1+\alpha^2}\delta_1^2}{2\left(\alpha^2+\frac{3}{2}d_2\right)^{3/2}}+\frac{\sqrt{1+\alpha^2}\vartheta}{3\alpha^2}\delta_1^3\notag\\
    \leq & -\left(\delta_1^2\vartheta\sqrt{1+\alpha^2}\right)\left[\frac{\alpha^2}{24d_2^{3/2}}-\frac{\delta_1}{3\alpha^2} \right]\notag\\
    =& -\frac{67\delta_1^2\vartheta\alpha^2\sqrt{1+\alpha^2}}{270d_2^{3/2}}
    =-\epsilon_1.\label{eq:tail_epsilon1}
\end{align}

\paragraph{Focus on \cref{assump:third_der_modules1}}
We may take
\begin{align}\label{eq:m_2_hat_1}
\thirddertwo=0\quad\text{and}\quad\sup_{\theta_1}\frac{1}{n}\|\loglikelihood_1'''(\theta_1)\|_{op}\leq \frac{2\sqrt{1+\alpha^2}\vartheta}{\alpha^2}=:\thirdderone,
\end{align}
where the last inequality follows from \cref{eq:third_der_l1}.

\paragraph{Focus on \cref{assump:third_der1}}
Since $\partial_1^3\log\int e^{L_2(\theta_1,\theta)}e^{-50\|\theta\|^2}d\theta
=0$ and $\thirddertwo=\thirddermixed=\thirddermixedone=\partial_1^3\loglikelihood_2(\theta_1,\theta_2)=0$,  we can set any $\thirdder$ that satisfies $\thirdder\geq \thirdderone\|\postfisherinformation(\mle)^{-1}\|_{op}$. Now, we note that
\[
\partial_{11}\log\int e^{\loglikelihood_2(\theta_1,\theta_2)}e^{-50\|\theta_2\|^2}d\theta_2 = -\frac{100n}{100+n}I_{d_1}\quad\text{and}\quad \partial_{11}\loglikelihood_2(\theta_1,\theta_2)=-nI_{d_1}.
\]
It follows that
\begin{align}\label{eq:j_bar_biased_data}
\postfisherinformation(\mle)=\begin{pmatrix}
   -\frac{1}{n}\loglikelihood_1''(\mle_1)+\left(1-\frac{100}{n+100}+\frac{1}{n}\right)I_{d_1}&I_{d_2}\\
    I_{d_1}&\left(1+\frac{100}{n}\right)I_{d_2}
\end{pmatrix}.
\end{align}
Now, recall from \cref{eq:huber_l1_upper_bound} that, for $v$ with $\|v\|=1$,
\begin{align}
    -\frac{1}{n}v^T\loglikelihood_1''(\mle_1)v\geq &\frac{\vartheta \alpha^2\sqrt{1+\alpha^2}}{2(\alpha^2+\frac{3}{2}d_2)^{3/2}}>0.\label{eq:hessian_l1}
\end{align}
It follows that
\[
\left\|\left(-\frac{1}{n}\loglikelihood_1''(\mle_1)+\left(1-\frac{100}{n+100}+\frac{1}{n}\right)I_{d_2}\right)^{-1}\right\|_{op}\leq \left[1+\frac{\vartheta \alpha^2\sqrt{1+\alpha^2}}{4(\alpha^2+\frac{3}{2}d_2)^{3/2}}\right]^{-1}\leq 1
\]
and 
\begin{align}
    &\|\postfisherinformation(\mle)^{-1}\|_{op}\leq \max\left\{\left[1+\frac{\vartheta \alpha^2\sqrt{1+\alpha^2}}{4(\alpha^2+\frac{3}{2}d_2)^{3/2}}\right]^{-1},\frac{n}{n+100}\right\}\leq  \min\left\{1,\frac{4(\alpha^2+\frac{3}{2}d_2)^{3/2}}{\vartheta \alpha^2\sqrt{1+\alpha^2}}\right\}.\label{eq:op_norm_inverse}\\
    &\hspace{-0.4cm} \text{So, we may take }\thirdder=\min\left\{\frac{4\sqrt{1+\alpha^2}\vartheta}{\alpha^2},\frac{4\left(\alpha^2+\frac{3}{2}d_2\right)^{3/2}}{\alpha^4}\right\}.\label{eq:thirdder_biased_data}\\
    &\hspace{-0.4cm}\text{We also note that }\thirdder\sqrt{\delta_1^2+\delta_2^2}=\sqrt{10}M_2\delta_1<\frac{9}{10}.\label{eq:thirdder_delta_biased_data}
\end{align}

\paragraph{Focus on \cref{assump:fisher_inform1}} 
Recall \cref{eq:j_bar_biased_data}
and that
\[
\fisherinformation(\mle)=-\frac{\loglikelihood''(\mle)}{n}=\begin{pmatrix}
   -\frac{1}{n}\loglikelihood_1''(\mle_1)+I_{d_2}&I_{d_2}\\
    I_{d_2}&I_{d_2}
\end{pmatrix}.
\]
It now follows immediately from \cref{eq:hessian_l1} (using Schur's complement) that \cref{assump:fisher_inform1} is satisfied.

\subsubsection{Trace of the inverse Fisher information matrix}\label{sec:trace_biased_data}
As before, all the inequalities below hold for large enough $n$ and with probability at least $1-\frac{c}{n}$ for some absolute constant $c$. We note that, by \cref{eq:op_norm_inverse}, 
\begin{align*}
    \Tr\left[\postfisherinformation(\mle)^{-1}\right]\leq 2d_2\left\|\postfisherinformation(\mle)^{-1}\right\|_{op}\leq 2d_2
\end{align*}
In order to apply \cref{thm:without_correction1}, we require that
\[
\delta_1^2n>\Tr\left[\postfisherinformation(\mle)^{-1}\right],
\]
which holds for large enough $n$, since
$\frac{d_2^{11/2}}{n}\to 0.$  In particular,
\begin{align*}
\delta_1\sqrt{n}-\sqrt{\Tr\left[\postfisherinformation(\mle)^{-1}\right]}\geq \frac{\sqrt{n}\alpha^4}{180d_2^{3/2}}-\sqrt{2d_2}=&\frac{\sqrt{n}\alpha^4-180\sqrt{2}(d_2)^{2}}{180d_2^{3/2}}\\
\geq& \frac{\sqrt{n}\alpha^4}{200d_2^{3/2}}.
\end{align*}
Furthermore, it follows from \cref{eq:op_norm_inverse}  that $\postminevaluemle\geq 1$
and hence
\begin{align}
\decayhatbar=&\exp\left[-\frac{1}{2}
        \left((\delta_1\wedge \delta_2)\sqrt{n}-        \sqrt{\Tr(\postfisherinformation(\mle)^{-1})}\right)^2\postminevaluemle
        \right]\notag\\
        \leq& \exp\left[-
        \left(\delta_1\sqrt{n}-        \sqrt{\Tr(\postfisherinformation(\mle)^{-1})}\right)^2
        \right]
        \leq  \exp\left[-\frac{\alpha^{8}n}{4\cdot 10^4d_2^{3}}
        \right].\label{eq:decayhatbar_biased_data}
\end{align}
Moreover,
\[
\jhatplus:=-\frac{1}{n}\partial_{11}\loglikelihood_1(\mle)+\frac{\delta_1\thirdderone}{3}I_{d_1}
\]
and, using \cref{eq:huber_l1_upper_bound,eq:m_2_hat_1,eq:two_mle_bounds},
\[
\minevalmleplus\geq \frac{\alpha^2\sqrt{1+\alpha^2}\vartheta}{2\left(\alpha^2+\frac{3}{2}d_2\right)^{3/2}}+\frac{\alpha^2\sqrt{1+\alpha^2}\vartheta}{270d_2^{3/2}}\geq \frac{\alpha^2\sqrt{1+\alpha^2}\vartheta}{36d_2^{3/2}}.
\]
It follows that
$
\Tr[\jhatplus^{-1}]\leq \frac{{36d_2^{5/2}}}{\alpha^2\sqrt{1+\alpha^2}\vartheta}
$
and
\[
\delta_1\sqrt{n}-\sqrt{\Tr[\jhatplus^{-1}]}\geq\frac{\sqrt{n}\alpha^4}{180d_2^{3/2}}-\frac{6d_2^{5/4}}{\alpha(1+\alpha^2)^{1/4}\vartheta^{1/2}}\geq\frac{\sqrt{n}\alpha^4}{200(d_2)^{3/2}}.
\]
Hence,
\begin{align}
\decayhatplus\leq \exp\left[-\frac{\vartheta\alpha^{10}\sqrt{1+\alpha^2}n}{144\cdot 10^4(d_2)^{9/2}} \right].\label{eq:decayhatplus_biased_data}
\end{align}
Also, using \cref{eq:huber_l1_upper_bound,eq:m_2_hat_1},
\begin{align}
    \left\|\jhatplus\right\|_{op}\leq &\frac{\delta_1\thirdderone}{3}+\frac{\vartheta\alpha^2\sqrt{1+\alpha^2}}{2(\alpha^2+\frac{3}{2}d_2)^{3/2}}\leq \frac{\vartheta\alpha^2\sqrt{1+\alpha^2}}{270\cdot (d_2)^{3/2}}+\frac{\vartheta\alpha^2\sqrt{1+\alpha^2}}{2(\alpha^2+\frac{3}{2}d_2)^{3/2}}\notag\\
    \leq& \frac{\vartheta \sqrt{1+\alpha^2}}{\alpha},\notag\\
    &\hspace{-3cm}\text{and hence }\det{\jhatplus}^{1/2}\leq \left(\frac{\vartheta \sqrt{1+\alpha^2}}{\alpha}\right)^{d_2/2}.\label{eq:jhatplus_det_biased_data}
\end{align}

\subsubsection{Gradient of the generalized log-likelihood}\label{sec:gradient_biased_data}
We note that
\begin{align*}
\loglikelihood'(\mle)=&\partial_1\loglikelihood_2(\mle)-\partial_1\log\int e^{\loglikelihood_2(\cdot,\theta_2)}e^{-50\|\theta_2\|^2}d\theta_2\Bigg|_{\cdot=\mle_1}\\
=& \partial_1\left(\frac{1}{2}\sum_{j=1}^{n}\|x_{2,j}-\theta_1\|^2- \frac{n^2}{4\left(50 + \frac{n}{2}\right)} \left\| \bar{x}_2 - \theta_1 \right\|^2\right)\Bigg|_{\theta_1=\mle_1}\\
=&\frac{-100n(\bar{x}_2-\mle_1)}{n+100}
\end{align*}
and so, by \cref{eq:two_mle_bounds}, with high probability, and for large enough $n$,
\begin{align*}
    \left\|\loglikelihood'(\mle)\right\|\leq 100\|\bar{x}_2-\mle_1\|\lesssim \frac{(d_2\vee\log n)^2}{\alpha^3\vartheta^{1/2}\sqrt{n}}+\sqrt{\frac{d_2\vee\log n}{n}}\lesssim \frac{(d_2\vee\log n)^2}{\alpha^3\sqrt{n}}.
\end{align*}
Also, it follows from \cref{eq:op_norm_inverse} that 
\begin{align}\label{eq:fisher_inverse_root_operator}
    \|\postfisherinformation(\mle)^{-1/2}\|_{op}\leq 1.
\end{align}
Therefore,
\begin{align}\label{eq:gradient_bound_biased_data}
    \|\postfisherinformation(\mle)^{-1/2}\loglikelihood'(\mle)\|\lesssim \frac{(d_2\vee\log n)^2}{\alpha^3\sqrt{n}}.
\end{align}

\subsubsection{Establishing the final bound in \cref{prop:biased_data}}\label{sec:final_biased_data}
We shall now calculate the bound in \cref{thm:without_correction1}. Again, all the inequalities below hold for large enough $n$ and with probability at least $1-\frac{c}{n}$ for some absolute constant $c$. Recall that, by \cref{eq:thirdder_biased_data,eq:gradient_bound_biased_data,eq:decayhatbar_biased_data,eq:thirdder_delta_biased_data,eq:prior_five_bounded_data} , we have, for large enough $n$,
\begin{align}
    &\thirdder\lesssim 1;\qquad\sqrt{\delta_1^2+\delta_2^2}\thirdder<\frac{9}{10};\qquad \|\postfisherinformation(\mle)^{-1/2}\loglikelihood'(\mle)\|\lesssim \frac{(d_2\vee\log n)^2}{\alpha^3\sqrt{n}};\notag\\
    &\decayhatbar\leq \exp\left[-\frac{\alpha^{8}n}{4\cdot 10^4d_2^{3}}
        \right];\qquad\priorboundfive \lesssim \frac{(d_2\vee\log n)^2}{\sqrt{n}}+\frac{1}{d_2^{3/2}}.\label{eq:d_n_bar_biased}
\end{align}
Hence,
\begin{align}\label{eq:c_1_biased_data}
(C_1+C_2)n^{-1/2}\lesssim d_2n^{-1/2}.
\end{align}
Moreover, it follows from \cref{eq:prior_two_biased_data,eq:epsilons_biased_data,eq:decayhatplus_biased_data,eq:jhatplus_det_biased_data} that
\begin{align*}
&\priorboundtwo\lesssim (2\pi)^{d_1/2};\qquad\det{\jhatplus}^{1/2}\leq \left(\frac{\vartheta \sqrt{1+\alpha^2}}{\alpha}\right)^{d_2/2};\\
&\decayhatplus\leq \exp\left[-\frac{\vartheta\alpha^{10}\sqrt{1+\alpha^2}n}{144\cdot 10^4(d_2)^{9/2}} \right];\qquad\epsilon_1=\frac{67\vartheta\alpha^{10}\sqrt{1+\alpha^2}}{8748000d_2^{9/2}}.
\end{align*}
Therefore,
\begin{align}\label{eq:c_2_biased_data}
C_3n^{d_1/2}e^{-n\epsilon_1}\lesssim n^{d_2/2}\exp\left[-\frac{67\vartheta\alpha^{10}\sqrt{1+\alpha^2}n}{8748000d_2^{9/2}}\right].
\end{align}
Furthermore, by \cref{eq:bound_epsilon_firstder_biased_data,eq:prior_three_biased_data,eq:delta_3_biased_data}
\[
\priorboundthree\lesssim \left(\frac{\pi}{50}\right)^{d_2/2};\qquad\epsilon_2-(\delta_1+\delta_3)\firstder\geq \frac{\epsilon_2}{7}=\frac{\alpha^8}{113400d_2^3};\qquad\delta_3\geq \frac{\alpha^4}{1260d_2^{3/2}}.
\]
Hence,
\begin{align}\label{eq:c_3_biased_data}
C_4n^{d_2/2}e^{-n(\epsilon_2-(\delta_1+\delta_3)\firstder)}\lesssim n^{d_2/2} \exp\left[-\frac{\alpha^8n}{113400d_2^3} \right].
\end{align}
Finally, we note that $\|J_{n}(\mle)-\postfisherinformation(\mle)\|_{op}\lesssim n^{-1}$. Hence, by \cite[Theorem 1.1]{devroye2023} and by an argument analogous to the one of \cref{sec:proof_gaussian_comparison}, and by \cref{eq:op_norm_inverse}, we have that
\begin{align}\label{eq:gaussian_comparison_gaussian_data}
\mathrm{TV}\left(
	\mathcal{N}(0,\postfisherinformation(\mle)^{-1}), \mathcal{N}(0,J_{n}(\mle)^{-1})
	\right)\leq \frac{3}{2}\sqrt{d_2}\|\postfisherinformation(\mle)^{-1}\|_{op}\|J_{n}(\mle)-\postfisherinformation(\mle)\|_{op}\lesssim d_2^{1/2}n^{-1}.
\end{align}
The bound in \cref{prop:biased_data} now follows from \cref{thm:without_correction1,eq:c_1_biased_data,eq:c_2_biased_data,eq:c_3_biased_data,eq:gaussian_comparison_gaussian_data,eq:d_n_bar_biased}.

\subsubsection{Computing the bound in \cref{lem:gradient_bound_new}}\label{sec:grad_bound_biased_data}
We note that
\begin{align}\label{eq:new_js_biased_data}
&\jbarplus=\jbarminus=\jbarminustwo=\jbartwo=\left(1+\frac{100}{n}\right)I_{d_2},
\end{align}
because we may take $\thirddertwo=0$ and $\priorboundthirdder=0$, by \cref{eq:priorboundthirdder,eq:m_2_hat_1}.
Moreover, recall that $\seconddermixed=1$, by \cref{eq:seconddermixed_biased_data}.
Furthermore, with probability at least $1-\frac{c}{n}$ for some absolute constant $c>0$, and for large enough $n$,
\begin{align}
    \decaybarplus=&\exp\left[-\frac{1}{2}\left(\delta_2\sqrt{n}-\sqrt{\frac{d_2n}{n+100}}-\frac{100\sqrt{n}}{n+100}\|\mle_2\|\right)^2\frac{n+100}{n}  \right]\notag\\
    =&\exp\left[-\frac{1}{2}\left(\delta_2\sqrt{n+100}-\sqrt{d_2}-\frac{100\|\mle_2\|}{\sqrt{n+100}} \right)^2 \right]\leq \exp\left[-\frac{\alpha^4n}{8\cdot 10^4d_2^{3}}\right],\label{eq:decaybarplus_biased_data}
\end{align}
by \cref{eq:deltas_biased_data,eq:two_mle_bounds}.
So, looking at the bound in \cref{lem:gradient_bound_new}, and using \cref{eq:thirddermixed_biased_data,eq:two_mle_bounds,eq:epsilons_biased_data,eq:prior_three_biased_data,eq:firstdersquared_biased_data,eq:firstder_biased_data,eq:fisher_inverse_root_operator}, we see that, for large enough $n$ and with high probability,
\begin{align*}
    &D_1'=0;\quad D_2'=0;\quad
    D_4'n^{d_2/2+1}e^{-n\epsilon_2}\in\mathcal{O}\left(n^{d_2/2+1}\exp\left[ -\frac{n\alpha^8}{16200d_2^{3}}\right]\right);  \\
    &D_3'\lesssim\|\mle_2\|d_2\exp\left[\frac{7\cdot 10^4\|\mle_2\|^2}{2(n+100)}\right]\in\mathcal{O}\left(\frac{d_2(d_2\vee\log n)^2}{\sqrt{n}}\right).
\end{align*}
It follows from \cref{eq:fisher_inverse_root_operator} that
\begin{align}
\|\postfisherinformation(\mle)^{-1/2}\loglikelihood'(\mle)\|\in\mathcal{O}\left(\frac{d_2(d_2\vee\log n)^2}{\sqrt{n}}\right).
\end{align}
If we were to use this bound instead of the one arising from \cref{sec:gradient_biased_data}, then  the bound in \cref{prop:biased_data}, calculated from \cref{thm:without_correction1}, would remain unchanged.

\subsubsection{Computing the bound of \cref{lem:covariance_comparison_new,prop:gaussian_comparison}}\label{sec:cov_compaison_biased_data}
Note that
\begin{align}\label{eq:decaytilde_biased_data}
\decaybartwo=\exp\left[-\frac{1}{2}\left(\delta_2\sqrt{n}-\sqrt{\frac{d_2n}{n+100}} \right)^2\frac{n+100}{n}\right].
\end{align}
Hence, with high probability, and for large enough $n$,
\begin{align*}
    &E_1'=0,\text{ by \cref{eq:thirddermixed_biased_data}};\\
    &E_2'n^{(d_2+3)/2}e^{-n\epsilon_2}\in\mathcal{O}\left(d_2n^{d_2/2+3/2}\exp\left[ -\frac{n\alpha^8}{16200d_2^{3}}\right]\right),\text{ by \cref{eq:firstdersquared_biased_data,eq:secondderone_biased_data,eq:secondderintegral_biased_data,eq:prior_three_biased_data,eq:firstder_biased_data,eq:epsilons_biased_data,eq:seconddermixed_biased_data}}; \\
    &E_3'=0,\text{ by \cref{eq:thirddermixed_biased_data}};\\
    &E_4'\in\mathcal{O}\left(d_2^{3/2}\|\mle_2\|\right)=\mathcal{O}\left(d_2^{3/2}(d_2\vee\log n)^2n^{-1/2} \right),\text{ by \cref{eq:two_mle_bounds,eq:new_js_biased_data,eq:m_2_hat_1,eq:priorboundthirdder,eq:seconddermixed_biased_data,eq:m_2_hat_1}}; \\
    &E_5'\in\mathcal{O}\left( \exp\left[-\frac{\alpha^8n}{7300 d_2^3} \right]nd_2^{-3/2}\right),\text{ by \cref{eq:new_js_biased_data,eq:decaybarplus_biased_data,eq:decaytilde_biased_data,eq:deltas_biased_data}}.
\end{align*}
Using also the analysis of \cref{sec:grad_bound_biased_data}, as well as \cref{eq:op_norm_inverse}, it follows from \cref{prop:gaussian_comparison} that
\begin{align*}
\mathrm{TV}\left(
	\mathcal{N}(0,\postfisherinformation(\mle)^{-1}), \mathcal{N}(0,J_{n}(\mle)^{-1})
	\right)\in\mathcal{O}\left(d_2^{2}(d_2\vee\log n)^2n^{-1}\right).
\end{align*}
If we were to use this bound, rather than \cref{eq:gaussian_comparison_gaussian_data}, then the bound in \cref{prop:biased_data}, calculated from \cref{thm:without_correction1} would change as long as $d_2>\log n$. In that case, the term of order $d_2n^{-1/2}$ in \cref{prop:biased_data} would need to be replaced with $d_2^{5/4}n^{-1/2}$.

\subsubsection{Verifying the assumptions of \cref{thm:bvm_cut}}\label{sec:verify_assumptions_example}
We note that the two-step estimator is consistent by \cref{sec:concentration_two_step_mle}, with $\truepar_1=0$ and $\truepar_2=0.$ The Gaussian priors satisfy \cref{assump:priors_asymptotic}. \Cref{assump:tails_asymptotic} is satisfied by \cref{eq:tail_epsilon2,eq:tail_epsilon1,eq:deltas_biased_data,eq:epsilons_biased_data}. Moreover, \cref{assump:third_der_asymptotic} holds by \cref{eq:m_2_hat_1,eq:thirddermixed_biased_data} and the fact that $\partial_1^3\loglikelihood_2(\theta_1,\theta_2)=0$ for all $(\theta_1,\theta_2)\in\mathbbm{R}^{d_1}\times\mathbbm{R}^{d_2}$. \Cref{assump:finite_hessian_asymptotic} holds by the fact that $f_2(\cdot|\theta_1,\theta_2,x_1)$ is a Gaussian density with mean $\theta_1+\theta_2$, and by \cref{eq:l_1_hessian,eq:l_1_hessian_bound}, which yield control over the second derivatives of $\theta_1\mapsto f_1(x_{1}|\theta_1)$, for any data point $x_1$. 
Furthermore, \cref{assump:positive_definite_asymptotic} is satisfied again by \cref{eq:l_1_hessian,eq:l_1_hessian_bound}, which imply that $v^T\loglikelihood(\theta_1)v<0$ for any $v\in\mathbbm{R}^{d_1}$ and any $\theta_1$, and by the fact that $f_2(\cdot|\theta_1,\theta_2,x_1)$ is a Gaussian density. \Cref{assump:der_bounds_asymptotic} holds by \cref{eq:thirddermixed_biased_data,eq:seconddermixed_biased_data,eq:firstder_biased_data,eq:secondderone_biased_data} and  \cref{assump:integrability_asymptotic} holds by \cref{eq:firstdersquared_biased_data,eq:secondderintegral_biased_data}.

\section{Posterior bootstrap for modular inference}\label{suppl:proof_main_thms}

\subsection{Setup and notation}
Let $\ell_1(x_{1,j},\theta_1) = \log  f_1(x_{1,j}|\theta_1)$ and $\ell_2(x_{1,j},x_{2,j},  \theta_1, \theta_2) = \log  f_2(x_{2,j}| \theta_1, \theta_2,x_{1,j})$. 
We define 
\begin{align}\label{eq:L_w_def}
L_{1}^w(\theta_1) &= \sum_{j=1}^{n}w_j \ell_1(x_{1,j},\theta_1) + w_0\log \pi_1(\theta_1),\\
\label{eq:L_v_def}
L_{2}^w(\theta_1, \theta_2) &= \sum_{j=1}^{n}w_j\ell_2(x_{1,j},x_{2,j},\theta_1, \theta_2) + w_0 \log \pi_2(\theta_2).
\end{align}

\subsection{Assumptions of \cref{thm:dependent}}

\begin{assumption}\label{assump:log_lik} \normalfont {(Parameter space and generalised log-likelihood functions)}. The parameter spaces $\Theta_1,\Theta_2$ are compact. The generalised log-likelihood functions $\ell_1(x_1, \theta_1)$ and $\ell_2(x_1,x_2, \theta_1, \theta_2) $ are bounded from above and measurable in $x_1,x_2$ for all $\theta_1 \in \Theta_1$, $\theta_2 \in \Theta_2$. The functions $\theta_1\mapsto\ell_1(x_1,\param_1),\, (\theta_1,\theta_2)\mapsto\ell_2(x_1,x_2,\param_1,\param_2)$ are continuous for each fixed $(x_1,x_2)\in\mathbbm{R}^{d_1}\times\mathbbm{R}^{d_2}$. Moreover, there exist some $\epsilon>0$ and an open ball $B\subseteq \Theta_1$ containing $\theta_1^*$, such that
$$E_{p_1^*} \left[\sup_{\theta_1\in\Theta_1}\big|\ell_1(X_1, \theta_1) \big|^{1+\epsilon}\right] < \infty\quad\text{ and }\quad E_{p*} \left[\sup_{\theta_1\in\overline{B},\theta_2\in\Theta_2}\big|\ell_2(X_1,X_2, \theta_1, \theta_2) \big|^{1+\epsilon}\right] < \infty,$$ 
where $\overline{B}$ is the closure of $B$.
\end{assumption}

\begin{assumption}\label{assump:identifiability}\normalfont (Identifiability). 
The pseudo-true parameters $(\truepar_1,\truepar_2)$ given by \cref{eq:pseudotrue} are unique and belong to the interior of $\Theta_1\times\Theta_2$.
	Moreover, there is an open ball $B\subseteq \Theta_1$ containing $\theta_1^*$ such that, for each $\theta_1 \in B$, we have a unique maximizer
	$$\theta_2^*(\theta_1) = \arg\max_{\theta_2 \in \Theta_2} E_{p^*} \left\{\ell_2(X_1,X_2, \theta_1, \theta_2)\right\}.
	$$
    Furthermore, for all $\epsilon>0$,
    \begin{align*}
    &\sup_{\param_1\,:\,\|\param_1-\truepar_1\|>\epsilon}\mathbbm{E}_{p_1^*}\left[\ell_1(X_1,\theta_1)\right]<\mathbbm{E}_{p_1^*}\left[\ell_1(X_1,\truepar_1)\right],\\
    &\sup_{\param_2\,:\,\|\param_2-\truepar_2\|>\epsilon}\mathbbm{E}_{p^*}\left[\ell_1(X_1,X_2,\truepar_1,\param_2)\right]<\mathbbm{E}_{p^*}\left[\ell_1(X_1,X_2,\truepar_1,\truepar_2)\right].
    \end{align*}
\end{assumption}

\begin{assumption}
	\normalfont\label{assump:log_lik_smoothness} (Smoothness of the generalised log-likelihood functions). There is an open ball $B_1 \subseteq \Theta_1$ containing $\theta_1^*$ such that  $\ell_1(x_1, \theta_1)$ is three times continuously differentiable with respect to $\theta_1 \in B_1$ almost surely on $x_1$ with respect to $p_1^*$. Moreover, there is an open ball $B_2 \subseteq \Theta_2$ containing $\theta_2^*$ such that  $\ell_2(x_1,x_2, \theta_1, \theta_2)$ is three times continuously differentiable with respect to $\theta_1$ and $\theta_2$ for $(\theta_1, \theta_2) \in B_1 \times B_2$, almost surely on $(x_1,x_2)$ with respect to $p^*$. Furthermore, there exist measurable functions $G_k$ for $k= 1, \ldots, 6$
  such that for $(\theta_1, \theta_2) \in B_1 \times B_2$ we have
	\begin{align*}
	\Big|\partial_{[j]} \ell_1(x_1,\theta_1)\Big| &\leq G_{1}(x_1), &1\leq j \leq d_1,&
	& E_{p_1^*} \left\{G_1(X_1)^2\right\} &< \infty,\\
	\Big|\partial_{[jk]}\ell_1(x_1,\theta_1)  \Big| &\leq G_{2}(x_1), &1\leq j \leq k \leq d_1,& 
	& E_{p_1^*}\left\{ G_{2}(X_1)^2 \right\} &< \infty,\\
	\Big|\partial_{[jkl]} \ell_1(x_1,\theta_1)  \Big| &\leq G_{3}(x_1), &1\leq j \leq k \leq l \leq d_1&
	& E_{p_1^*} \left\{G_{3}(X_1)\right\} &< \infty \\
		\Big|\partial_{[j]} \ell_2(x_1,x_2,\theta_1, \theta_2) \Big| &\leq G_{4}(x_1,x_2), &1\leq j \leq(d_1 + d_2),&
	& E_{p^*} \left\{G_4(X_1,X_2)^2\right\} &< \infty,\\
	\Big| \partial_{[jk]}\ell_2(x_1,x_2,\theta_1, \theta_2)  \Big| &\leq G_{5}(x_1,x_2), &1\leq j \leq k \leq (d_1 + d_2),& 
	& E_{p^*}\left\{ G_{5}(X_1,X_2)^2 \right\} &< \infty,\\
	\Big| \partial_{[jkl]} \ell_2(x_1,x_2,\theta_1, \theta_2)   \Big| &\leq G_{6}(x_1,x_2), &1\leq j \leq k \leq l \leq (d_1 + d_2),&
	& E_{p^*} \left\{G_{6}(X_1,X_2)\right\} &< \infty,
	\end{align*}
	where in the above equations $\partial_{[m]}$ denotes the partial derivative with respect to the $m$-th coordinate of either $\theta_1\in\mathbb{R}^{d_1}$ or $\theta = (\theta_1, \theta_2)\in\mathbb{R}^{d_1+d_2}$.
\end{assumption}

\begin{assumption}\label{assump:positive_definite}
	\normalfont (Positive definiteness of information matrices). For $B_i$ as in Assumption \ref{assump:log_lik_smoothness}, $i =1,2$, the matrices 
	$I_i$ and $J_i$, defined as
	\begin{align*}
    &I_1(\param_1) = \mathbbm{E}_{X_{1}\sim p_1^*}\left[{\partial_1\ell_1(X_{1},\param_1)}\cdot {\partial_1\ell_1(X_{1},\param_1)}^T   \right]\notag,\\
    &I_2(\param_1,\param_2) = \mathbbm{E}_{(X_{1},X_{2})\sim p^*}\left[{\partial_2\ell_2(X_{1},X_2,\,\param_1, \param_2)} \cdot {\partial_2\ell_2(X_1,X_{2},\,\param_1, \param_2)}^T   \right]\notag,\\
    &J_1(\theta_1)=-\mathbbm{E}_{X_{1}\sim p_1^*}\left[{\partial_{11}\ell_1(X_{1},\param_1)}   \right]\notag,\\
    &J_2(\theta_1,\theta_2)=-\mathbbm{E}_{(X_{1},X_{2})\sim p^*}\left[{\partial_{22}\ell_2(X_{1},X_2,\,\param_1, \param_2)}   \right]\notag,
\end{align*}
	are all positive definite for $\theta_i \in B_i$, with all elements finite.
\end{assumption}
\begin{assumption}\label{assump:log_prior_smoothness}\normalfont (Smoothness of log prior densities).
    For $i=1,2$, the function $\log\pi_i(\theta_i)$ is measurable, bounded on $\Theta_i$ and three times continuously differentiable at $\truepar_i$ with its first, second and third derivatives all bounded on $B_i$ as in \Cref{assump:log_lik_smoothness}. 
\end{assumption}

\subsection{Proof of Theorem \ref{thm:dependent}}
The proof of \cref{thm:dependent} follows the same logic as the proofs of \cite[Chapter 3, Theorem 7]{newton1991weighted} and \cite[Theorem 1]{lyddon2019general}.
Below, we outline the main steps.
\subsubsection*{Step 1.}
By \cref{assump:log_lik,assump:log_prior_smoothness} and by \cite[Theorem 2.2]{spencer:2024}, we have, for open ball $B\subseteq \Theta_1$ containing $\theta_1^*$,
\begin{align}\label{eq:spencer_result}
    \begin{split}
    &\sup_{\theta_1\in\Theta_1}\left|\frac{L_1^w(\theta_1)}{n} - \mathbbm{E}_{p_1^*}\left[\ell_1(X_1,\theta_1)\right]\right|\xrightarrow[n\to\infty]{a.s.}0\quad\text{and}\\
    &\sup_{\theta_1\in \overline{B},\,\theta_2\in\Theta_2}\left|\frac{L_{2}^w(\theta_1, \theta_2)}{n}-\mathbbm{E}_{p^*}\left[\ell_1(X_1,X_2,\theta_1,\theta_2)\right]\right|\xrightarrow[n\to\infty]{a.s.}0.
    \end{split}
\end{align}
Hence, by \cref{assump:identifiability,eq:spencer_result} and by \cite[Theorem 5.7]{van2000asymptotic},  
\begin{align}\label{eq:bootstrap_consistency}
\tilde{\theta}_1\xrightarrow[n\to\infty]{\mathbbm{P}}\truepar_1\quad\text{and}\quad\tilde{\theta}_2\xrightarrow[n\to\infty]{\mathbbm{P}}\truepar_2.
\end{align}
Similarly, by \cref{assump:log_lik,assump:identifiability} and by \cite[Theorem 5.7]{van2000asymptotic}, it follows that 
\begin{align}\label{eq:mle_consistency}
\mle_1\xrightarrow[n\to\infty]{\mathbbm{P}}\truepar_1\quad\text{and}\quad\mle_2\xrightarrow[n\to\infty]{\mathbbm{P}}\truepar_2.
\end{align}
Now, note that
\begin{align}
 &{\partial_1 L_1^w(\hat{\theta}_1)} = \sum_{j=1}^{n} (w_j-1) {\partial_1 \ell_1(x_{1,j},\hat{\theta}_1)} +   w_0 \;{\partial_1\log \pi_1(\hat{\theta}_1)};\notag\\
 &\partial_2 L_2^w(\hat{\theta}_1, \hat{\theta}_2) = \sum_{j=1}^{n} (w_j-1)\partial_2 \ell_2(x_{2,j},\hat{\theta}_1, \hat{\theta}_2) + w_0 \partial_2 \log \pi_2(\hat\theta_2).\label{eq:weight_log_lik}
\end{align}
By Taylor's theorem, 
\begin{align*}
    0=\partial_1 L_1^w(\tilde{\theta}_1)= \partial_1 L_1^w(\hat{\theta}_1)+\left[\int_0^1 \partial_{11}L_1^w(\mle_1+t(\tilde{\theta}_1-\mle_1)) dt\right](\tilde{\theta}_1-\mle_1).
\end{align*}
Hence, by \cref{assump:positive_definite,assump:identifiability,eq:bootstrap_consistency,eq:mle_consistency}, and the Uniform Law of Large Numbers \cite[Theorem 2.2]{spencer:2024}, the matrix
$$
\left[-\frac{1}{n}\int_0^1 \partial_{11} L_1^w(\mle_1+t(\tilde{\theta}_1-\mle_1))dt\right]
$$
is invertible with probability converging to one and
\begin{align*}
    \sqrt{n}(\tilde{\theta}_1-\mle_1)=\left[-\frac{1}{n}\int_0^1\partial_{11} L_1^w(\mle_1+t(\tilde{\theta}_1-\mle_1))dt\right]^{-1}\left[\frac{1}{\sqrt{n}} \partial_1 L_1^w(\hat{\theta}_1)\right],
\end{align*}
whence, using additionally \cref{eq:weight_log_lik}, we obtain
\begin{align}\label{eq:bootstrap_growth1}
    \sqrt{n}\|\tilde{\theta}_1-\mle_1\|\in\mathcal{O}_{\mathbbm{P}}(1).
\end{align}
Similarly,
\begin{align*}
    0=\partial_2 L_2^w(\tilde{\theta}_1,\tilde{\theta}_2) = \partial_2 L_2^w(\tilde{\theta}_1,\mle_2)+\left[\int_0^1 {\partial_{22} L_2^w(\tilde{\theta}_1,\mle_2+t(\tilde{\theta}_2-\mle_2))}dt\right](\tilde{\theta}_2-\mle_2),
\end{align*}
and
\begin{align*}
    \sqrt{n}(\tilde{\theta}_2-\mle_2)=\left[-\frac{1}{n}\int_0^1{\partial_{22} L_2^w(\tilde{\theta}_1,\mle_2+t(\tilde{\theta}_2-\mle_2))}dt\right]^{-1}\left[\frac{1}{\sqrt{n}} {\partial_2 L_2^w(\tilde{\theta}_1,\mle_2)}\right],
\end{align*}
whence, using \cref{eq:weight_log_lik},
\begin{align}\label{eq:bootstrap_growth2}
    \sqrt{n}\|\tilde{\theta}_2-\mle_2\|\in\mathcal{O}_{\mathbbm{P}}(1).
\end{align}
\subsubsection*{Step 2.}
By \cref{assump:log_lik_smoothness,assump:log_prior_smoothness},
\begin{equation*}
	0 = \partial_1 L_{1}^w(\tilde{\theta}_1)  = \partial_1  L_1^w(\hat{\theta}_1) + {\partial_{11} L_1^w(\hat{\theta}_1)}(\tilde{\theta}_1 - \hat{\theta}_1) +   \mathcal{O}_{\mathbbm{P}} \left( n \|\tilde{\theta}_1 - \hat{\theta}_1 \|^2\right),
	\end{equation*}
and therefore 
\begin{equation}\label{eq:proof_first_module}
  - \frac{1}{\sqrt{n}}  \partial_1 L_1^w(\hat{\theta}_1) = \frac{1}{n}{\partial_{11} L_1^w(\hat{\theta}_1)} \sqrt{n}(\tilde{\theta}_1 - \hat{\theta}_1) + \mathcal{O}_{\mathbbm{P}} \left( \sqrt{n}\|\tilde{\theta}_1 - \hat{\theta}_1 \|^2\right).
\end{equation}
Similarly, for the second module, by \cref{assump:log_lik_smoothness,assump:log_prior_smoothness}, we have
\begin{equation*}
	0 =  {\partial_2 L_{2}^w( \tilde{\theta}_1, \tilde{\theta}_2)} ={\partial_2  L_2^w(\hat{\theta}_1, \hat{\theta}_2) } + {\partial_{12} L_2^w(\hat{\theta}_1, \hat{\theta}_2)} (\tilde{\theta}_1 - \hat{\theta}_1)  + {\partial_{22} L_2^w(\hat{\theta}_1, \hat{\theta}_2)}(\tilde{\theta}_2 - \hat{\theta}_2) +   \mathcal{O}_{\mathbbm{P}} \left( n\|\tilde{\theta} - \hat{\theta} \|^2\right),
\end{equation*}
	where $\tilde{\theta} = (\tilde{\theta}_1, \tilde{\theta}_2)$ and $\hat{\theta} = (\hat{\theta}_1, \hat{\theta}_2)$. Hence, we have
	\begin{equation}\label{eq:proof_second_module}
  - \frac{1}{\sqrt{n}} \partial_2 L_2^w(\hat{\theta}_1, \hat{\theta}_2) - \frac{1}{n} {\partial_{12} L_2^w(\hat{\theta}_1, \hat{\theta}_2)} \sqrt{n}(\tilde{\theta}_1 - \hat{\theta}_1) = {\partial_{22} L_2^w(\hat{\theta}_1,\hat{\theta}_2)} \frac{\sqrt{n}}{n}(\tilde{\theta}_2 - \hat{\theta}_2) + \mathcal{O}_{\mathbbm{P}} \left( \sqrt{n}\|\tilde{\theta} - \hat{\theta} \|^2\right).
\end{equation}
By \cref{eq:mle_consistency,assump:positive_definite,assump:identifiability,assump:log_lik_smoothness,assump:log_prior_smoothness}, and \cite[Theorem 2]{spencer:2024},
$$
-\frac{1}{n}  {\partial_{11} L_1^w(\hat{\theta}_1)}, \quad
  -\frac{1}{n} {\partial_{12} L_2^w( \hat{\theta}_1, \hat{\theta}_2)}, \quad  -\frac{1}{n}{\partial_{22} L_2^w(\hat{\theta}_1, \hat{\theta}_2)}
$$ 
are invertible with probability going to one. Using \eqref{eq:proof_first_module} and \eqref{eq:proof_second_module}, for large enough $n$, 
\begin{align}\label{eq:proof_repeat_line}
\begin{split}
  &\sqrt{n}(\tilde{\theta}_2 - \hat{\theta}_2)  \\
  = &  \left(-\frac{1}{n}{\partial_{22} L_2^w(\hat{\theta}_1, \hat{\theta}_2)}\right)^{-1}  \frac{1}{\sqrt{n}} \partial_2 L_2^w(\hat{\theta}_1, \hat{\theta}_2) \\
  &-   \left(-\frac{1}{n}{\partial_{22} L_2^w(\hat{\theta}_1, \hat{\theta}_2)}\right)^{-1}  \left( -\frac{1}{n} {\partial_{12} L_2^w( \hat{\theta}_1, \hat{\theta}_2)}\right)^{-1} \sqrt{n}(\tilde{\theta}_1 - \hat{\theta}_1) +  \mathcal{O}_{\mathbbm{P}} \left( \sqrt{n} \|\tilde{\theta} - \hat{\theta} \|^2\right) \\
  = &  \left(-\frac{1}{n}{\partial_{22} L_2^w(\hat{\theta}_1, \hat{\theta}_2)}\right)^{-1}  \frac{1}{\sqrt{n}} \partial_2 L_2^w(\hat{\theta}_1, \hat{\theta}_2)\\
  &-    \left(-\frac{1}{n}{\partial_{22} L_2^w(\hat{\theta}_1, \hat{\theta}_2)}\right)^{-1}  \left( -\frac{1}{n} {\partial_{12} L_2^w( \hat{\theta}_1, \hat{\theta}_2)}\right)^{-1} \left( -\frac{1}{n}  {\partial_{11} L_1^w(\hat{\theta}_1)}\right)^{-1} \frac{1}{\sqrt{n}} \partial_1 L_1^w( \hat{\theta}_1)\\
  &+  \mathcal{O}_{\mathbbm{P}} \left( \sqrt{n} \|\tilde{\theta} - \hat{\theta} \|^2\right),
  \end{split}
\end{align}
and
\begin{align}\label{eq:proof_repeat_line1}
    \begin{split}
        \sqrt{n}(\tilde{\theta}_1 - \hat{\theta}_1) = \left(-\frac{1}{n}{\partial_{11} L_1^w(\hat{\theta}_1)}\right)^{-1}\frac{1}{\sqrt{n}}  \partial_1 L_1^w(\hat{\theta}_1)+\mathcal{O}_{\mathbbm{P}} \left( \sqrt{n}\|\tilde{\theta}_1 - \hat{\theta}_1 \|^2\right).
    \end{split}
\end{align}

\subsubsection*{Step 3.}
We now find  the asymptotic distribution of 
\begin{equation}\label{eq:L1_L2_limiting}
  \left(\frac{1}{\sqrt{n_2}}  \partial_1 L_1^w(\hat{\theta}_1),  \frac{1}{\sqrt{n_2}} { \partial_2 L_2^v(\hat{\theta}_1, \hat{\theta}_2)} \Big| x_{1,1:n_1}, x_{2, 1:n_2} \right),
\end{equation}
using the Lindeberg-Feller Central Limit Theorem and the Cramér-Wold device.
Let
$$C:=
\begin{pmatrix}
 I_1^* &  R_I^*\\
 R_I^{*T} & I_2^*
\end{pmatrix},
$$
fix a unit vector $z\in\mathbbm{R}^d$ and let
\begin{align*}
t_n(z):=&z^T\left(\frac{1}{\sqrt{n}}  \partial_1 L_1^w(\hat{\theta}_1),  \frac{1}{\sqrt{n}} \partial_2 L_2^w( \hat{\theta}_1,\hat{\theta}_2) \right)\\
\stackrel{\cref{eq:weight_log_lik}}=&\sum_{k_1=1}^{d_1}\frac{z_{k_1}}{\sqrt{n}}\left(w_0 \partial_{[k_1]} \log\pi_1(\mle_1) +\sum_{j=1}^n(w_j-1) \partial_{[k_1]} l_1(x_{1,j},\mle_1) \right)\\
&+\sum_{k_2=1}^{d_2}\frac{z_{d_1+k_2}}{\sqrt{n}}\left(w_0 \partial_{[k_2]} \log\pi_2(\mle_2) + \sum_{j=1}^n(w_j-1) \partial_{[k_2]} l_2(x_{2,j},\mle_1,\mle_2) \right)\\
=&\frac{a_{0,n}w_0}{\sqrt{n}}+\frac{1}{\sqrt{n}}\sum_{j=1}^na_{j,n}(w_j-1),
\end{align*}
where the partial derivative $\partial_{[k]}$ is defined as in \cref{assump:log_lik_smoothness}, and 
\begin{align*}
    &a_{0,n}=\sum_{k_1=1}^{d_1}z_{k_1} \partial_{[k_1]} \log \pi_1(\mle_1) + \sum_{k_2=1}^{d_2}z_{d_1+k_2} \partial_{[k_2]} \log \pi_2(\mle_2);\\ &a_{j,n}=\sum_{k_1=1}^{d_1}z_{k_1} \partial_{[k_1]} l_1(x_{1,j},\mle_1) + \sum_{k_2=1}^{d_2}z_{d_1+k_2} \partial_{[k_2]} l_2(x_{2,j},\mle_1,\mle_2),\quad \text{for }j\geq 1.
\end{align*}
Let $a_n^2=\max_{1\leq i\leq n} a_{i,n}^2$ and $s_n^2=\frac{1}{n}\sum_{j=1}^n a_{j,n}^2$. We have that $s_n^2\xrightarrow[n\to\infty]{}z^TCz>0$
 in $p^*$-probability and $\frac{1}{n}a_n^2\xrightarrow[n\to\infty]{}0$ in $p^*$-probability \cite[Chapter 3, Lemma 8]{newton1991weighted}. Hence, for any $\epsilon>0$,
\begin{align*}
    &\frac{1}{ns_n^2}\sum_{j=1}^n\mathbbm{E}\left[a_{j,n}^2(w_j-1)^2\mathbbm{1}_{\{|a_{j,n}(w_j-1)|^2>\epsilon ns_n^2\}}\,\Big|\,x_{1,1:n},x_{2,1:n}\right]\\
    \leq  &\mathbbm{E}\left[(w_1-1)^2\mathbbm{1}_{\{|w_1-1|^2>\epsilon ns_n^2/a_n^2\}}\,\Big|\,x_{1,1:n},x_{2,1:n}\right]\xrightarrow[n\to\infty]{}0\quad\text{in }p^*\text{-probability}.
\end{align*}
It follows from the Lindeberg-Feller Central Limit Theorem that $\mathcal{L}(t_n(z)\,|\,x_{1,1:n},x_{2,1:n})$ converges weakly, in $p^*$-probability to $\mathcal{N}(0,z^TCz)$. The Cramér-Wold device implies that the distribution of
\begin{equation*}
  \left(\frac{1}{\sqrt{n}}  \partial_1 L_1^w(\hat{\theta}_1),  \frac{1}{\sqrt{n}} \partial_2 L_2^w( \hat{\theta}_1,\hat{\theta}_2) \Big| x_{1,1:n}, x_{2,1:n} \right)
\end{equation*}
converges weakly to the Normal distribution with mean 0 and covariance matrix $C$ in $p^*$-probability.

\subsubsection*{Conclusion.}
Note that, by \cref{eq:mle_consistency,assump:positive_definite,assump:identifiability,assump:log_lik_smoothness,assump:log_prior_smoothness} and the uniform law of large numbers \cite[Theorem 2]{spencer:2024},
$$
-\frac{1}{n}  {\partial_{11} L_1^w(\hat{\theta}_1)}, \quad
  -\frac{1}{n} {\partial_{12} L_2^w( \hat{\theta}_1, \hat{\theta}_2)}, \quad  -\frac{1}{n}{\partial_{22} L_2^w(\hat{\theta}_1, \hat{\theta}_2)}
$$ 
converge in probability to $J_1^*$, $R_J^*$ and $J_2^*$, respectively. Using Slutsky's
theorem together with the conclusion of \textbf{Step 3} and \cref{eq:proof_repeat_line,eq:proof_repeat_line1,eq:bootstrap_growth1,eq:bootstrap_growth2}, we get that the law of $$\left(\sqrt{n}(\tilde{\theta}_1 - \hat{\theta}_1,\tilde{\theta}_2 -
\hat{\theta}_2)\,\Big|\,x_{1,1:n_1}, x_{2,1:n_2}\right)$$ converges weakly  to the Normal distribution with mean 0 and
covariance matrix~\eqref{eq:asymptotic2sM}, in $p^*$-probability.
\qed

\section{Prediction under PBMI versus
cut posteriors}\label{suppl:prediction}

In \cref{section:algorithms:asymptotics} we mentioned briefly the predictive
performance of 
the Weighted Likelihood Bootstrap relative to a simple Bayesian model corresponding to the first module of \cref{eq:general_model}, considered in isolation. Formally
this can be analysed by
considering the following risk  function, defined using the notation of \cref{section:two_scenarios}:
\begin{align}\label{eq:risk_function}
  \mathbbm{E}_{x_{1:n}} \left[D_{KL}\left\{p^*||\hat{p}(\cdot|x_{1:n})\right\} \right]
  = \int p^*(x_{1:n}) \int p^*(x_{n+1}) \log \frac{p^*(x_{n+1})}{\hat{p}(x_{n+1}|x_{1:n})} \mathrm{d}x_{n+1}  \mathrm{d}x_{1:n},
\end{align}
where $\hat{p}(\cdot|x_{1:n})$ is the predictive distribution given data
$x_{1:n}$ under a given method of inference. Asymptotically the risk in
\eqref{eq:risk_function} is smaller under Weighted Likelihood Bootstrap than
under Bayesian inference \citep{fushiki2005bootstrap, fushiki2010bayesian,
lyddon2018nonparametric}. 

In what follows we first find the asymptotic approximation of the risk
\eqref{eq:risk_function} for prediction on $x_2$, under Posterior Bootstrap
for Modular Inference, and under cut posteriors. 
We then design an example showing that each of the two approaches may yield better
prediction for $x_{2}$, depending on the exact setting.

\subsection{Informal asymptotic expansion of the risk}
Let $\hat{p}_B$ and $\hat{p}_{PB}$ denote the predictive distribution on
$x_{2}$  given $x_{1, 1:n_1}, x_{2, 1:n_2}$ under the cut posterior and under
Algorithm~\ref{alg:pb_cut_dependent} (or its modified version given in \cref{rmk:modified_algorithm}). Recall that $p_2^*$ denotes the density of the true
data-generating distribution of $x_{2,1:n_2}$. Our (informal) expansion of the risk is based on \citet{shimodaira2000improving}.
We define 
\begin{align}\label{eq:a_b_a_pb}
\begin{split}
a_{B} =&  \lim_{n_2\to\infty} n_2 \int (\theta - \hat{\theta})\pi_{\text{cut}}(\theta| x_{1,1:n_1},  x_{2, 1:n_2}) \mathrm{d} \theta,\\
a_{PB} =& \lim_{n_2\to\infty} n_2 \int (\theta - \hat{\theta})p_{PB}(\theta| x_{1,1:n_1}, x_{2, 1:n_2}) \mathrm{d} \theta,
\end{split}
\end{align}
where the above limits are taken in $p^*$-probability, and $p_{PB}(\theta| x_{1,1:n_1}, x_{2, 1:n_2})$
refers to the distribution of the output of Algorithm \ref{alg:pb_cut_dependent} (or its version from \cref{rmk:modified_algorithm}). We assume that those limits exist and are deterministic (see, e.g. \cite[Theorem 3.1]{johnson1970asymptotic}). For simplicity, we assume that $f_2(x_{2}|\theta_1,\theta_2,x_{1})$, introduced in \cref{section:two_scenarios} can be written as $f_2(x_{2}|\theta_1,\theta_2)$, i.e. it does not depend on $x_1$.
We also introduce the matrices
\begin{align}
\begin{split}\label{eq:I_J_f_2}
I_{f_2}^{*} =\mathbbm{E}_{p_2^*} \left\{ \partial\log  f_2(x_{2}| \theta_1^*, \theta_2^*) \cdot  \left(\partial\log  f_2(x_{2}| \theta_1^*, \theta_2^*) \right)^T \right\}, \quad J_{f_2}^{*} =  -\mathbbm{E}_{p_2^*} \left\{ \partial^2 \log f_2(x_{2}| \theta_1^*, \theta_2^*) \right\}.
\end{split}
\end{align}

In order to
simplify the notation, we denote $(x_{1,1:n_1}, x_{2,1:n_2})$ by $x$.
Analogously to \eqref{eq:risk_function}, we define the risk associated with
prediction for the second module, as
\begin{equation*}
	\mathbbm{E}_{p^*} \left[D_{KL}\left\{p_2^*||\hat{p}(\cdot|x)\right\} \right] = \int p^*(x) \left\{\int p_2^*(x_{2}) \log \frac{p_2^*(x_{2})}{\hat{p}(x_{2}|x)} \mathrm{d}x_{2} \right\} \mathrm{d}x,
\end{equation*}
for a predictive distribution $\hat{p}$, which in our case is either $\hat{p}_B$ or $\hat{p}_{PB}$. Firstly, notice that 
\begin{equation}\label{eq:risk_function_cut}
\mathbbm{E}_{p^*} \left[D_{KL}\left\{p_2^*||\hat{p}(\cdot|x)\right\} \right] = - \int p^*(x) \left\{\int p_2^*(x_2) \log {\hat{p}(x_{2}|x)} \mathrm{d}x_{2} \right\} \mathrm{d}x + c,
\end{equation}
where the constant $c$ does not depend on the prediction method.

In order to find asymptotic expansions of  $\mathbbm{E}_{p^*}
\left[D_{KL}\left\{p_2^*||\hat{p}_B(\cdot|x)\right\} \right]$ and $\mathbbm{E}_{p^*}
\left[D_{KL}\left\{p_2^*||\hat{p}_{PB}(\cdot|x)\right\} \right]$ up to
$o(n_2^{-1})$, we follow the steps of \citet{fushiki2010bayesian,shimodaira2000improving}.
We have 
\begin{equation}\label{eq:predictive_cut_Bayesian}
    \hat{p}_B(x_{2}| x) = \int f_2(x_{2}| \theta_1, \theta_2) \pi_{\text{cut}}(\theta_1,\theta_2| x) \mathrm{d} \theta_1\mathrm{d} \theta_2,
\end{equation}
and 
\begin{equation*}
    \hat{p}_{PB}(x_{2}| x) = \int f_2(x_{2}| \theta_1, \theta_2) p_{PB}(\theta_1,\theta_2| x) \mathrm{d} \theta_1\mathrm{d} \theta_2.
\end{equation*}
We start by expanding asymptotically equation \eqref{eq:predictive_cut_Bayesian} around $(\hat{\theta}_1, \hat{\theta}_2)$. We suppose that the cut posterior satisfies the assumptions of \cref{thm:bvm_cut} and that the convergence holds not only with respect to the total variation distance but that the first and second moments also converge.
Setting $\theta = (\theta_1, \theta_2)$, we then have
\begin{align*}
\hat{p}_B(x_{2}| x) = &  f_2(x_{2}| \hat{\theta}_1, \hat{\theta}_2) \int\pi_{\text{cut}}(\theta| x) \mathrm{d} \theta + \partial f_2(x_{2}| \hat{\theta}_1, \hat{\theta}_2)^T  \int (\theta - \hat{\theta})\pi_{\text{cut}}(\theta| x) \mathrm{d} \theta \\
& \quad  +\int (\theta - \hat{\theta})^T  \partial^2 f_2(x_{2}| \hat{\theta}_1, \hat{\theta}_2) (\theta -\hat{\theta}) \pi_{\text{cut}}(\theta| x)  \mathrm{d} \theta  + o_P(n_2^{-1})\\
= &  f_2(x_{2}| \hat{\theta}_1, \hat{\theta}_2) + \partial f_2(x_{2}| \hat{\theta}_1, \hat{\theta}_2)^T \int (\theta - \hat{\theta})\pi_{\text{cut}}(\theta| x) \mathrm{d} \theta \\
& \quad + \frac{1}{2n_2} \text{tr} \left\{\partial^2 f_2(x_{2}| \hat{\theta}_1, \hat{\theta}_2) \Sigma_B \right\}+ o_{\mathbbm{P}}(n_2^{-1}),
\end{align*}
where $\Sigma_B=H^{-1}$ for $H$ given in \cref{eq:Hinverse}. Therefore,
\begin{align*}
\begin{split}
\hat{p}_B(x_{2}| x) = &  f_2(x_{2}| \hat{\theta}_1, \hat{\theta}_2) \Bigg[1 +  \partial\log  f_2(x_{2}| \hat{\theta}_1, \hat{\theta}_2)^T \int (\theta - \hat{\theta})\pi_{\text{cut}}(\theta| x) \mathrm{d} \theta \\
& + \quad  \frac{1}{2n_2} \text{tr} \left\{\frac{1}{f_2(x_{2}| \hat{\theta}_1, \hat{\theta}_2)}\partial^2 f_2(x_{2}| \hat{\theta}_1, \hat{\theta}_2) \Sigma_B\right\}\Bigg] + o_{\mathbbm{P}}(n_2^{-1}).
\end{split}
\end{align*}
Following \citet{shimodaira2000improving}, we use the identity 
\begin{equation*}
    \frac{1}{g} {\partial^2 g} =  {\partial \log g}  \cdot {\partial \log g}^T + \partial^2 \log g,
\end{equation*}
combined with the Taylor expansion of $\log(1+y)$ around $y=0$ to get
\begin{align*}
\begin{split}
\log \hat{p}_B(x_{2}| x) = & \log  f_2(x_{2}| \hat{\theta}_1, \hat{\theta}_2)    +  \partial\log  f_2(x_{2}| \hat{\theta}_1, \hat{\theta}_2)^T \int (\theta - \hat{\theta})\pi_{\text{cut}}(\theta_1, \theta_2| x) \mathrm{d} \theta_1 \mathrm{d} \theta_2 \\
& + \quad  \frac{1}{2n_2} \text{tr} \left[ \left\{I_{f_2}^{\hat{\theta}}(x_2) - J_{f_2}^{\hat{\theta}}(x_2) \right\}\Sigma_B \right] + o_P(n_2^{-1}),
\end{split}
\end{align*}
where
$$
I_{f_2}^{\theta}(x_2) = \partial\log  f_2(x_{2}| \theta_1, \theta_2) \cdot \left(\partial\log  f_2(x_{2}| \theta_1, \theta_2)\right)^T, \quad J_{f_2}^{\theta}(x_2) =  - \partial^2 \log f_2(x_{2}| \theta_1, \theta_2).
$$
Finally, we use \eqref{eq:risk_function_cut} to get that
\begin{align*}
\mathbbm{E}_{p^*} \left[D_{KL}\left\{p_2^*||\hat{p}_B(\cdot|x)\right\} \right] = &\mathbbm{E}_{p^*} \left[D_{KL}\left\{p_2^*|| f_2(\cdot |\hat{\theta}_1, \hat{\theta}_2)\right\}\right] -  \mathbbm{E}_{p_2^*}\left\{\partial\log  f_2(x_{2}| {\theta}_1^*, {\theta}_2^*) \right\}^T \frac{a_B}{n_2}  \\
& \quad -  \frac{1}{2n_2} \text{tr} \left\{ \left(I_{f_2}^{*}
- J_{f_2}^{*}\right)\Sigma_B \right\} + o(n_2^{-1}),
\end{align*}
for matrices $I_{f_2}^{*}$ and $J_{f_2}^{*}$ defined in \eqref{eq:I_J_f_2}.
Now, assume that the assumptions of \cref{thm:dependent} hold and that, additionally to the convergence in \cref{thm:dependent}, convergence of first and second moments holds.
Then, performing analogous calculations for Posterior Bootstrap, we obtain
\begin{align*}
\mathbbm{E}_{p^*} \left[D_{KL}\left\{p_2^*||\hat{p}_{PB}(\cdot|x)\right\} \right] = &\mathbbm{E}_{p^*} \left[D_{KL}\left\{p_2^*|| f_2(\cdot |\hat{\theta}_1, \hat{\theta}_2)\right\}\right] -  \mathbbm{E}_{p_2^*}\left\{{\partial\log  f_2(x_{2}| {\theta}_1^*, {\theta}_2^*)}^T \right\}\frac{a_{PB}}{n_2}  \\
& \quad -  \frac{1}{2n_2} \text{tr} \left\{ \left(I_{f_2}^{*}
- J_{f_2}^{*}\right\}\Sigma_{PB} \right) + o(n_2^{-1}),
\end{align*}
where $\Sigma_{PB}$ is equal to $\Sigma$ given in \cref{eq:asymptotic2sM}.

\subsection{Example}
We use the above informal expansions to show heuristically that Posterior
Bootstrap in some settings yields asymptotically better prediction for the
second module, whereas in other settings Bayesian modular inference performs
better. Let us consider the following misspecified model:
\begin{align}\label{eq:counterexample_model}
\begin{split}
x_{1,1:n} &\sim N(\theta_1,1), \quad\theta_1 \sim N(0,1), \\
 x_{2,1:n}& \sim N\left((\theta_1, \theta_2), \Sigma_1\right), \quad \theta_2 \sim N(0,1),
\end{split}
\end{align}
where 
$$
\Sigma_1 = \begin{pmatrix}
 4/3 & -2/3 \\
 -2/3 & 4/3
\end{pmatrix} \quad \text{and therefore} \quad \Sigma_1^{-1} = \begin{pmatrix}
 1 & 0.5 \\
 0.5 & 1
\end{pmatrix}.
$$
The data are generated independently according to
\begin{align}\label{eq:counterexample_data_generating}
\begin{split}
x_{1,1:n} \sim N(0, \sigma^2), \quad
x_{2,1:n} \sim N(0, \Sigma_2),
\end{split}
\end{align}
 where   
$$
\Sigma_2 = \Sigma_1 \begin{pmatrix}
0.8 & 0.3\\
0.3 & 0.8 
\end{pmatrix} \Sigma_1.
$$
We consider below different scenarios for the value of $\sigma$.

Firstly, note that $(\theta_1^*, \theta_2^*) = (0,0)$, and that the pseudo-true parameter for $\theta_1$ with respect to $f_2$ is also equal to 0, thus
$$
E_{p_2^*}\left\{{\partial\log  f_2(x_{2}| {\theta}_1^*, {\theta}_2^*)}\right\} = 0.
$$

Consequently
\begin{align*}
& \mathbbm{E}_{p^*} \left[D_{KL}\left\{p_2^*||\hat{p}_{PB}(\cdot|x)\right\} \right] - E_{p^*} \left[D_{KL}\left\{p_2^*||\hat{p}_{B}(\cdot|x)\right\} \right]
=  \frac{1}{2n}\text{tr} \left\{ \left(I_{f_2}^{*}
- J_{f_2}^{*}\right) \left(\Sigma_{B} -
\Sigma_{PB}\right) \right\} + o(n^{-1}).
\end{align*}

For this example we have
$$
I_{f_2}^{*} = \begin{pmatrix}
0.8 & 0.3\\
0.3 & 0.8 
\end{pmatrix}, \quad
J_{f_2}^{*} = \begin{pmatrix}
  1 &  0.5 \\
0.5 & 1
\end{pmatrix},
$$
and $R_J^* = 0.5$. Thus $J_1^{*-1} = 1$, $J_2^{*-1} = 1$, $I_1^* =\sigma^2$ and $I_1^* =2.5$. We also have
$$
\Sigma_B = \begin{pmatrix}
 1 &  -0.5 \\
-0.5  & 1.25
\end{pmatrix}, \quad 
\Sigma_{PB} = \begin{pmatrix}
\sigma^2 &  - 0.5 \sigma^2\\
-0.5 \sigma^2  & 0.8 + 0.25 \sigma^2
\end{pmatrix}.
$$
Since
$$
I_{f_2}^{*} - J_{f_2}^{*}= \begin{pmatrix}
 - 0.2 &  -0.2\\
-0.2  & -0.2
\end{pmatrix},
$$
we have $\text{tr} \{ (I_{f_2}^{*} - J_{f_2}^{*})\Sigma_B \} = -0.25$. As for the corresponding trace for prediction based on Posterior Bootstrap,  for $\sigma = 1$ we have $\text{tr} \{ (I_{f_2}^{*} - J_{f_2}^{*})\Sigma_{PB} \} = -0.21$, whereas for $\sigma = 2$ the value of the trace is $-0.36$.

To summarize, we have shown that Algorithm~\ref{alg:pb_cut_dependent} in its modified version from \cref{rmk:modified_algorithm}, in model \eqref{eq:counterexample_model}, does not necessarily yield asymptotically better prediction for the second module than the cut posterior, and that it depends on the value of $\sigma^2$ in the data-generating mechanism \eqref{eq:counterexample_data_generating}.

\end{document}

%% file: shortcuts.tex
\newcommand{\fisherinformation}{\hat{J}_{n}}

\newcommand{\postfisherinformation}{\bar{J}_{n}}
\newcommand{\param}{\theta}
\newcommand{\logposterior}{\overline{L}}
\newcommand{\posterior}{\Pi_n}
\newcommand{\mle}{\hat{\theta}}
\newcommand{\truepar}{\theta^*}

\newcommand{\prior}{\pi}
\newcommand{\loglikelihood}{L}
\newcommand{\loglikelihoodone}{L_1}
\newcommand{\loglikelihoodtwo}{L_2}
\newcommand{\Tr}{\text{Tr}}

\newcommand{\thirdder}{M_{2}}

\newcommand{\asymptdermixed}{M_4(X_1,X_2)}
\newcommand{\asymptthirdder}{M_1(X_1,X_2)}
\newcommand{\asymptintegrals}{M_5(X_1,X_2)}

\newcommand{\priorboundone}{\hat{M}_1}
\newcommand{\priorboundtwo}{\hat{M}_1^{(1)}}
\newcommand{\priorboundthree}{\hat{M}_1^{(2)}}
\newcommand{\priorboundfour}{\widetilde{M}_1}
\newcommand{\priorboundfive}{M_1}
\newcommand{\priorboundsix}{\widetilde{M}_1^{(2)}}
\newcommand{\priorboundseven}{M_1^{(2)}}
\newcommand{\thirdderone}{\hat{M}_2^{(1)}}
\newcommand{\thirddertwo}{\hat{M}_2^{(2)}}
\newcommand{\firstder}{\widetilde{M}_2}
\newcommand{\secondderone}{\overline{M}_2^{(1)}}
\newcommand{\secondder}{\widetilde{M}_2^{(2)}}
\newcommand{\thirddermixed}{\overline{M}_2^{(2)}}
\newcommand{\seconddermixed}{\overline{M}_2}
\newcommand{\firstdersquaredintegral}{M_3}
\newcommand{\secondderintegral}{\overline{M}_3}

\newcommand{\thirddermixedone}{\widetilde{M}_2^{(1)}}
\newcommand{\priorboundthirdder}{\overline{M}_1}

\newcommand{\normone}[1]{\ensuremath{\|#1\|_{d_1}}}
\newcommand{\normtwo}[1]{\ensuremath{\|#1\|_{d_2}}}

\newcommand{\innerint}[1]{\int_{\tiny\begin{cases}\normone{#1}\leq\delta_1\sqrt{n}\\\normtwo{#1}\leq \delta_2\sqrt{n}\end{cases}}}

\def\jhat{\fisherinformation(\mle)}
\def\jbartwo{\bar{J}_{2,n}(\mle)}

\def\jbarplus{\bar{J}^{p}_{2,n}(\mle)}
\def\jhatplus{\hat{J}^{p}_{n}(\mle)}

\def\jtildeplus{\tilde{J}^{p}_{n}(\mle)}

\def\jtilde{\hat{J}_{2,n}(\mle)}
\def\jbreve{\hat{J}_{1,n}(\mle_1)}

\def\jbarminus{\bar{J}^{m}_{2,n}(\mle)}
\def\jbarminustwo{\bar{J}^{m,(2)}_{n}(\mle)}

\def\det#1{\left|\mathrm{det}\left(#1\right)\right|}

\def\decaybartwo{\bar{\mathscr{D}}^{(2)}(n)}

\def\decaybarplus{\bar{\mathscr{D}}^{p}(n)}

\def\decayhatbar{\bar{\mathscr{D}}(n)}

\def\decayhatplus{\hat{\mathscr{D}}^{p}(n)}

\def\expplus{{\mathscr{E}_n^p}}
\def\expminus{{\mathscr{E}_n^m}}

\def\pminus{\mathscr{P}_n^{m}}

\newcommand{\postminevaluemle}{\bar{\lambda}_{\min}(\mle)}

\def\minevalmleplus{\hat{\lambda}^{p}_{\min}(\mle)}

\def\minevalpost{\tilde{\lambda}_{\min}^{\text{post}}(\mle)}
\def\minevalbarplus{\bar{\lambda}^p_{\min}(\mle)}

\crefname{equation}{}{}

%% file: numerics_input.tex
\section{Illustrations}\label{section:illustrations_modified}




\subsection{Toy example}\label{section:toy_example_v2}

We consider the following model, defined conditionally on covariates $x_1, \ldots, x_n$:
\begin{align}\label{model:toy_example_v2}
  \begin{pmatrix}
    z_i\\
    y_i
  \end{pmatrix} | x_i, \theta_1, \theta_2
  &\overset{i.i.d}{\sim}
   \mathrm{Normal}\left(
    \begin{pmatrix}
      \theta_1\\
      \theta_1 + \theta_2 x_i
    \end{pmatrix},
    \begin{pmatrix}
      1 & 0\\
      0 & 1
    \end{pmatrix}
  \right), \quad i = 1, \ldots, n.
\end{align}
The first module refers to the marginal model for $z_i$ given $\theta_1$, 
and the second module to the conditional model for $y_i$ given $\theta_1,\theta_2,x_i$.
Independent $\mathrm{Normal}(0, 10^2)$ priors are set on $\theta_1$ and $\theta_2$.

The MLE in the first module is $\hat{\theta}_1 = n^{-1}\sum_{i=1}^n z_i$, converging to $\mathbb{E}[z]$.
Given $\theta_1$, the MLE in the second module is $\hat{\theta}_2 = \sum_{i=1}^n (y_i - \theta_1) x_i/\sum_{i=1}^n x_i^2$, converging to ${\mathbb{E}[(y-\theta_1)x]}/{\mathbb{E}[x^2]}$. In the limit, we obtain
$\theta_1^* = \mathbb{E}[z]$, and $\theta_2^* = \{\mathbb{E}[yx] - \mathbb{E}[x]\theta_1^*\}/\{\mathbb{E}[x^2]\}$.
In the joint model, the MLE $(\hat{\theta}_1^{\text{joint}}, \hat{\theta}_2^{\text{joint}})$ can also be obtained analytically. In general the two-stage and joint MLEs differ, and their limits differ as well.
The limit of the joint model MLE is given by
\begin{align*}
  \theta_1^{\text{joint}*} &= \frac{\mathbb{E}[z] + \mathbb{E}[y] - \mathbb{E}[x]\mathbb{E}[xy]/\mathbb{E}[x^2]}{2\left\{1 - {\mathbb{E}[x]^2}/\{2\mathbb{E}[x^2]\}\right\} },\;\;
  \theta_2^{\text{joint}*} = \frac{\mathbb{E}[x y] - \mathbb{E}[x]\theta_1^{\text{joint}*}}{\mathbb{E}[x^2]}.
\end{align*}

\begin{figure}
  \centering
  \begin{subfigure}{0.45\textwidth}
    \includegraphics[width=\textwidth]{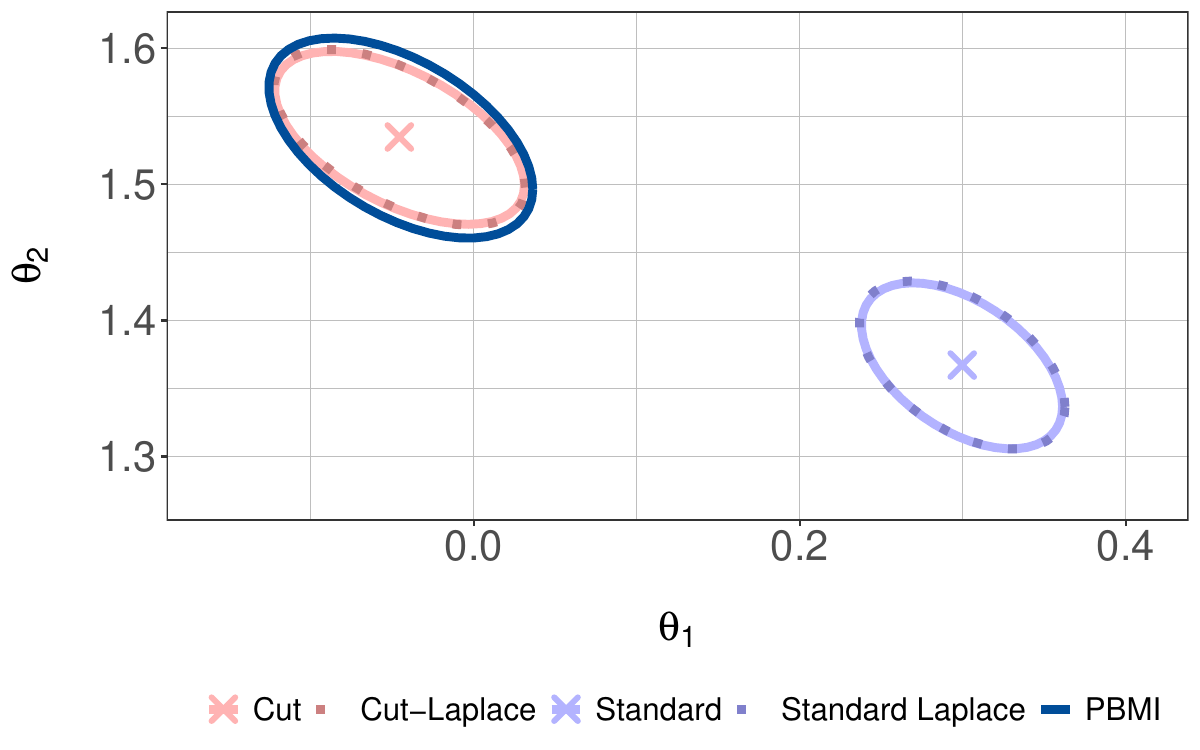}
    \caption{Scenario 1: $\rho = 0, \sigma^2 = 1$.}
    \label{fig:subfig1}
  \end{subfigure}
  \hfill
  \begin{subfigure}{0.45\textwidth}
    \includegraphics[width=\textwidth]{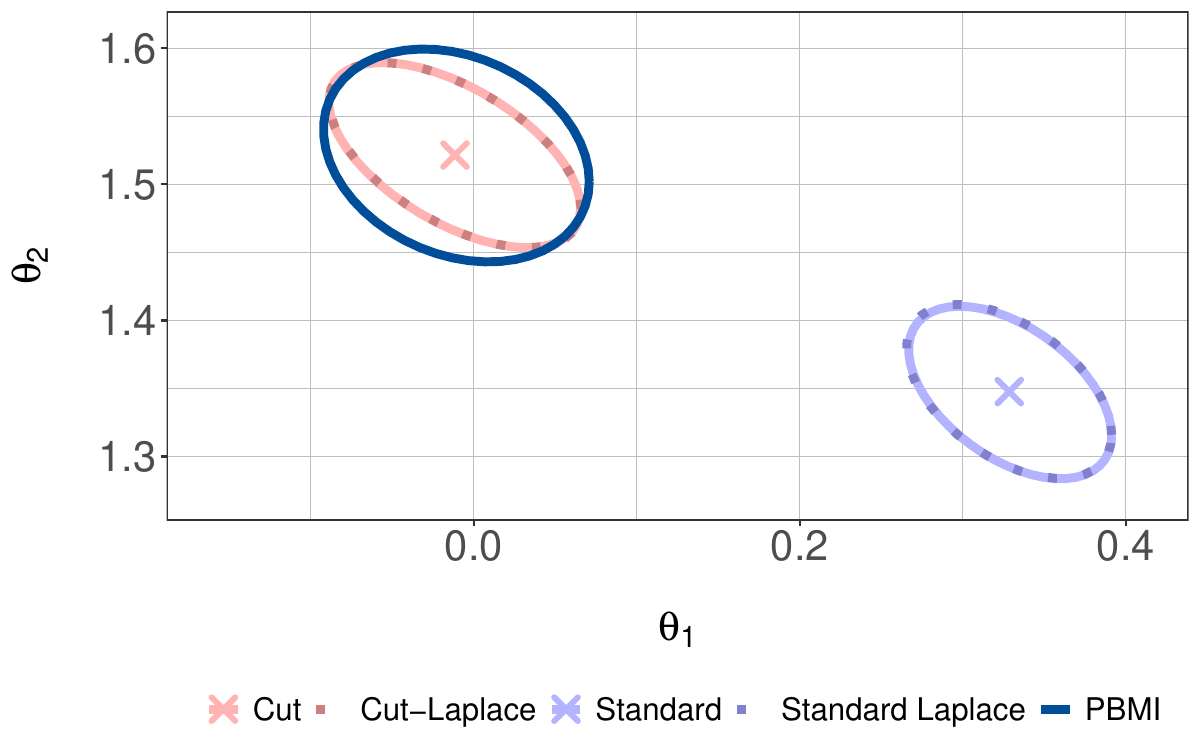}
    \caption{Scenario 2: $\rho = 0.5, \sigma^2 = 2$.}
    \label{fig:subfig2}
  \end{subfigure}
  \caption{\label{fig:toy_example_v2}
  Standard Posterior distribution, its Laplace approximation, cut posterior distribution, and its Laplace approximation, and PBMI for model \eqref{model:toy_example_v2} under different scenarios.}
\end{figure}

We generate $x_1, \ldots, x_n \sim \mathrm{Normal}(1, 1)$, 
with $n = 1000$, 
and generate $(z_i, y_i)$ from a bivariate Normal distribution with mean $(0, 1+x_i)$ and covariance matrix
$$
\begin{pmatrix}
  1 & \rho\\
  \rho & \sigma^2
\end{pmatrix}.
$$
The model is misspecified.
We consider two different values of $\rho$ and $\sigma^2$. In Scenario 1, we set $\rho = 0$ to make $z_i$ and $y_i$ independent given $x_i$,
and set $\sigma^2 = 1$ to make the second module well-specified given $\theta_1$. In Scenario 2, we instead set 
$\rho = 0.5$ and $\sigma^2 = 2$. 
Figures \ref{fig:subfig1}-\ref{fig:subfig2} display the standard posterior and the cut posterior. 
They concentrate on noticeably different regions of the parameter space, in both scenarios, 
for both $\theta_1$ and $\theta_2$.
We also observe that the Laplace approximations are accurate.
Finally, we note that the distributions are very similar in both scenarios, so that neither the introduction of correlations between $z_i$ and $y_i$,
nor the misspecification of the second module, have a strong impact on the standard and the two-stage posteriors.

The results of PBMI (Algorithm~\ref{alg:pb_cut_dependent}, with $w_0=v_0=1$) are shown in \cref{fig:subfig1}-\cref{fig:subfig2}. We observe that PBMI and the cut posterior yield similar results in Scenario 1, where $z_i$ and $y_i$ are independent given $x_i$. In Scenario 2, where there is dependence between $z_i$ and $y_i$ given $x_i$, the difference between PBMI and the cut posterior is noticeable, with PBMI reflecting increased uncertainty about $\theta_2$.
These observations align with Theorems~\ref{thm:bvm_cut} and \ref{thm:dependent}, because in Scenario 1, $I_1^* = J^*_1$ and $R_I^*=0$, therefore PBMI and cut posteriors have the same asymptotic variance, while they do not in Scenario 2.

 

In Table~\ref{table:coverage_toy_example_v2} we report the coverage of 
95\% credible intervals for $\theta_1$, $\theta_2$, and $\theta_1 + \theta_2$. The numbers
are obtained by repeatedly generating datasets as above,
and computing the frequency of the event where the two-stage pseudo-true parameter $(\theta_1^*, \theta_2^*)$ is contained in the credible intervals, for each method. Here, the cut posterior is approximated by its Laplace approximation. In Scenario 1, both the cut posterior and PBMI provide close to nominal coverage for all three targets of inference.
In Scenario 2, the cut posterior undercovers $\theta_2$ and $\theta_1 + \theta_2$, while PBMI provides close to nominal coverage. The coverage for the standard posterior is not reported as it concentrates on a different region of the parameter space, as seen in Figure~\ref{fig:toy_example_v2}.

\begin{table}[H]
\begin{center}
    \begin{subfigure}{0.48\textwidth}
    \input{toy_example_v2_coverage_scenario1.tex}
    \caption{Scenario 1: $\rho = 0, \sigma^2 = 1$.}
    \label{fig:subtable:scenario1}
  \end{subfigure}
  \hfill
    \begin{subfigure}{0.48\textwidth}
    \input{toy_example_v2_coverage_scenario2.tex}
    \caption{Scenario 2: $\rho = 0.5, \sigma^2 = 2$.}
    \label{fig:subtable:scenario2}
  \end{subfigure}
  \end{center}
     \caption{\label{table:coverage_toy_example_v2} Coverage of 0.95-credible intervals for $\theta_1$, $\theta_2$ and $\theta_1 + \theta_2$ in model \eqref{model:toy_example_v2}, for the two-stage pseudo-true parameter. Based on 1000 datasets with $n=1000$.}
\end{table}

\subsection{Causal inference with propensity scores\label{section:causal_inference_example_v2}}

\begin{figure}[t]
  \centering
  \begin{subfigure}{0.48\textwidth}
    \includegraphics[width=\textwidth]{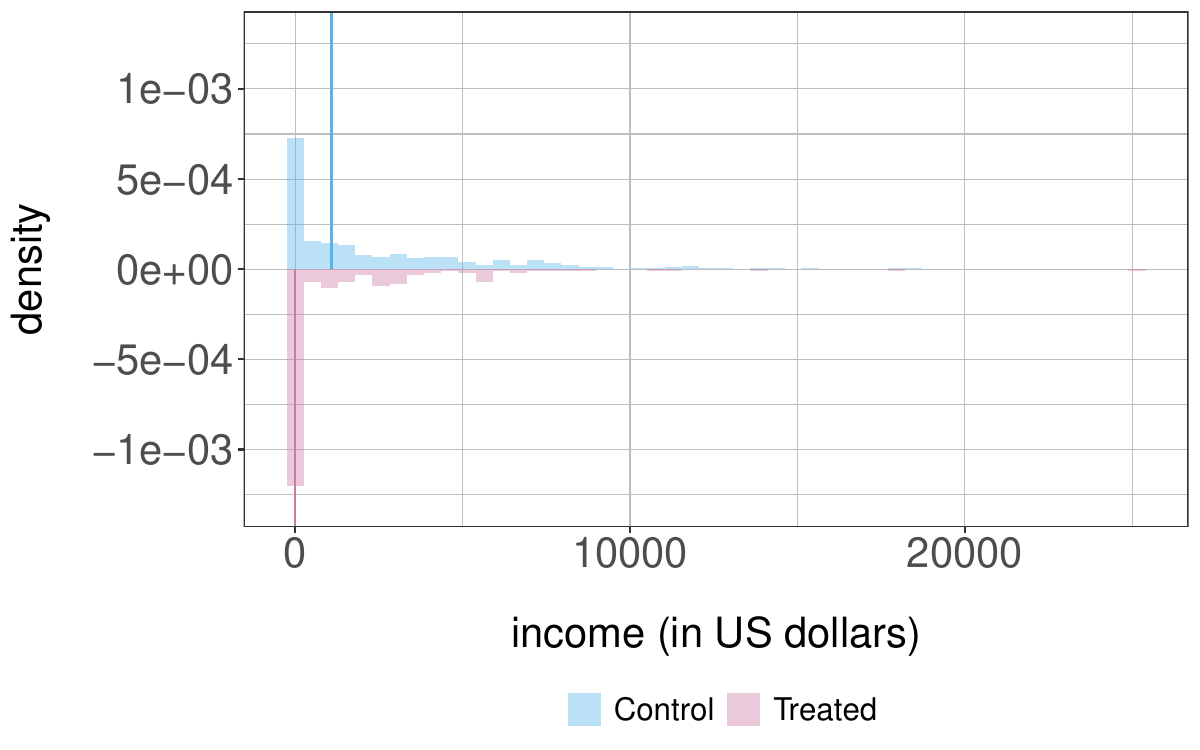}
    \caption{Pre-intervention income distribution.}
    \label{fig:income_histogram_v2:subfig1}
  \end{subfigure}
  \hfill
  \begin{subfigure}{0.48\textwidth}
    \includegraphics[width=\textwidth]{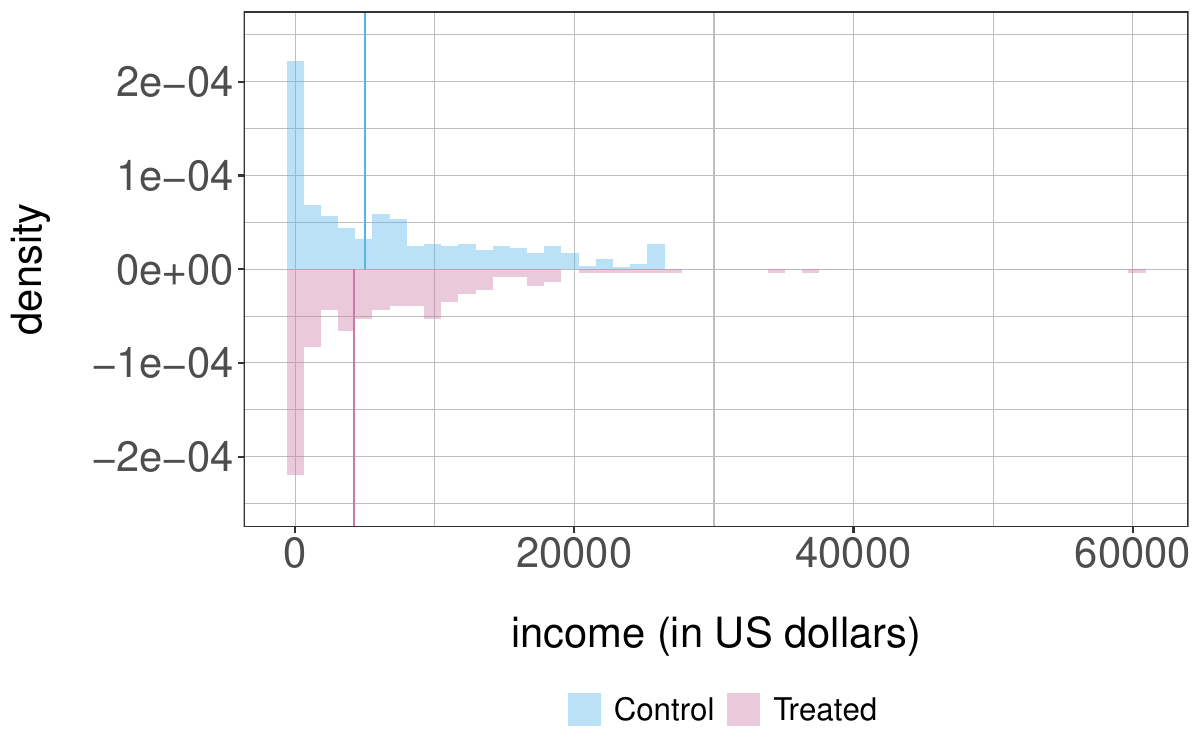}
    \caption{Post-intervention income distribution.}
    \label{fig:income_histogram_v2:subfig2}
  \end{subfigure}
  \caption{\small Histograms of income of the participants of the study, pre-intervention (left panel) and post-intervention (right panel) for the treated and the non-treated group. The histograms are normalized by group. Vertical segments represent the median of each group. \label{fig:income_histogram_v2}}
\end{figure}

To illustrate Algorithm~\ref{alg:pb_cut_dependent} on causal
inference models described in Section~\ref{section:two_scenarios}, we revisit
the example studied by \cite{lalonde1986evaluating} and later by
\cite{dehejia1999causal}, who measured the effect of labor training on income. The dataset comes from  National Supported Work
Demonstration and  the Population Survey of Income Dynamics, and is
available as the \texttt{lalonde} dataset in \texttt{MatchIt} \texttt{R}
package.  The treatment assignment was not randomized, in particular the
assignment to the intervention was highly dependent on the pre-intervention
income. This can be seen on \cref{fig:income_histogram_v2},
which shows the income distribution per treatment group, before and after intervention.
For example, the median pre-intervention income for the treated group is
0, while it is around 1000 US dollars for the non-treated group. The average income per group is respectively 1532 and 2466 US dollars.

Following \cite{dehejia1999causal}, we define the first module as a logistic
regression with treatment as the dependent variable, and the following
covariates: age, age squared, number of years of schooling, race (black,
Hispanic or white), an indicator for being married, an indicator for whether the
individual has a high school degree, income in 1974 and income in 1975. 
We
standardize all continuous variables so that they have mean 0 and variance 1,
and add a column of ones to the covariates to account for the intercept. 
The second module is a linear regression model, where the outcome variable
$Y_i$ is the income of individual $i$ after the intervention (in 1978),
standardized to have expectation 0 and variance 1. 
The covariates are
the treatment indicator and quintiles of propensity scores obtained in the
first module. Thus:
\begin{align}\label{model:linear_causal_v2}
    \mathrm{Module\ 1:}  \quad \text{logit} P(Z_i =1|X_i) &=  X_i^T \theta_1, \quad \mathrm{Module\ 2:} \quad Y_i = \beta_{Z} Z_i + \mathrm{ps}_i^T \gamma + \epsilon_i,
\end{align}
where $\epsilon_i \sim \mathrm{Normal}(0, \sigma^2)$,  $\mathrm{ps}_i$
is a 5-dimensional vector with first entry equal to one, and, for $k\in\{2,3,4,5\}$, $k$-th entry equal to one if
$X_i^{\top}\theta_1 \in [q_{k}, q_{k+1})$, where $q_k$
is the $k$th empirical quintile (i.e. $0.2\times k$-th empirical quantile) of $(X_i^{\top}\theta_1)_{i=1}^n$, and $q_6=+\infty$. 
We employ a $\mathrm{Normal}(0,1000^2)$ prior for all regression coefficients
and a $\mathrm{Gamma}(0.01,0.01)$ prior on the standard deviation $\sigma$ of the residuals.

This model does not fit in the formalism of \cref{section:two_scenarios}, 
because the quintile indicators are computed using the entire set of $(X_i^T \theta_1)_{i=1}^n$,
whereas we assume that $f_2$ only depends on data from individual $i$. The model would fit if the indicators were computed using prespecified thresholds.
Furthermore, the second log-posterior density is discontinuous and piecewise constant in $\theta_1$ because of the discretization of $(X_i^T \theta_1)$ into quintiles. Thus, the Laplace approximation of \cref{section:nonasymptotic_control} is ill-defined in this setting. In effect, the matrix $\partial_{12} \bar{L}_2(\hat{\theta})$ appearing in \eqref{eq:j_n_simplified} would be equal to zero almost surely, which would result in the variance of the second stage parameter being estimated simply as $(-\partial_{22} \bar{L}_2(\hat{\theta}))^{-1}$, thus ignoring the uncertainty about $\theta_1$.

\begin{figure}[t]
  \centering
  \begin{subfigure}{0.48\textwidth}
    \includegraphics[width=\textwidth]{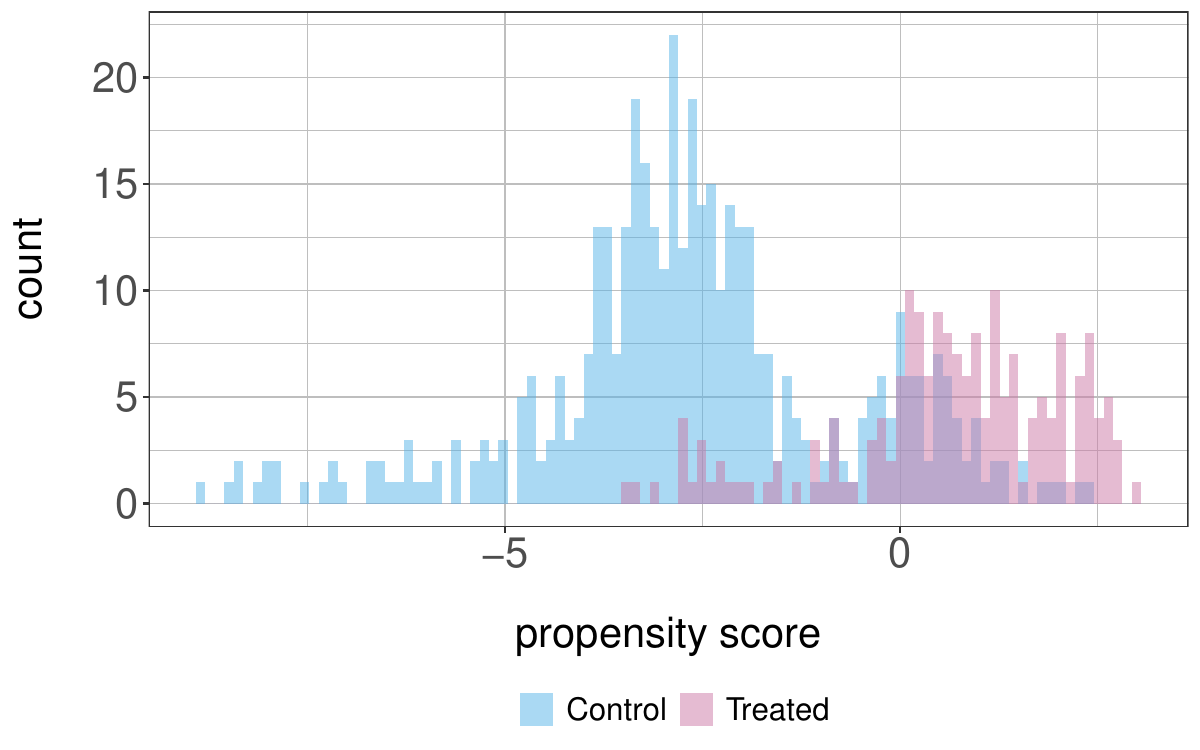}
    \caption{Estimated propensity scores.}
    \label{fig:causal:subfig21}
  \end{subfigure}
  \hfill
  \begin{subfigure}{0.48\textwidth}
    \includegraphics[width=\textwidth]{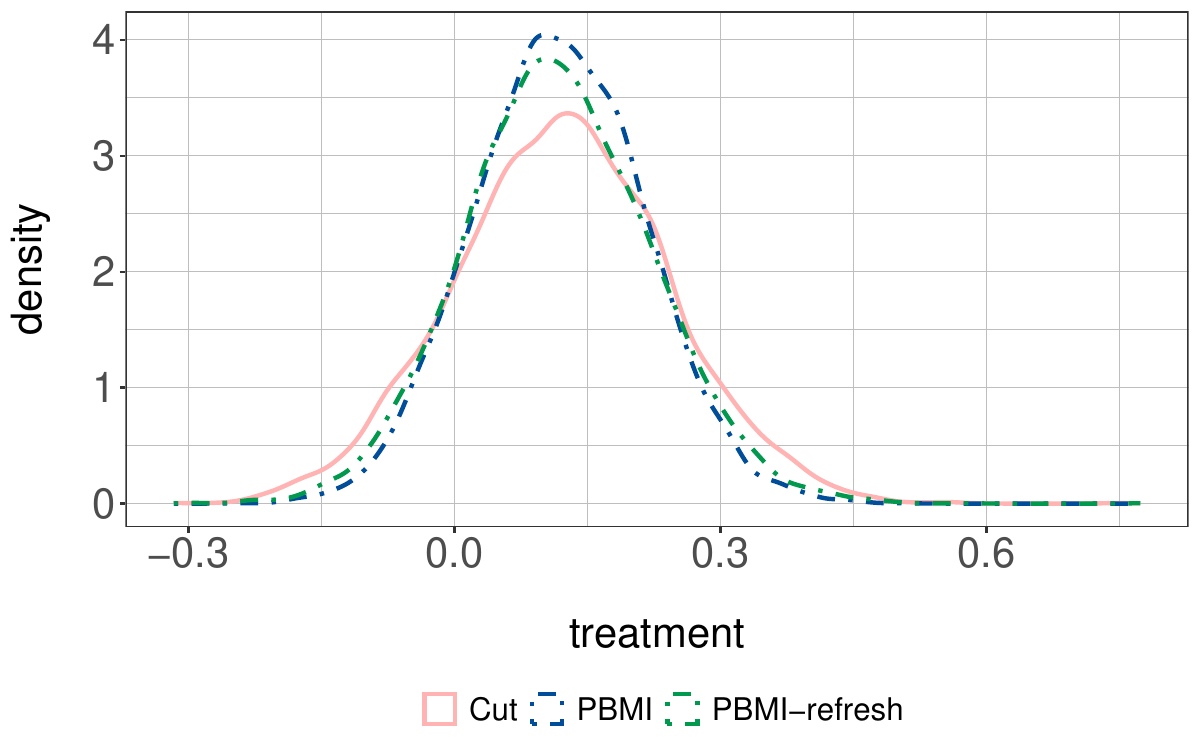}
    \caption{Coefficient for treatment variable.}
    \label{fig:causal:subfig22}
  \end{subfigure}
  \caption{\small Left: propensity scores $(X_i^T \theta_1)$ where $\theta_1$ is set to the MAP in the first module. Right: density plot of $\beta_{Z}$, the regression coefficient associated with the effect
  of treatment, drawn from the cut
  posterior and PBMI with and without weight refreshment, for model \eqref{model:linear_causal_v2}.}
  \label{fig:lalonde_example_results_v2}
\end{figure}

Figure~\ref{fig:causal:subfig21} shows the estimated propensity scores $(X_i^T \theta_1)$ when setting $\theta_1$ as the MAP estimate in the first module,
and colored by treatment group.
We compare the distribution of $\beta_Z$ obtained using the cut distribution and PBMI (Algorithm~\ref{alg:pb_cut_dependent}), with and without redrawing the weights between the two stages. 
In PBMI we set $w_0 = v_0 =0$. The
distributions of $\beta_Z$ are shown in Figure
\ref{fig:causal:subfig22}. We notice that PBMI yields slightly narrower credible regions around the mean, but otherwise provides similar inference to the cut posterior.
Whether to redraw the weights or not between the two stages has little impact on PBMI in this example. Note that PBMI provides a viable alternative to nested MCMC here,
irrespective of the discontinuity of the second module log-likelihood with respect to $\theta_1$.

\subsection{Epidemiological study\label{section:epidemiological_study_v2}}

This example
was used in \citet{plummer2015cuts}
and concerns the relation between human papillomavirus (HPV) prevalence and cervical cancer incidence.
Consider two modules, based on two different data
sources.  The high-risk HPV prevalence dataset was obtained via a series of
surveys carried out in 13 different countries.  Incidence data come from cancer
registries in the same populations.   In each population $i$, for $i = 1,
\ldots, 13$, we let $Z_i$ be the number of women infected with high-risk HPV in
a sample of size $n_{1,i}$ from the same population. Moreover, the outcome
$Y_i$ denotes the number of cancer cases arising from $T_i$ woman-years of
follow-up. We consider the following model.
\begin{align}\label{eq:model_epidemiology_v2}
  \begin{split}
    Z_i \sim &\; \text{Binomial}(n_{1,i}, \theta_{1,i}),\\
    \log \mu_i =&  \;\theta_{2,1} + \theta_{2,2} \theta_{1,i} + T_i, \quad
    Y_i \sim \; \text{Poisson}(\mu_i).
  \end{split}
\end{align}
We set independent priors $\theta_{1,i} \sim \text{Beta}(1,1)$ for $i =1,
\ldots, 13$. The prior for $\theta_2$ is $\theta_{2,1} \sim \mathrm{Normal}(0, 10^3)$,
$\theta_{2,2} \sim \mathrm{Normal}(0, 10^3)$. As mentioned by \cite{plummer2015cuts}, the
log-linear relationship $\log \mu_i =  \theta_{2,1} + \theta_{2,2} \theta_{1,i}
+ T_i$ is speculative, which suggests the misspecification of the second
module. 

\begin{figure}
  \centering
  \begin{subfigure}{0.48\textwidth}
    \includegraphics[width=\textwidth]{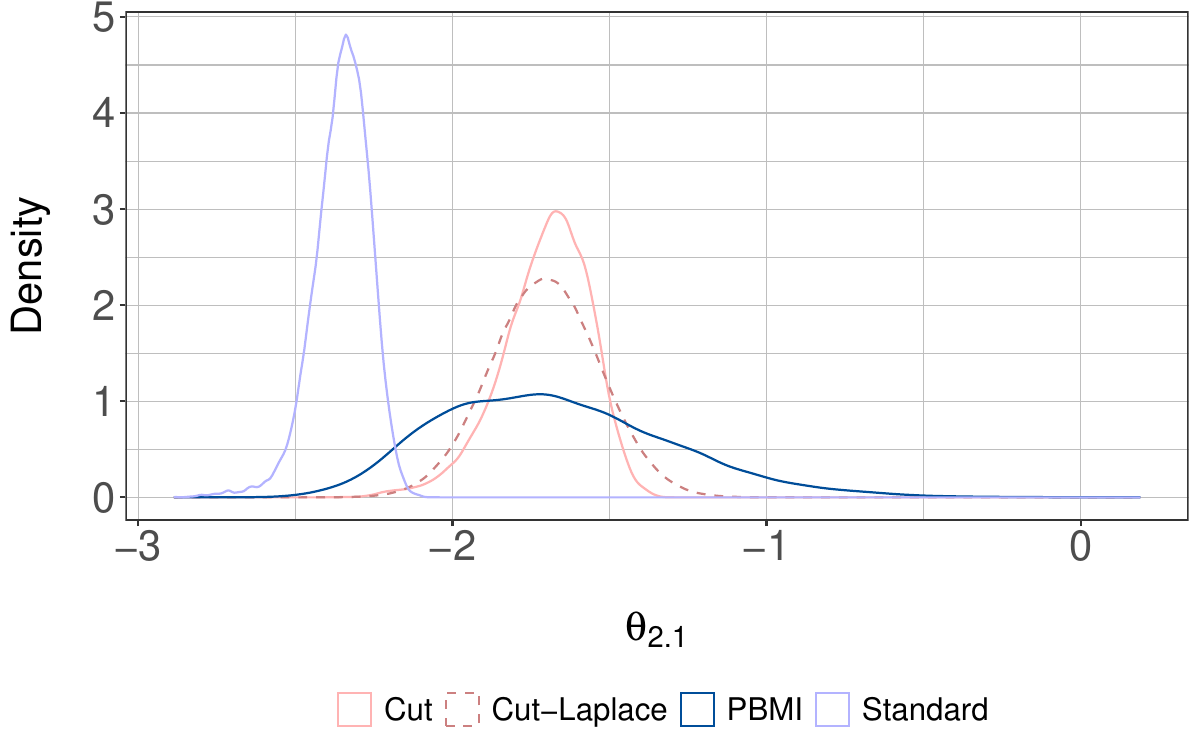}
    \caption{}
    \label{fig:epidemio:posterior_marginal1}
  \end{subfigure}
  \hfill
  \begin{subfigure}{0.48\textwidth}
    \includegraphics[width=\textwidth]{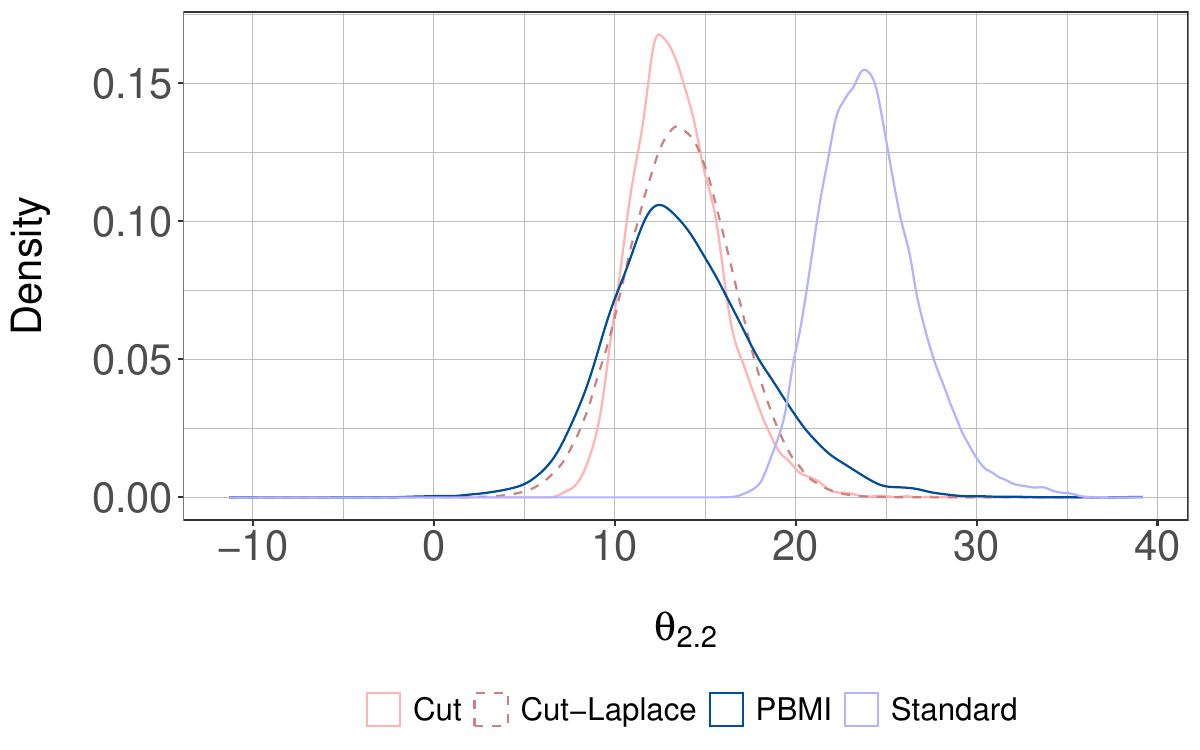}
    \caption{}
    \label{fig:epidemio:posterior_marginal2}
  \end{subfigure}
  \caption{\small Standard Posterior distribution, cut posterior, Cut-Laplace and PBMI, for $\theta_{2,1}$ (left) and $\theta_{2,2}$ (right) in the epidemiological example of \cref{section:epidemiological_study_v2}.}
  \label{fig:epidemio:posteriors}
\end{figure}

The small number of countries ($n_2 =13$) makes it unlikely that the estimators have reached
their asymptotic regime. Furthermore, the MAP estimate in the first module are located on the boundary of the parameter space, which is $[0,1]^{13}$,
 due to $Z_i$ being zero for one country.
Thus, we compute the Laplace approximation using the posterior mean in the first module, instead of the MAP.
The Cut-Laplace approximation appears reasonably accurate, yet \cref{fig:epidemio:posteriors} shows that the cut posterior is skewed for both $\theta_{2,1}$ and $\theta_{2,2}$, which is not captured by the Normal approximation.

The PBMI distribution in \cref{fig:epidemio:posteriors}
corresponds to an algorithm where the first stage optimization is done exactly (with $w_0=0$), and new random weights are drawn for the second stage, in which we set $v_0=1$.
The credible regions for the cut distribution are smaller than the credible regions under PBMI in this example.
Skewness is also visible in the PBMI marginal distributions.
We observe again that the standard posterior
 puts its mass on a different region compared to the two-stage approaches.

%% file: toy_example_v2_coverage_scenario1.tex
\begin{tabular}{lrr}
\toprule
Parameter & Cut-Laplace & PBMI\\
\midrule
$\theta_1$ & 0.95 & 0.95\\
$\theta_2$ & 0.93 & 0.95\\
$\theta_1 + \theta_2$ & 0.93 & 0.94\\
\bottomrule
\end{tabular}

%% file: toy_example_v2_coverage_scenario2.tex
\begin{tabular}{lrr}
\toprule
Parameter & Cut-Laplace & PBMI\\
\midrule
$\theta_1$ & 0.95 & 0.95\\
$\theta_2$ & 0.90 & 0.94\\
$\theta_1 + \theta_2$ & 0.84 & 0.94\\
\bottomrule
\end{tabular}